\documentclass[12pt]{article}

\usepackage{amsmath,amssymb,amsthm}
\usepackage{graphicx}
\usepackage[top=0.99in, bottom=0.99in, left=0.99in, right=0.99in]{geometry}
\usepackage{algorithm}
\usepackage{algpseudocode}
\usepackage{fmtcount}
\usepackage{setspace}
\usepackage[pagewise]{lineno}
\usepackage[dvipsnames,usenames]{color}
\usepackage[normalem]{ulem}
\usepackage[comma,authoryear]{natbib}
\usepackage{tabularx}
\usepackage{subfigure}
\usepackage{rotating}
\usepackage{enumitem}

\usepackage{multibib}
\newcites{sec}{References}

\makeatletter
\renewcommand*\env@matrix[1][c]{\hskip -\arraycolsep
  \let\@ifnextchar\new@ifnextchar
  \array{*\c@MaxMatrixCols #1}}
\makeatother

\newtheorem{thm}{Theorem}
\newtheorem{lemma}{Lemma}
\theoremstyle{remark} 
\newtheorem{rem}{Remark}

\newcommand{\ones}{\mathbf 1}
\DeclareMathOperator{\est}{est}

\DeclareMathOperator*{\argmin}{arg\,min}

\DeclareMathOperator*{\sign}{sign}

\DeclareMathOperator*{\tr}{tr}
\newcommand{\cur}{\mbox{\scriptsize\,current}}
\newcommand{\old}{\mbox{\scriptsize\,old}}

\newcommand{\spc}{\mbox{\scriptsize spc}}
\newcommand{\sym}{\mbox{\scriptsize sym}}
\newcommand{\con}{\mbox{\scriptsize con}}
\newcommand{\spl}{\mbox{\scriptsize spl}}
\DeclareMathOperator{\uni}{uni}

\newcommand{\tagarray}{%
  \mbox{}\refstepcounter{equation}%
  $(\theequation)$%
}

\allowdisplaybreaks

\title{A convex pseudo-likelihood framework for high
  dimensional partial correlation estimation with convergence
  guarantees}

\author{ Kshitij Khare, \emph{University of Florida, USA}\\
  Sang-Yun Oh, \emph{Stanford University, USA}\\
  Bala Rajaratnam, \emph{Stanford University, USA}} \date{}

\begin{document}

\pagenumbering{gobble}
\vspace*{-70pt}
    {\let\newpage\relax\maketitle}

\begin{abstract}
  \onehalfspacing

  Sparse high dimensional graphical model selection is a topic of much
  interest in modern day statistics. A popular approach is to apply
  $\ell_1$-penalties to either (1) parametric likelihoods, or, (2)
  regularized regression/pseudo-likelihoods, with the latter having
  the distinct advantage that they do not explicitly assume
  Gaussianity. As none of the popular methods proposed for solving
  pseudo-likelihood based objective functions have provable
  convergence guarantees, it is not clear if corresponding estimators
  exist or are even computable, or if they actually yield correct
  partial correlation graphs. This paper proposes a new
  pseudo-likelihood based graphical model selection method that aims
  to overcome some of the shortcomings of current methods, but at the
  same time retain all their respective strengths. In particular, we
  introduce a novel framework that leads to a convex formulation of
  the partial covariance regression graph problem, resulting in an
  objective function comprised of quadratic forms. The objective is
  then optimized via a coordinatewise approach. The specific
  functional form of the objective function facilitates rigorous
  convergence analysis leading to convergence guarantees; an important
  property that cannot be established using standard results, when the
  dimension is larger than the sample size, as is often the case in
  high dimensional applications. These convergence guarantees ensure
  that estimators are well-defined under very general conditions, and
  are always computable. In addition, the approach yields estimators
  that have good large sample properties and also respect
  symmetry. Furthermore, application to simulated/real data, timing
  comparisons and numerical convergence is demonstrated. We also
  present a novel unifying framework that places all graphical
  pseudo-likelihood methods as special cases of a more general
  formulation, leading to important insights.

  \vspace{0.5cm}
  
\end{abstract}

\newpage

\pagenumbering{arabic}

\section{Introduction}

\noindent
One of the hallmarks of modern day statistics is the advent of
high-dimensional datasets arising particularly from applications in
the biological sciences, environmental sciences and finance. A central
quantity of interest in such applications is the covariance matrix
$\Sigma$ of high dimensional random vectors. It is well known that the
sample covariance matrix $\mathbf{S}$ can be a poor estimator of
$\Sigma$, especially when $p/n$ is large, where $n$ is the sample size
and $p$ is the number of variables in the dataset. Hence $\mathbf{S}$
is not a useful estimator for $\Sigma$ for high-dimensional datasets,
where often either $p \gg n$ (``large $p$, small $n$'') or when $p$ is
comparable to $n$ and both are large (``large $p$, large $n$''). The
basic problem here is that the number of parameters in $\Sigma$ is of
the order $p^2$. Hence in the settings mentioned above, the sample
size is often not large enough to obtain a good estimator.

For many real life applications, the quantity of interest is the inverse 
covariance/partial covariance matrix $\Omega = \Sigma^{-1}$. In 
such situations, it is often reasonable to assume that there
are only a few significant partial correlations and the other partial
correlations are negligible in comparison. In mathematical terms, this
amounts to making the assumption that the inverse covariance matrix $\Omega =
\Sigma^{-1} = ((\omega_{ij}))_{1 \leq i, j \leq p}$ is sparse, i.e.,
many entries in $\Omega$ are zero. Note that $\omega_{ij} = 0$ is
equivalent to saying that the partial correlation between the $i^{th}$
and $j^{th}$ variables is zero (under Gaussianity, this 
reduces to the statement that the $i^{th}$ and $j^{th}$ variables are
conditionally independent given the other variables). The zeros in 
$\Omega$ can be conveniently represented by partial
correlation graphs. The assumption of a sparse graph is often deemed
very reasonable in applications. For example, as \cite{Peng2009} point
out, among 26 examples of published networks compiled by
\cite{Newman2003}, 24 networks had edge density less than 4\%.

A number of methods have been proposed for identifying sparse partial
correlation graphs in the penalized likelihood and penalized
regression based framework
\citep{Meinshausen2006,Friedman2008,Peng2009,Friedman2010}. The main
focus here is estimation of the sparsity pattern. Many of these
methods do not necessarily yield positive definite estimates of
$\Omega$. However, once a sparsity pattern is established, a positive
definite estimate can be easily obtained using efficient methods (see
\cite{HastieESL,Speed1986}).

The penalized likelihood approach induces sparsity by minimizing the
(negative) log-likelihood function with an $\ell_1$ penalty on the
elements of $\Omega$. In the Gaussian setup, this approach was pursued
by \cite{Banerjee2008} and others. \cite{Friedman2008} proposed the
\emph{graphical lasso} (``Glasso'') algorithm for the above
minimization problem, and is substantially faster than earlier
methods. In recent years, many interesting and useful methods have
been proposed for speeding up the performance of the graphical lasso
algorithm (see \cite{Mazumder2012} for instance). It is worth noting
that for these methods to provide substantial improvements over the
graphical lasso, certain assumptions are required on the number and
size of the connected components of the graph implied by the zeros in
$\hat{\Omega}$ (the minimizer).

Another useful approach introduced by \cite{Meinshausen2006} estimates
the zeros in $\Omega$ by fitting separate lasso regressions for each
variable given the other variables. These individual lasso fits give
neighborhoods that link each variable to others. \cite{Peng2009}
improve this neighborhood selection (NS) method by taking the natural
symmetry in the problem into account (i.e., $\Omega_{ij} =
\Omega_{ji}$), as not doing so could result in less efficiency and
contradictory neighborhoods.

In particular, the SPACE (Sparse PArtial Correlation Estimation)
method was proposed by \cite{Peng2009} as an effective alternative to
existing methods for sparse estimation of $\Omega$. The SPACE
procedure iterates between (1) updating partial correlations by a
joint lasso regression and (2) separately updating the partial
variances. As indicated above, it also accounts for the symmetry in
$\Omega$ and is computationally efficient. \cite{Peng2009} show that
under suitable regularity conditions, SPACE yields consistent
estimators in high dimensional settings. All the above properties make
SPACE an attractive regression based approach for estimating sparse
partial correlation graphs. In the examples presented in
\cite{Peng2009}, the authors find that empirically the SPACE algorithm
seems to converge really fast. It is however not clear if SPACE will
converge in general. Convergence is of course critical so that the
corresponding estimator is always guaranteed to exist and is therefore
meaningful, both computationally and statistically. In fact, as we
illustrate in Section \ref{spacecnvgc}, the SPACE algorithm might fail
to converge in simple cases, for both the standard choices of weights
suggested in \cite{Peng2009}. Motivated by SPACE, \cite{Friedman2010}
present a coordinate-wise descent approach (the ``Symmetric lasso''),
which may be considered as a symmetrized version of the approach in
\cite{Meinshausen2006}. As we show in Section \ref{sect:symlasso}, it
is also not clear if the Symmetric lasso will converge.

In this paper, we present a new method called the CONvex CORrelation
selection methoD (CONCORD) algorithm for sparse estimation of
$\Omega$. The algorithm obtains estimates of $\Omega$ by minimizing an
objective function, which is jointly convex, but more importantly
comprised of quadratic forms in the entries of $\Omega$. The
subsequent minimization is performed via coordinate-wise descent. The
convexity is strict if $n \geq p$, in which case standard results
guarantee the convergence of the coordinate-wise descent algorithm to
the unique global minimum.  If $n < p$, the objective function may not
be strictly convex. As a result, a unique global minimum may not
exist, and existing theory does not guarantee convergence of the
sequence of iterates of the coordinate-wise descent algorithm to a
global minimum. In Section \ref{cnvgcproof}, by exploiting the
quadratic forms present in the objective, it is rigorously
demonstrated that the sequence of iterates does indeed converge
to a global minimum of the objective function regardless of the
dimension of the problem. Furthermore, it is shown in Section
\ref{sect:largesample} that the CONCORD estimators are asymptotically
consistent in high dimensional settings under regularity assumptions
identical to \cite{Peng2009}.  Hence, our method preserves all the
attractive properties of SPACE, while also providing a theoretical
guarantee of convergence to a global minimum. In the process CONCORD
yields an estimator $\hat\Omega$ that is well-defined and is always
computable.  The strengths of CONCORD are further illustrated in the
simulations and real data analysis presented in Section
\ref{simulation}. A comparison of the relevant properties of different
estimators proposed in the literature is provided in Table
\ref{tbl:algcomparison} (Neighborhood selection (NS) by
\cite{Meinshausen2006}, SPACE by \cite{Peng2009}, Symmetric lasso
(SYMLASSO) by \cite{Friedman2010}, SPLICE by \cite{Rocha2008} and
CONCORD). The table shows that the CONCORD algorithm preserves all the
attractive properties of existing algorithms, while also providing
rigorous convergence guarantees. Another major contribution of the
paper is the development of a unifying framework that renders the
different pseudo-likelihood based graphical model selection procedures
as special cases. This general formulation facilitates a direct
comparison between the above pseudo-likelihood based methods and gives
deep insights into their respective strengths and weaknesses.

\begin{table}
  \centering
  \begin{tabular}{|c|c|c|c|c|c|}\cline{2-6}
    \multicolumn{1}{c|}{} & \multicolumn{5}{c|}{\bf METHOD} \\ \hline
    {\bf Property}
    & \begin{sideways} NS \end{sideways}
    & \begin{sideways} SPACE \end{sideways}
    & \begin{sideways} SYMLASSO\,\,\end{sideways}
    & \begin{sideways} SPLICE \end{sideways}
    & \begin{sideways} CONCORD \end{sideways}
    \\\hline
    \multicolumn{1}{|l|}{Symmetry} & & + & + & + & + \\\hline
    \multicolumn{1}{|l|}{Convergence guarantee (fixed $n$)} & N/A & &
    & & + \\\hline
    \multicolumn{1}{|l|}{Asymptotic consistency ($n,p\rightarrow\infty$)} 
    & + & + & & & + \\\hline
  \end{tabular}
  \caption{Comparison of regression based graphical model selection methods. A ``+" 
    indicates that a specified method has the given property. A blank space indicates
    the absence of a property. ``N/A" stands for not applicable.}
  \label{tbl:algcomparison}
\end{table}

The remainder of the paper is organized as follows. Section
\ref{spacecnvgc} briefly describes the SPACE algorithm and presents
examples where it fails to converge. This section motivates our work
and also analyzes other regression-based or pseudo-likelihood methods
that have been proposed.  Section \ref{qalgorithm} introduces the
CONCORD method and presents a general framework that unifies recently
proposed pseudo-likelihood methods. Section \ref{cnvgcproof}
establishes convergence of CONCORD to a global minimum, even if $n <
p$. Section \ref{simulation} illustrates the performance of the
CONCORD procedure on simulated and real data. Comparisons to SPACE and
Glasso are provided. When applied to gene expression data, the results
given by CONCORD are validated in a significant way by a recent
extensive breast cancer study.  Section \ref{sect:largesample}
establishes large sample properties of the CONCORD
approach. Concluding remarks are given in Section
\ref{sect:conclusion}. The supplemental document contains proofs of
some of the results in the paper.

\section{The SPACE algorithm and convergence properties}
\label{spacecnvgc}

\noindent
Let the random vector ${\bf Y}^k = \left( y_1^k, y_2^k, \cdots, y_p^k
\right)'$, $k = 1,2, \cdots, n$ denote \emph{i.i.d.} observations
from a multivariate distribution with mean vector ${\bf 0}$ and
covariance matrix $\Sigma$. Let $\Omega = \Sigma^{-1} =
((\omega_{ij}))_{1 \leq i,j \leq p}$ denote the inverse covariance
matrix, and let $\boldsymbol{\rho} = (\rho^{ij})_{1 \leq i < j \leq
  p}$ where
$\rho^{ij}=-\frac{\omega_{ij}}{\sqrt{\omega_{ii}\omega_{jj}}}$ denotes
the partial correlation between the $i^{th}$ and $j^{th}$ variable for
$1\leq i \neq j \leq p$. Note that $\rho^{ij}=\rho^{ji}$ for $i\neq
j$. Denote the sample covariance matrix by $\mathbf{S}$, and the sample
corresponding to the $i^{th}$ variable by ${\bf Y}_i = (y_i^1, y_i^2,
\cdots, y_i^n)'$.

\subsection{The SPACE algorithm}

\cite{Peng2009} propose the following novel iterative algorithm to
estimate the partial correlations $\{\rho^{ij}\}_{1 \leq i < j \leq
  p}$ and the partial covariances $\{\omega_{ii}\}_{1 \leq i \leq p}$
corresponding to $\Omega$ (see Algorithm \ref{spcalgrthm}).

\subsection{Convergence Properties of SPACE} \label{sect:convergence} 

\noindent
From empirical studies, \cite{Peng2009} find that the SPACE algorithm
converges quickly. As mentioned in the introduction, it is not
immediately clear if convergence can be established theoretically. In
an effort to understand such properties, we now place the SPACE algorithm
in a useful optimization framework.

\begin{lemma}\label{lemma1}
  For the choice of weights, $w_i = \omega_{ii}$, the SPACE algorithm
  corresponds to an iterative partial minimization procedure (IPM)
  for the following objective function:
  \begin{align}
    Q_{\spc} (\Omega) &= \frac{1}{2}
    \sum_{i=1}^p \left( -n\log \omega_{ii} +
      \omega_{ii} \| {\bf Y}_i - \sum_{j \neq i} 
      \rho^{ij}\sqrt{\frac{\omega_{jj}}{\omega_{ii}}} {\bf Y}_j \|^2\right) + \lambda
    \sum_{1 \leq i < j \leq p} \left| \rho^{ij} \right|\nonumber\\
    &= \frac{1}{2}
    \sum_{i=1}^p -n\log \omega_{ii} + \frac{1}{2} 
      \omega_{ii} \| {\bf Y}_i + \sum_{j \neq i} 
      \frac{\omega_{ij}}{\omega_{ii}} {\bf Y}_j \|^2 + \lambda
    \sum_{1 \leq i < j \leq p} \left| \rho^{ij} \right|. \label{eq1}
  \end{align}
\end{lemma}

\begin{figure}[H]
  \centering
  \begin{minipage}[t]{0.9\textwidth}
    \alglanguage{pseudocode}
    \begin{algorithm}[H]
      \caption{(SPACE pseudocode)}\label{spcalgrthm}
      \begin{algorithmic}
        \State Input: Standardize data to have mean zero and standard deviation one
        \State Input: Fix maximum number of iterations: $r_{max}$
        \State Input: Fix initial estimate:
        ($\hat{\omega}_{ii}^{(0)}=1/s_{ii}$ as suggested)
        \State Input: Choose weights\footnote{\cite{Peng2009} suggest
          two natural choices of weights $w_i$: (1) uniform
          weights $w_i=1$ for all $i=1,2,\dots,p$ (ii) partial 
          variance weights $w_i=\omega_{ii}$.}: $w_i$
        ($w_i=\omega_{ii}$ or $w_i=1$)
        \State Set $r \gets 1$
        
        \Repeat
        \Statex\hspace{\algorithmicindent}{\tt \#\# Update partial correlations}
        \State Update $\hat{\boldsymbol{\rho}}^{(r)}$ by minimizing
        (with current estimates $\{\hat\omega_{ii}^{(r-1)}\}_{i=1}^p$ as fixed)
        \begin{align}
          \frac{1}{2}
          \sum_{i=1}^p \left( w_i \| {\bf Y}_i - \sum_{j \neq i} \rho^{ij}
            \sqrt{\frac{\hat\omega_{jj}^{(r-1)}}{\hat\omega_{ii}^{(r-1)}}} {\bf
              Y}_j \|_2^2\right) + \lambda \sum_{1 \leq i < j \leq p} \left| \rho^{ij}
          \right|\label{eq:spaceobjective}
        \end{align}

        \Statex
        \Statex\hspace{\algorithmicindent}{\tt \#\# Update conditional variances}
        \State Update $\{\omega_{ii}^{(r)}\}_{i=1}^p$ by computing (with fixed 
        $\hat{\rho}_{ij}^{(r-1)}$ and $\hat{\omega}_{ii}^{(r-1)}$ for
        all $i$ and $j$)
        \begin{align}
          \frac{1}{\hat\omega_{ii}^{(r)}} = \frac{1}{n}\|{\bf Y}_i - \sum_{j
            \neq i} (\hat\rho^{ij})^{(r-1)}
          \sqrt{\frac{\hat\omega_{jj}^{(r-1)}}{\hat\omega_{ii}^{(r-1)}}} {\bf
            Y}_j\|_2^2\label{eq:updatepvar}
        \end{align}
        \State for $i=1,\dots,p$.

        \Statex
        \State $r \gets r + 1$
        \State Update weights: $w_i$
        \Until{$r==r_{max}$}

        \State Return $(\hat{\boldsymbol\rho}^{(r_{max})},\{\hat\omega_{ii}^{(r_{max})}\}_{i=1}^p)$
      \end{algorithmic}
    \end{algorithm}
  \end{minipage}  
\end{figure}

\noindent {\it Proof}\,: Note that when fixing
the diagonals $\{\omega_{ii}\}_{i=1}^p$, the minimization in \eqref{eq:spaceobjective} 
 in the SPACE algorithm (with weights $w_i=\omega_{ii}$), corresponds to minimizing
$Q_{\spc}$ with respect to $\boldsymbol\rho$. Now, let $\hat\omega_{ii}$ be the
minimizer of $Q_{\spc}$ with respect to $\omega_{ii}$, fixing
$\{\beta_{ij}\}_{1\leq i\neq j\leq p}$ (where
$\beta_{ij}=\rho^{ij}\sqrt{\frac{\omega_{jj}}{\omega_{ii}}}=-\frac{\omega_{ij}}{\omega_{ii}}$). Then,
it follows that
\begin{align}
  \hat\omega_{ii} = \left(\frac{1}{n}\|{\bf Y}_i - \sum_{j
    \neq i} \beta_{ij} {\bf
    Y}_j\|_2^2\right)^{-1}\label{eq:omegaupdate}
\end{align}

\noindent
The result follows by comparing \eqref{eq:omegaupdate} with the
updates in \eqref{eq:updatepvar}. \hfill$\Box$

\noindent
Although Lemma \ref{lemma1} identifies SPACE as an IPM, existing
theory for iterative partial minimization (see for example
\cite{Zangwill1969}, \cite{Jensen1991}, \cite{Lauritzen1996}, etc)
only guarantees that every accumulation point of the sequence of
iterates is a stationary point of the objective function
$Q_{\spc}$. To establish convergence, one needs to prove that every
contour of the function $Q_{\spc}$ contains only finitely many
stationary points. It is not clear if this latter condition holds for
the function $Q_{\spc}$. Moreover, for choice of weights $w_i = 1$,
the SPACE algorithm does not appear to have an iterative partial
minimization interpretation.

To improve our understanding of the convergence properties of SPACE,
we started by testing the algorithm on simple examples. On some examples, 
SPACE converges very quickly; however, examples can be found where SPACE
does not converge when using the two possible choices for weights: partial
variance weights ($w_i = \omega_{ii}$) and uniform weights ($w_i =
1$). We now give an example of the lack of convergence. 


\noindent
{\it Example 1}: Consider the following population covariance and inverse 
covariance matrices:  
\begin{align}
  \Omega=\begin{pmatrix}[r]
    3.0 & 2.1 & 0.0 \\
    2.1 & 3.0 & 2.1 \\
    0.0 & 2.1 & 3.0 \\
  \end{pmatrix},\
  \Sigma=\Omega^{-1}= 
  \begin{pmatrix}[r]
      8.500 & -11.667 &   8.167 \\
    -11.667 &  16.667 & -11.667 \\
      8.167 & -11.667 &   8.500 \\
  \end{pmatrix}\label{eq:sigmadef}
\end{align}

\noindent
A sample of $n = 100$ \emph{i.i.d.} vectors was generated from the corresponding 
$\mathcal{N} ({\bf 0}, \Sigma)$ distribution. The data was
standardized and the SPACE algorithm was run with choice of weights
$w_i = \omega_{ii}$ and $\lambda = 160$. After the first few iterations 
successive SPACE iterates alternate between the 
following two matrices:
\begin{align}
  \begin{pmatrix}[r]
    29.009570 & 27.266460 & 0.000000 \\ 
    27.266460 & 51.863320 & 24.680140 \\ 
    0.000000 & 24.680140 & 26.359350 \\ 
  \end{pmatrix} 
  \mbox{ and } 
  \begin{pmatrix}[r] 
    28.340040 & 27.221520 & -0.705390 \\ 
    27.221520 & 54.255190 & 24.569900 \\ 
    -0.705390 & 24.569900 & 25.753040 \\ 
  \end{pmatrix},\label{eq:example 1 matrices}
\end{align}

\noindent
thereby establishing non-convergence of the SPACE algorithm in this
example (see also Figure \ref{fig:weightsW}). Note that the two 
matrices in \eqref{eq:example 1 matrices} have different sparsity 
patterns. A similar example of non-convergence of SPACE with 
uniform weights is provided in 
Supplemental Section \ref{sect:nonconvergence of space}. 


A natural question to ask is whether the non-convergence of SPACE is 
pathological or whether is it widespread in settings of interest. To this 
end, the following simulation study was undertaken. 

\smallskip

\noindent {\it Example 2}: We created a sparse $100 \times 100$ matrix $\Omega$ with 
edge density 4\% and a condition number of 100. A total of 100 multivariate Gaussian
    datasets (with $n=100$) having mean vector zero and covariance
    matrix $\Sigma=\Omega^{-1}$ were generated. Table \ref{table1}
    summarizes the number of times (out of 100) SPACE1
    (SPACE with uniform weights) and SPACE2 (SPACE with partial
    variance weights) do not converge within 1500 iterations.  When
    they do converge, the mean number of iterations are 22.3 for
    SPACE1 and 14.1 for SPACE2 (note that since the original implementation of SPACE
    by \cite{Peng2009} was programmed to stop after $3$ iterations,
    we modified the implementation to allow for more iterations in order to
    check for convergence of parameter estimates). It is clear from
    Table \ref{table1} that both variations of SPACE, using unit
    weights as well as $\omega_{ii}$ weights, exhibit extensive
    non-convergence behavior. Our simulations suggest that the
    convergence problem is exacerbated as the condition number of
    $\Omega$ increases.

    \begin{table}[H]
      \centering
      \begin{tabular}{||rrr||rrr||}
        \hline
        \multicolumn{3}{||c||}{SPACE1 ($w_i=1$)} 
        & \multicolumn{3}{c||}{SPACE2 ($w_i=\omega_{ii}$)}\\
        \multicolumn{1}{||c}{$\lambda^*$} & \multicolumn{1}{c}{NZ} & NC & 
        \multicolumn{1}{|c}{$\lambda^*$} & \multicolumn{1}{c}{NZ} & NC \\ 
        \hline
        0.026 & 60.9\% &  92 & 0.085 & 79.8\% & 100 \\ 
        0.099 & 19.7\% & 100 & 0.160 & 28.3\% &   0 \\ 
        0.163 &  7.6\% & 100 & 0.220 & 10.7\% &   0 \\ 
        0.228 &  2.9\% & 100 & 0.280 &  4.8\% &   0 \\ 
        0.614 &  0.4\% &   0 & 0.730 &  0.5\% &  97 \\ 
        \hline
      \end{tabular}
      \caption{Number of simulations (out of 100) that do not converge within 1500
      iterations (NC) for select values of penalty parameter ($\lambda^* = \lambda/n$). 
      Average percentage of non-zeros (NZ) in $\hat\Omega$ are also shown.} 
      \label{table1}
    \end{table}

\subsection{Symmetric lasso}\label{sect:symlasso}
The Symmetric lasso algorithm was proposed as a useful alternative to
SPACE in a recent work by \cite{Friedman2010}. Symmetric
lasso minimizes the following (negative) pseudo-likelihood:
\begin{align}
  Q_{\sym} (\boldsymbol{\alpha}, \breve\Omega) = \frac{1}{2}\sum_{i=1}^p
  \left[ n \log \alpha_{ii} + \frac{1}{\alpha_{ii}} \| {\bf Y}_i +
    \sum_{j \neq i} \omega_{ij}\alpha_{ii} {\bf Y}_j \|^2\right] +
  \lambda \sum_{1 \leq i < j \leq p} \left| \omega_{ij}
  \right|. \label{eq:symlassoobjective}
\end{align}
where $\alpha_{ii}=1/\omega_{ii}$. Here $\boldsymbol{\alpha}$ denotes
the vector with entries $\alpha_{ii}$ for $i=1,\dots,p$ and
$\breve\Omega$ denotes the matrix $\Omega$ with diagonal entries set
to zero. A comparison of \eqref{eq1} and \eqref{eq:symlassoobjective}
shows a deep connection between SPACE (with $w_i=\omega_{ii}$) and
Symmetric lasso objective functions. In particular, the $Q_{\sym}
(\boldsymbol{\alpha}, \breve\Omega)$ objective function in
\eqref{eq:symlassoobjective} is a reparameterization of \eqref{eq1}:
the only difference is that the $\ell_1$ penalty on the elements of
$\boldsymbol\rho$ is replaced by a penalty on the elements of $\Omega$
in \eqref{eq:symlassoobjective}. The minimization of the objective
function in \eqref{eq:symlassoobjective} is performed by
coordinate-wise descent on
$({\boldsymbol\alpha},\breve\Omega)$. Symmetric lasso is indeed a
useful and computationally efficient procedure. However, theoretical
properties such as convergence or asymptotic consistency have not yet
been established. The following lemma investigates the properties of
the objective function used in Symmetric lasso.

\begin{lemma} 
  \label{convexity of symlasso} 
  The Symmetric Lasso objective function in
  \eqref{eq:symlassoobjective} is a non-convex function of
  $(\boldsymbol{\alpha}, \breve\Omega)$.
\end{lemma}

The proof of Lemma 2 is given in Supplemental Section \ref{sect:proof
  convexity}. The arguments in the proof of Lemma 2 demonstrate that
the objective function used in Symmetric lasso is not convex, or even
bi-convex in the parameterization used above. However, it can be shown
that the SYMLASSO objective function is jointly convex in the elements
of $\Omega$ (see \cite{leehstmdml} and Supplemental section
\ref{sect:symlasso:convexity:omega}). It is straightforward to check
that the coordinatewise descent algorithms for both parameterizations
are exactly the same. However, unless a function is strictly convex,
there are no general theoretical guarantees of convergence for the
corresponding coordinatewise descent algorithm. Indeed, when $n < p$,
the SYMLASSO objective function is not strictly convex. Therefore, it
is not clear if the coordinate descent algorithm converges in
general. We conclude this section by remarking that both SPACE and
symmetric lasso are useful additions to the graphical model selection
literature, especially because they both respect symmetry and give
computationally fast procedures.

\subsection{The SPLICE algorithm} \label{sect:splice}

The SPLICE algorithm (Sparse Pseudo-Likelihood Inverse Covariance
Estimates) was proposed by \cite{Rocha2008} as an alternative means to
estimate $\Omega$. In particular, the SPLICE formulation uses an
$\ell_1$-penalized regression based pseudo-likelihood objective
function parameterized by matrices $\mathbf{D}$ and $\mathbf{B}$ where
$\Omega=\mathbf{D}^{-2}(\mathbf{I}-\mathbf{B})$. The diagonal matrix
$\mathbf{D}$ has elements $d_{jj}=1/\sqrt{\omega_{jj}}$,
$j=1,\dots,p$. The (asymmetric) matrix $\mathbf{B}$ has as columns the
vectors of regression coefficients, $\beta_j\in \mathbb{R}^p$. These
coefficients, $\beta_j$, arise when regressing ${\bf Y}_j$ on the
remaining variables. A constraint on each $\beta_{j}$ is imposed so
that regression of ${\bf Y}_j$ is performed without including itself
as a predictor variable: i.e., $\beta_{jj}=0$. Based on the above
properties, the $\ell_1$-penalized pseudo-likelihood objective
function of SPLICE algorithm (without the constant term) is given by
\begin{align}
  Q_{\spl}(\mathbf{B},\mathbf{D}) &= {n\over
    2}\sum_{i=1}^p\log(d_{ii}^2) + {1\over 2}\sum_{i=1}^p {1\over
    d_{ii}^2}\| {\bf Y}_i - \sum_{j\neq i}\beta_{ij}{\bf Y}_j\|^2 +
  \lambda \sum_{i<j}|\beta_{ij}|.
  \label{eq:spliceobjective}
\end{align}

In order to optimize \eqref{eq:spliceobjective} with respect to
$\mathbf{B}$ and $\mathbf{D}$, \cite{Rocha2008} also propose an
iterative algorithm that alternates between maximizing $\mathbf{B}$
fixing $\mathbf{D}$, followed by maximizing $\mathbf{D}$ fixing
$\mathbf{B}$. As with other regression-based graphical model selection
algorithms, a proof of convergence of SPLICE is not available. 
The following lemma gives the convexity properties of the SPLICE
objective function.

\begin{lemma} 
  \label{lemma:splice}
  i) The SPLICE objective function $Q_{\spl}(\mathbf{B},\mathbf{D})$
  is not jointly convex in $(\mathbf{B},\mathbf{D})$.

  
  \ \,\quad\qquad ii) Under the transformation
  $\mathbf{C}=\mathbf{D}^{-1}$, $Q_{\spl}(\mathbf{B},\mathbf{C})$ is
  bi-convex.
\end{lemma}

The proof of Lemma \ref{lemma:splice} is given in Supplemental Section
\ref{sect:proof splice}.  The convergence properties of the SPLICE
algorithm is not immediately clear since its objective function is
non-convex. Furthermore, it is not clear whether the SPLICE solution
yields a global optimum.

\section{CONCORD: A convex pseudo-likelihood framework for sparse
  partial covariance estimation} \label{qalgorithm}

\noindent
The two pseudo-likelihood based approaches, SPACE and Symmetric lasso,
have several attractive properties such as computational efficiency,
simplicity and use of symmetry. They also do not directly depend on
the more restrictive Gaussian assumption. Additionally,
\cite{Peng2009} also establish (under suitable regularity assumptions)
consistency of SPACE estimators for distributions with sub-Gaussian
tails. However, none of the existing pseudo-likelihood based
approaches yield a method that is provably convergent. In Section
\ref{sect:convergence}, we showed that there are instances where SPACE
does not converge. As explained earlier, convergence is critical as
this property guarantees well defined estimators which always exist,
and are computable regardless of the data at hand. An important
research objective therefore is the development of a pseudo-likelihood
framework which preserves all the attractive properties of SPACE and
SYMLASSO, and at the same time, leads to theoretical guarantees of
convergence.  It is however not clear immediately how to achieve this
goal. A natural approach to take is to develop a convex formulation of
the problem. Such an approach can yield many advantages, including 1)
Guarantee of existence of a global minimum, 2) Better chance of
convergence using convex optimization algorithms, 3) Deeper
theoretical analysis of the properties of the solution and
corresponding algorithm. As we have shown, the SPACE objective
function is not jointly convex in the elements of $\Omega$ (or any
natural reparameterization). Hence, one is not in a position to
leverage tools from convex optimization theory for understanding its
behavior. The SYMLASSO objective function is jointly convex in the
elements of $\Omega$. However, unless a function is strictly convex,
there are no general guarantees of convergence for the corresponding
coordinatewise descent algorithm. Indeed, when $n < p$, the SYMLASSO
objective function is not strictly convex, and it is not clear if the
corresponding coordinatewise descent algorithm converges.

In this section, we introduce a new approach for estimating $\Omega$,
called the CONvex CORrelation selection methoD (CONCORD) that aims to
achieve the above objective. The CONCORD algorithm constructs sparse
estimators of $\Omega$ by minimizing an objective function that is
jointly convex in the entries of $\Omega$. We start by introducing the
objective function for the CONCORD method and then proceed to derive
the details of the corresponding coordinate-wise descent
updates. Convergence is not obvious, as the function may not be
strictly convex if $n < p$. It is proved in Section \ref{cnvgcproof}
that the corresponding coordinate-wise descent algorithm does indeed
converge to a global minimum. Computational complexity and running
time comparisons for CONCORD are given in Sections
\ref{sect:complexity} and \ref{sect:simulated data},
respectively. Subsequently, large sample properties of the resulting
estimator are established in Section \ref{sect:largesample} in order
to provide asymptotic guarantees in the regime when both the dimension
$p$ and the sample size $n$ tend to infinity. Thereafter, the
performance of CONCORD on simulated data, and real data from
biomedical and financial applications is demonstrated. Such analysis
serves to establish that CONCORD preserves all the attractive
properties of existing pseudo-likelihood methods and additionally
provides the crucial theoretical guarantee of convergence and
existence of a well-defined solution.

\subsection{The CONCORD objective function}

In order to develop a convex formulation of the pseudo-likelihood
graphical model selection problem let us first revisit the formulation
of the SPACE objective function in \eqref{eq1}  with arbitrary
weights $w_i$ instead of $\omega_{ii}$. 

\begin{align}
  Q_{\spc} (\Omega) &= \frac{1}{2} \sum_{i=1}^p \left( -n\log
    \omega_{ii} + w_{i} \| {\bf Y}_i - \sum_{j \neq i}
    \rho^{ij}\sqrt{\frac{\omega_{jj}}{\omega_{ii}}} {\bf Y}_j
    \|_2^2\right) + \lambda \sum_{1 \leq i < j \leq p} \left| \omega^{ij}
  \right| \label{reveq1}
\end{align}

Now note that the above objective is not jointly convex in the
elements of $\Omega$ since, 1) The middle term for the regression with
the choices $w_i = 1$ or $w_i = \omega_{ii}$ is not a jointly convex
function of the elements of $\Omega$.  2) The penalty term is on the
partial correlations $ \rho^{ij} = -\frac{\omega_{ij}}{
  \sqrt{\omega_{ii} \omega_{jj}} }$ and is hence not a jointly convex
function of the elements of $\Omega$.

Now note the following for the regression term:
\begin{align*} 
  w_i\| {\bf Y}_i - \sum_{j \neq
    i}\rho^{ij}\sqrt{\frac{\omega_{jj}}{\omega_{ii}}}{\bf Y}_j \|_2^2
  &= w_{i}\| {\bf Y}_i + \sum_{j \neq i}
  \frac{\omega_{ij}}{\omega_{ii}} {\bf Y}_j \|_2^2\qquad
  \left(\because \rho^{ij}=\frac{-\omega_{ij}}{\sqrt{\omega_{ii}\omega_{jj}}}\right)\\
  &= w_i \|\frac{1}{\omega_{ii}}( \omega_{ii}{\bf Y}_i +
  \sum_{j \neq i} \omega_{ij} {\bf Y}_j) \|_2^2\\
  &= \frac{w_i}{\omega_{ii}^2} \| \sum_{j=1}^p \omega_{ij} {\bf Y}_j
  \|_2^2\\
  &= \frac{w_i}{\omega_{ii}^2} \left(\omega_{\bullet i}'{\bf Y}'{\bf
    Y}\omega_{\bullet i}\right)
\end{align*}
The choice of weights $w_i = \omega_{ii}^{2}$ yields
\begin{align}
  w_{i} \| {\bf Y}_i - \sum_{j \neq i} \rho^{ij}
  \sqrt{\frac{\omega_{jj}}{\omega_{ii}}} {\bf Y}_j \|_2^2 =
  \omega_{\bullet i}'{\bf Y}'{\bf Y}\omega_{\bullet i} \geq
  0\label{eq:quadratic}
\end{align}

The above expression in \eqref{eq:quadratic} is a quadratic form (and
hence jointly convex) in the elements of $\Omega$. Putting
the $\ell_1$-penalty term on the partial covariances $ \omega_{ij}$
instead of on the partial correlations $ \rho^{ij}$ yields the
following jointly convex objective function:

\begin{align} 
  Q_{\con} (\Omega) &=: \mathcal{L}_{\con} (\Omega) +
  \lambda \sum_{1 \leq i < j \leq p} |\omega_{ij}| \nonumber\\
  &=: -\sum_{i=1}^p n\log \omega_{ii} + \frac{1}{2} \sum_{i=1}^p \|
  \omega_{ii} {\bf Y}_i + \sum_{j \neq i} \omega_{ij} {\bf Y}_j \|_2^2
  + \lambda \sum_{1 \leq i < j \leq p} |\omega_{ij}|. \label{eq2}
\end{align}

\noindent

The function $\mathcal{L}_{\con}(\Omega)$ can be regarded as a
pseudo-likelihood function in the spirit of \cite{Besag1975}. Since
$-\log x$ and $|x|$ are convex functions, and $\sum_{i=1}^p \|
\omega_{ii} {\bf Y}_i + \sum_{j \neq i} \omega_{ij} {\bf Y}_j \|^2$ is
a positive semi-definite quadratic form in $\Omega$, it follows that
$Q_{\con} (\Omega)$ is a jointly convex function of $\Omega$ (but not
necessarily strictly convex). As we shall see later, this particular
formulation above helps us establish theoretical guarantees of
convergence (see Section \ref{cnvgcproof}), and, consequently, yields
a regression based graphical model estimator that is well defined and
is always computable.  Note that the $n/2$ in \eqref{reveq1} has
been replaced by $n$ in \eqref{eq2}. The point is elaborated further
in Remark \ref{rem:corr}.  We now proceed to derive the details
of the coordinate-wise descent algorithm for minimizing $Q_{\con}
(\Omega)$. 

\subsection{A coordinatewise minimization algorithm for minimizing $Q_{\con} 
(\Omega)$}

\noindent
Let $\mathcal{A}_p$ denote the set of $p \times p$ real symmetric
matrices. Let the parameter space $\mathcal{M}$ be defined as
$$
\mathcal{M} := \{\Omega \in \mathcal{A}_p: \; \; \omega_{ii} > 0
\mbox{, for every } 1 
\leq i \leq p\}. 
$$

\noindent
Note that as in other regression based approaches (see \cite{Peng2009}), we 
have deliberately not restricted $\Omega$ to be positive definite as the 
main goal is to estimate the sparsity pattern in $\Omega$. As mentioned 
in the introduction, a positive definite estimator can be obtained by using 
standard methods (\cite{HastieESL}, \cite{Xu2011}) once a partial correlation 
graph has been determined. 

Let us now proceed to optimizing $Q_{\con}(\Omega)$. For $1 \leq i
\leq j \leq p$, define the function $T_{ij}: \mathcal{M} \rightarrow
\mathcal{M}$ by
\begin{align}
T_{ij} (\Omega) = \argmin_{\{\tilde\Omega: (\tilde{\Omega})_{kl}
= \omega_{kl} \; \forall (k,l) \neq (i,j)\}} Q_{\con}
(\tilde\Omega).\label{eq:ipm section}  
\end{align}

For each $(i,j)$, $T_{ij}(\Omega)$ gives the matrix where all the
elements of $\Omega$ are left as is except the $(i,j)^{th}$ element. The
$(i,j)^{th}$ element is replaced by the value that minimizes
$Q_{\con}(\Omega)$ with respect to $\omega_{ij}$ holding all other
variables $\omega_{kl},\; (k,l)\neq (i,j)$ constant. We now proceed to
evaluate $T_{ij}(\Omega)$ explicitly.

\begin{lemma}\label{ipm}
  The function $T_{ij}(\Omega)$ defined in \eqref{eq:ipm section} can
  be computed in closed form. In particular, for $1\leq i \leq p$,
\begin{align}
  \left( T_{ii} (\Omega) \right)_{ii} = \frac{- \sum_{j \neq i} \omega_{ij} s_{ij} + 
    \sqrt{\left( \sum_{j \neq i} \omega_{ij} s_{ij} \right)^2 + 4
      s_{ii}}}{2 s_{ii}}.\label{eq:concord diag update}
\end{align}
\noindent
For $1\leq i < j \leq p$,
\begin{align}
\left( T_{ij} (\Omega) \right)_{ij} =  \frac{S_{\lambda\over n}\left(-\left(\sum_{j' \neq j} 
\omega_{ij'} s_{jj'} + \sum_{i' \neq i} \omega_{i'j}
s_{ii'}\right)\right)} {s_{ii} + s_{jj}},\label{eq:concord offdiag update} 
\end{align}

\noindent
where $s_{ij}$ is the $(i,j)^{th}$ entry of $\frac{1}{n} {\bf Y}^T {\bf Y}$, and 
$S_{\lambda} (x) := \sign(x) (|x| - \lambda)_+$. 
\end{lemma}
The proof is given in Supplemental Section \ref{sect:proof ipm}. An
important contribution of Lemma \ref{ipm} is that it gives the
necessary ingredients for designing a coordinate descent approach to
minimizing the CONCORD objective function. More specifically,
\eqref{eq:concord diag update} can be used to update the partial
variance terms, and \eqref{eq:concord offdiag update} can be used to
update the partial covariance terms. The coordinate-wise descent
algorithm for CONCORD is summarized in Algorithm \ref{altalgrthm}. The
zeros in the estimated partial covariance matrix can then subsequently
be used to construct a partial covariance or partial correlation
graph. 

\begin{figure}
  \centering
  \begin{minipage}[t]{0.8\textwidth}
    \alglanguage{pseudocode}
    \begin{algorithm}[H]
      \caption{(CONCORD pseudocode)}\label{altalgrthm}
      \begin{algorithmic}
        \State Input: standardize data to have mean zero and standard deviation one
        \State Input: Fix maximum number of iterations: $r_{max}$
        \State Input: Fix initial estimate: $\hat{\Omega}^{(0)}$
        \State Input: Fix convergence threshold: $\epsilon$
        \State Set $r \gets 1$
        \State converged = FALSE
        \State Set $\hat{\Omega}^{\cur} \gets \hat{\Omega}^{(0)}$
        \Repeat
        
        \State $\hat\Omega^{\old} \gets \hat\Omega^{\cur}$
        \Statex\hspace{\algorithmicindent}{\tt \#\# Updates to partial
          covariances $\omega_{ij}$}    
        \For{$i \gets 1,2, \cdots, p-1$}
        \For{$j \gets i+1, \cdots, p$}
        \begin{align}
          \hat\omega^{\cur}_{ij} \gets (T_{ij}(\Omega^{\cur}))_{ij}
          \label{eq:update offdiag}
        \end{align}

        \EndFor
        \EndFor
        \Statex
        \Statex\hspace{\algorithmicindent}{\tt \#\# Updates to partial variances $\omega_{ii}$}
        \For{$i \gets 1,2,\cdots,p$}        
        \begin{align}
          \hat\omega^{\cur}_{ii} \gets (T_{ii}(\Omega^{\cur}))_{ii}
          \label{eq:update diag}
        \end{align}

        \EndFor
        \Statex
        \State $\hat{\Omega}^{(r)} \gets \hat{\Omega}^{\cur}$
        \Statex\hspace{\algorithmicindent}{\tt \#\# Convergence checking}
        \If{$\|\hat{\Omega}^{\cur} - \hat{\Omega}^{\old}\|_{\mbox{\scriptsize max}} < \epsilon$}
        \State converged = TRUE
        \Else
        \State $r \gets r + 1$
        \EndIf
        \Statex
        \Until{converged = TRUE or $r > r_{\max}$}
        \State Return final estimate: $\hat{\Omega}^{(r)}$ 
      \end{algorithmic}
    \end{algorithm}
  \end{minipage}  
\end{figure}

The following procedure can be used to select the penalty parameter $\lambda$. 
Define the residual sum of squares (RSS) for
  $i=1,\dots,p$ as
  $$ RSS_i(\lambda) = \sum_{k=1}^n \left(y_i^k - \sum_{j\neq i}
    \frac{\omega_{ij}}{\omega_{ii}}y_j^k \right)^2. $$ 
  Further, the $i$-th
  component of BIC type score can be defined as
  $$ BIC_i(\lambda) = n\log(RSS_i(\lambda)) + \log n\cdot |\{j: j\neq i,
  \omega_{ij,\lambda}\neq 0\}|. $$ The penalty parameter $\lambda$ can be
  chosen to minimize the sum $BIC(\lambda) = \sum_{i=1}^p
  BIC_i(\lambda)$.

\subsection{Computational complexity}\label{sect:complexity}

\noindent
We now proceed to show that the computational cost of each iteration
of CONCORD is $min\left(O(np^2),O(p^3)\right)$, that is, the CONCORD
algorithm is competitive with other proposed methods. The updates in
Equations in \eqref{eq:update offdiag} and \eqref{eq:update diag} are
implemented differently depending on whether $n\geq p$ or $n<p$.

\noindent
{\bf Case 1 ($n \geq p$)}: Let us first consider the case
when $n\geq p$. Note that both sums in \eqref{eq:concord offdiag
  update} are inner products between a row in $\hat\Omega$ and a row
in $\mathbf{S}$. Clearly, computing these sums require $O(p)$ operations
each. Similarly, the update in \eqref{eq:concord diag update} requires
$O(p)$ operations. Since there are $O(p^2)$ entries in $\Omega$, one
complete sweep of updates over all entries in $\hat\Omega$ would
require $O(p^3)$ operations.

\noindent
{\bf Case 2 ($n < p$)}: Let us now consider the case when
$n<p$. We show below that the updates can be performed in $O(np^2)$
operations. The main idea here is that the coordinate-wise
calculations at each iteration, which involves an inner product of two
$p\times 1$ vectors, can be reduced to an inner product calculation
involving auxiliary variables (residual variables to be more specific)
of dimension $n\times 1$.  The following lemmas are essential
ingredients in calculating the computational complexity in this
setting. In particular, Lemma \ref{eq:residuals} expresses the inner
product calculations in \eqref{eq:concord diag update} and
\eqref{eq:concord offdiag update} in terms of residual vectors.
\begin{lemma}\label{eq:residuals}
  For $1 \leq i,j \leq p$,
  \begin{align*}
    \sum_{k \neq j} \omega_{ik} s_{jk} &= -\omega_{ij} s_{jj} +
    \omega_{ii}{\bf Y}_j'{\bf r}_i,
  \end{align*}  
  where ${\bf Y}_j$ is the $j^{th}$ column of the data matrix ${\bf Y}$,
  and ${\bf r}_i={\bf Y}_i + \sum_{k\neq i}
  \frac{\omega_{ik}}{\omega_{ii}} {\bf Y}_k$ is an $n$-vector of
  residuals of regressing ${\bf Y}_i$ on the rest.
\end{lemma}

The following lemma now quantifies the computational cost of updating
the residual vectors during each iteration of the CONCORD algorithm.
\begin{lemma}\label{eq:residual updates}
  Define the residual vector ${\bf r}_m$ for $m=1,2,\dots,p$ as follows:
  \begin{align}
    {\bf r}_m = {\bf r}_m(\Omega) = {\bf Y}_m + \sum_{k\neq m}
    \frac{\omega_{mk}}{\omega_{mm}} {\bf Y}_k
  \end{align}
  where $\Omega=((\Omega_{ij}))_{1\leq i,j\leq p}$. Then,
  \begin{enumerate}
  \item For $m\neq k,l$, the residual vector ${\bf r}_m$ is functionally
    independent of $\omega_{kl}$. (The term $\omega_{kl}$ appears only
    in the expressions for the residual vectors ${\bf r}_k$ and ${\bf r}_l$.)
  \item Fix all the elements of $\Omega$ except $\omega_{kl}$. Suppose 
   $\omega_{kl}$ is changed to $\omega_{kl}^*$. Then, updating the 
   residual vectors ${\bf r}_k$ and ${\bf r}_l$ requires $O(n)$ operations. (Hence, 
   updating ${\bf r}_k$ and ${\bf r}_l$ after each update in 
   \eqref{eq:update offdiag} requires $O(n)$ operations.)
 \item For $m\neq k$, the residual vector ${\bf r}_m$ is functionally
   independent of $\omega_{kk}$. (The term $\omega_{kk}$ appears only
   in the expression for the residual vector ${\bf r}_k$.)
  \item Fix all elements of $\Omega$ except $\omega_{kk}$. Suppose 
   $\omega_{kk}$ is changed to $\omega_{kk}^*$. Then, updating the 
   residual vector ${\bf r}_k$ requires $O(n)$ operations. (Hence, 
   updating ${\bf r}_k$ after each update in \eqref{eq:update diag} requires 
   $O(n)$ operations). 
  \end{enumerate}
\end{lemma}

The proofs of Lemmas \ref{eq:residuals} and \ref{eq:residual updates}
are straightforward and are given in Supplemental Sections
\ref{sect:proof residual} and \ref{sect:proof residual updates}.  Note
that the inner product between ${\bf y}_j$ and ${\bf r}_i$ takes
$O(n)$ operations. Hence, by Lemma \ref{eq:residuals} the updates in
\eqref{eq:update offdiag} and \eqref{eq:update diag} require $O(n)$
operations. Also, after each update in \eqref{eq:update offdiag} and
\eqref{eq:update diag} the residual vectors need to be appropriately
modified. By Lemma \ref{eq:residual updates}, this modification can
also be achieved in $O(n)$ operations. As a result, one complete sweep
of updates over all entries in $\hat\Omega$ can be performed in
$O(np^2)$ operations.

Hence, we conclude that the computational complexity of the CONCORD
algorithm is competitive with the SPACE and Symmetric lasso
algorithms, which are also $\min\left(O(np^2),O(p^3)\right)$.

\subsection{A unifying framework for pseudo-likelihood based graphical
  model selection}
\label{sect:unification}

In this section, we provide a unifying framework which formally
connects the five pseudo-likelihood formulations considered in this
paper, namely, SPACE1, SPACE2, SYMLASSO, SPLICE and CONCORD
(counting two choices for weights in the SPACE algorithm as two
different formulations). Recall that the random vectors ${\bf Y}^k =
\left( y_1^k, y_2^k, \cdots, y_p^k \right)'$, $k = 1,2, \cdots, n$
denote \emph{i.i.d.}  observations from a multivariate distribution
with mean vector ${\bf 0}$ and covariance matrix $\Sigma$, the
precision matrix is given by $\Omega = \Sigma^{-1} =
((\omega_{ij}))_{1 \leq i,j \leq p}$, and $\mathbf{S}$ denotes the
sample covariance matrix. Let $\Omega_D$ denote the diagonal matrix
with $i^{th}$ diagonal entry given by $\omega_{ii}$.  Lemma
\ref{lemma:unify} below formally identifies the relationship between
all five of the regression-based pseudo-likelihood methods.

\begin{lemma}\label{lemma:unify}

  i) The (negative) pseudo-likelihood functions of CONCORD,
  SPACE1, SPACE2, SYMLASSO and SPLICE formulations can be expressed in
  matrix form as follows (up to reparameterization):

  \begin{table}[htb]
    \footnotesize
    \centering
    \begin{tabular}{|l|l|l|l|}\hline
      & Regression form & Matrix form &  \\\hline
      $\mathcal{L}_{\con} (\Omega)$ 
      & $\frac{1}{2} \sum_{i=1}^p \left[
        -n\log \omega_{ii}^2 + \quad \| \omega_{ii} {\bf Y}_i + \sum_{j
          \neq i} \omega_{ij} {\bf Y}_j \|_2^2 \right]$ 
      & $\frac{n}{2}\left[ -\log | \Omega_D^2 |
        + \tr(\mathbf{S}\Omega^2) \right]$ 
      & \tagarray\label{PL:concord} \\
      $\mathcal{L}_{\spc,1} (\Omega_D,\boldsymbol\rho)$ 
      & $\frac{1}{2} \sum_{i=1}^p \left[ -n\log \omega_{ii} + \quad \| {\bf
          Y}_i - \sum_{j \neq i} \rho^{ij} 
        \sqrt{\frac{\omega_{jj}}{\omega_{ii}}} {\bf Y}_j \|_2^2 \right]$ 
      & $\frac{n}{2}\left[ -\log | \Omega_D |
        + \tr(\mathbf{S}\Omega\Omega_D^{-2}\Omega) \right]$ 
      & \tagarray\label{PL:space1} \\
      $\mathcal{L}_{\spc,2} (\Omega_D,\boldsymbol\rho)$ 
      & $\frac{1}{2} \sum_{i=1}^p \left[ -n\log \omega_{ii} + \omega_{ii}\,
        \| {\bf Y}_i - \sum_{j \neq i} \rho^{ij} 
        \sqrt{\frac{\omega_{jj}}{\omega_{ii}}} {\bf Y}_j \|_2^2 \right]$ 
      & $\frac{n}{2}\left[ -\log | \Omega_D |
        + \tr(\mathbf{S}\Omega\Omega_D^{-1}\Omega) \right]$ 
      & \tagarray\label{PL:space2} \\
      $\mathcal{L}_{\sym} (\boldsymbol{\alpha}, \Omega_F)$ 
      & $\frac{1}{2} \sum_{i=1}^p \left[ \quad n\log \alpha_{ii} +
        \frac{1}{\alpha_{ii}} \| {\bf Y}_i + \sum_{j \neq i}
        \omega_{ij}\alpha_{ii} {\bf Y}_j \|^2
      \right]$ 
      & $\frac{n}{2}\left[ -\log | \Omega_D | 
        + \tr(\mathbf{S}\Omega\Omega_D^{-1}\Omega) \right]$ 
      & \tagarray\label{PL:symlasso} \\
      $\mathcal{L}_{\spl}(\mathbf{B},\mathbf{D})$ 
      & $\frac{1}{2} \sum_{i=1}^p \left[ \quad n\log(d_{ii}^2) + {1\over d_{ii}^2} \|
        {\bf Y}_i - \sum_{j\neq i}\beta_{ij}{\bf Y}_j\|_2^2
      \right]$ 
      & $\frac{n}{2}\left[ -\log | \Omega_D | 
        + \tr(\mathbf{S}\Omega\Omega_D^{-1}\Omega) \right]$ 
      & \tagarray\label{PL:splice} \\\hline
    \end{tabular}
  \end{table}

  ii) All five pseudo-likelihoods above correspond to a unified or
  generalized form of the Gaussian log-likelihood function
  \begin{align*}
    \mathcal{L}_{\uni}(G(\Omega),H(\Omega)) = \frac{n}{2}\left[
      -\log\det G(\Omega) + \tr(\mathbf{S} H(\Omega)) \right],
  \end{align*}
  where $G(\Omega)$ and $H(\Omega)$ are functions of $\Omega$. The
  functions $G$ and $H$ which characterize the pseudo-likelihood
  formulations corresponding to CONCORD, SPACE1, SPACE2, SYMLASSO and
  SPLICE are given as follows:
  \begin{align*}
    G_{\con}(\Omega) &= \Omega_D^2, & &H_{\con}(\Omega) = \Omega^2 \\
    G_{\spc,1}(\Omega) &= \Omega_D, & &H_{\spc,1}(\Omega) = \Omega\Omega_D^{-2}\Omega\\
    G_{\spc,2}(\Omega) = G_{\sym}(\Omega) =G_{\spl}(\Omega) &=
    \Omega_D, & &H_{\spc,2}(\Omega) = H_{\sym}(\Omega) =
    H_{\spl}(\Omega) = \Omega\Omega_D^{-1}\Omega
  \end{align*}
\end{lemma}

\noindent The proof of Lemma \ref{lemma:unify} is given in Supplemental
Section \ref{sect:proof unify}. The above lemma gives various useful
insights into the different pseudo-likelihoods that have been proposed
for the inverse covariance estimation problem. The following remarks
discuss these insights.

\begin{rem}
  Note that when $G(\Omega)=H(\Omega)=\Omega$,
  $\mathcal{L}(G(\Omega),H(\Omega))$ corresponds to the standard
  (negative) Gaussian log-likelihood function.
\end{rem}

\begin{rem}
  Note that $\Omega_D^{-1}\Omega$ is a re-scaling of $\Omega$ so as to
  make all the diagonal elements one (hence sparsity between $\Omega$
  and $\Omega_D^{-1}\Omega$ are the same). In this sense, the SPACE2,
  SYMLASSO and SPLICE algorithms make the same approximation to the
  Gaussian likelihood with the log determinant term, $\log|\Omega|$,
  replaced by $\log|\Omega_D|$. The trace term $\tr(\mathbf{S}\Omega)$
  is approximated by
  $\tr(\mathbf{S}\Omega\Omega_D^{-1}\Omega)$. Moreover, if $\Omega$ is
  sparse, then $\Omega_D^{-1}\Omega$ is close to the identity matrix,
  i.e., $\Omega_D^{-1}\Omega \approx I+{\bf C}$ for some ${\bf C}$. In
  this case, the term in the Gaussian likelihood
  $\tr(\mathbf{S}\Omega)$ is perturbed by an off-diagonal matrix
  $\mathbf{C}$ resulting in an expression of the form
  $\tr(\mathbf{S}\Omega(\mathbf{I}+\mathbf{C}))$.
\end{rem}
\begin{rem}
  Conceptually, the sole source of difference between the three
  regularized versions of the objective functions of SPACE2, SYMLASSO
  and SPLICE algorithms is in the way in which the $\ell_1$-penalties
  are specified. SPACE2 applies the penalty to the partial
  correlations, SYMLASSO to the partial covariances and SPLICE to the
  symmetrized regression coefficients.
\end{rem}

\begin{rem}\label{rem:corr}
  Note that the CONCORD method approximates the Normal likelihood by
  approximating the $\log|\Omega|$ term by $\log|\Omega_D^2|$, and
  $\tr(\mathbf{S}\Omega)$ by $\tr(\mathbf{S}\Omega^2)$. Hence, the
  CONCORD algorithm can be considered as a reparameterization of the
  Gaussian likelihood with the concentration matrix $\Omega^2$ (together
  with an approximation to the log determinant term). More
  specifically,
  \begin{align*} \mathcal{L}_{\con}(\Omega) =
    \mathcal{L}_{\uni}(\Omega_D^2,\Omega^2) = \frac{n}{2}
    \left(-\log\det\Omega_D^2 + \tr(\mathbf{S}\Omega^2)\right) =
    n\left(-\log\det\Omega_D +
      \frac{1}{2}\tr(\mathbf{S}\Omega^2)\right),
  \end{align*} and justifies the appearance of ``$n$'' as compared to
  ``$n/2$'' in the CONCORD objective in \eqref{eq2}.  In Supplemental
  Section \ref{sect:correction}, we illustrate the usefulness of this
  correction based on the insight from our unification framework, and
  show that it leads to better estimates of $\Omega$.
\end{rem}

\section{Convergence of CONCORD} \label{cnvgcproof}
\noindent
We now proceed to consider the convergence properties of the CONCORD
algorithm. Note that $Q_{\con} (\Omega)$ is not differentiable. Also,
if $n < p$, then $Q_{\con} (\Omega)$ is not necessarily strictly
convex. Hence, the global minimum may not be unique, and as discussed
below, the convergence of the coordinatewise minimization algorithm to 
a global minimum does not follow from existing theory. Note that 
although$Q_{\con} (\Omega)$ is not differentiable, it can be expressed 
as a sum of a smooth function of $\Omega$ and a separable function of
$\Omega$ (namely $\lambda \sum_{1 \leq i < j \leq p}
|\omega_{ij}|$). \cite{Tseng1988,Tseng2001} proves that under certain
conditions, every cluster point of the sequence of iterates of the
coordinatewise minimization algorithm for such an objective function is 
a stationary point of the objective function. However, if the function
is not strictly convex, there is no general guarantee that the
sequence of iterates has a unique cluster point, i.e., there is no 
theoretical guarantee that the sequence of iterates converges. The 
following theorem shows that the cyclic coordinatewise minimization 
algorithm applied to the CONCORD objective function converges to a 
global minimum. A proof of this result can be found in 
Supplemental Section \ref{sect:convergence:concord}.  
\begin{thm} \label{cnvconcord}
If $S_{ii} > {\bf 0}$ for every $1 \leq i \leq p$, the sequence of
  iterates $\left\{ \hat{\Omega}^{(r)} \right\}_{r \geq 0}$ obtained
  by Algorithm \ref{altalgrthm} converges to a global minimum of
  $Q_{\con} (\Omega)$. More specifically,
  $\hat\Omega^{(r)}\rightarrow\hat\Omega\in\mathcal{M}$ as
  $r\rightarrow\infty$ for some $\hat\Omega$, and furthermore 
  $Q_{\con}(\hat\Omega)\leq Q_{\con}(\Omega)$ for all $\Omega 
  \in \mathcal{M}$. 
\end{thm}

\begin{rem}
If $n \geq 2$, and none of the underlying $p$ marginal 
distributions (corresponding to the $p$-variate distribution for the 
data vectors) is degenerate, it follows that the diagonal entries of 
the data covariance matrix $S$ are strictly positive with probability $1$. 
\end{rem}

\begin{figure}
  \centering
  \subfigure[SPACE algorithm (partial variance weights) applied to
  dataset in Example 1.]{
    \includegraphics[width=0.47\textwidth,trim = 2mm 0mm 5mm 0mm, clip]{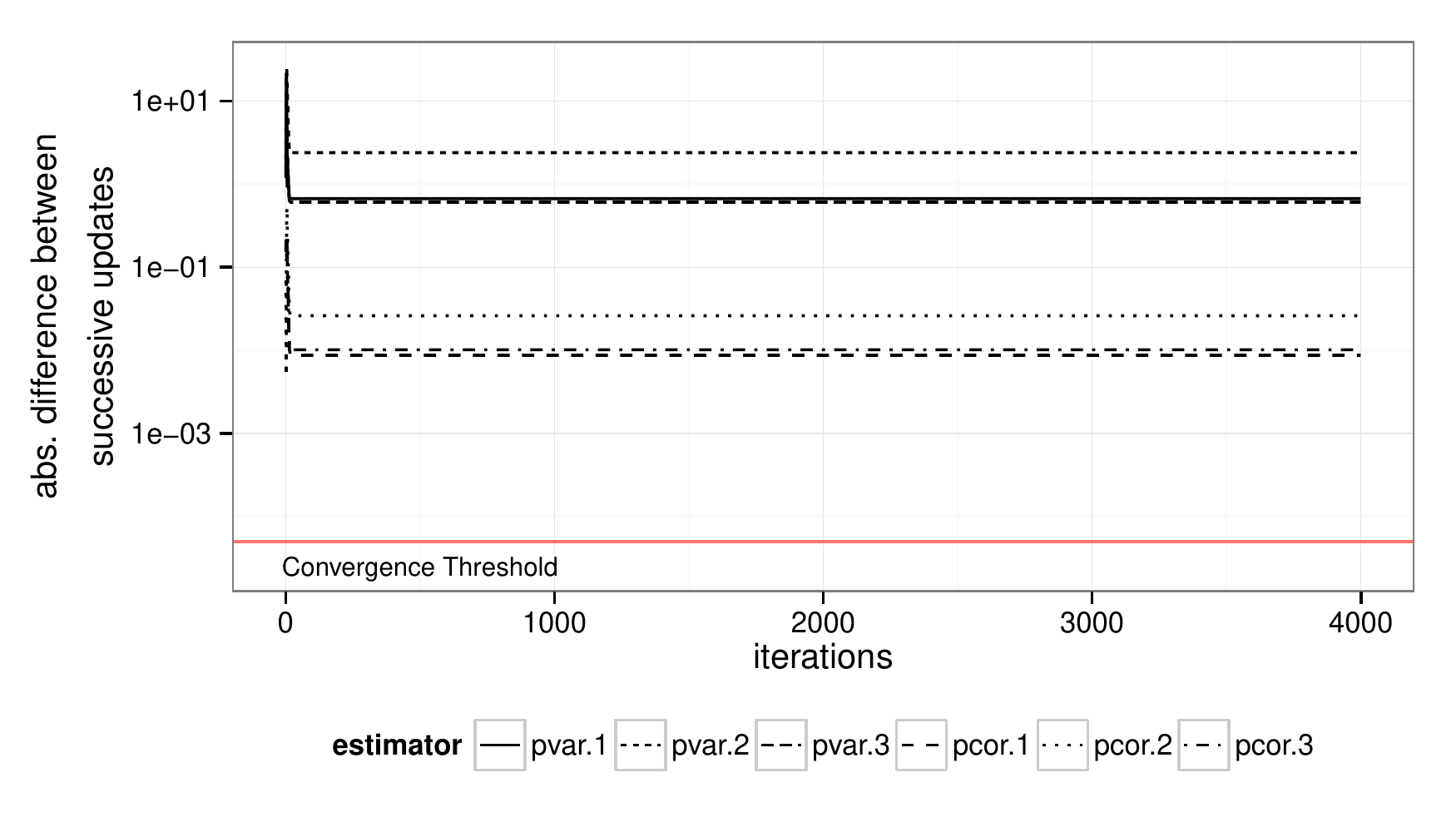}
    \label{fig:weightsW}
  }
  \subfigure[CONCORD algorithm applied to dataset in Example 1.]{
    \includegraphics[width=0.47\textwidth,trim = 2mm 0mm 5mm 0mm, clip]{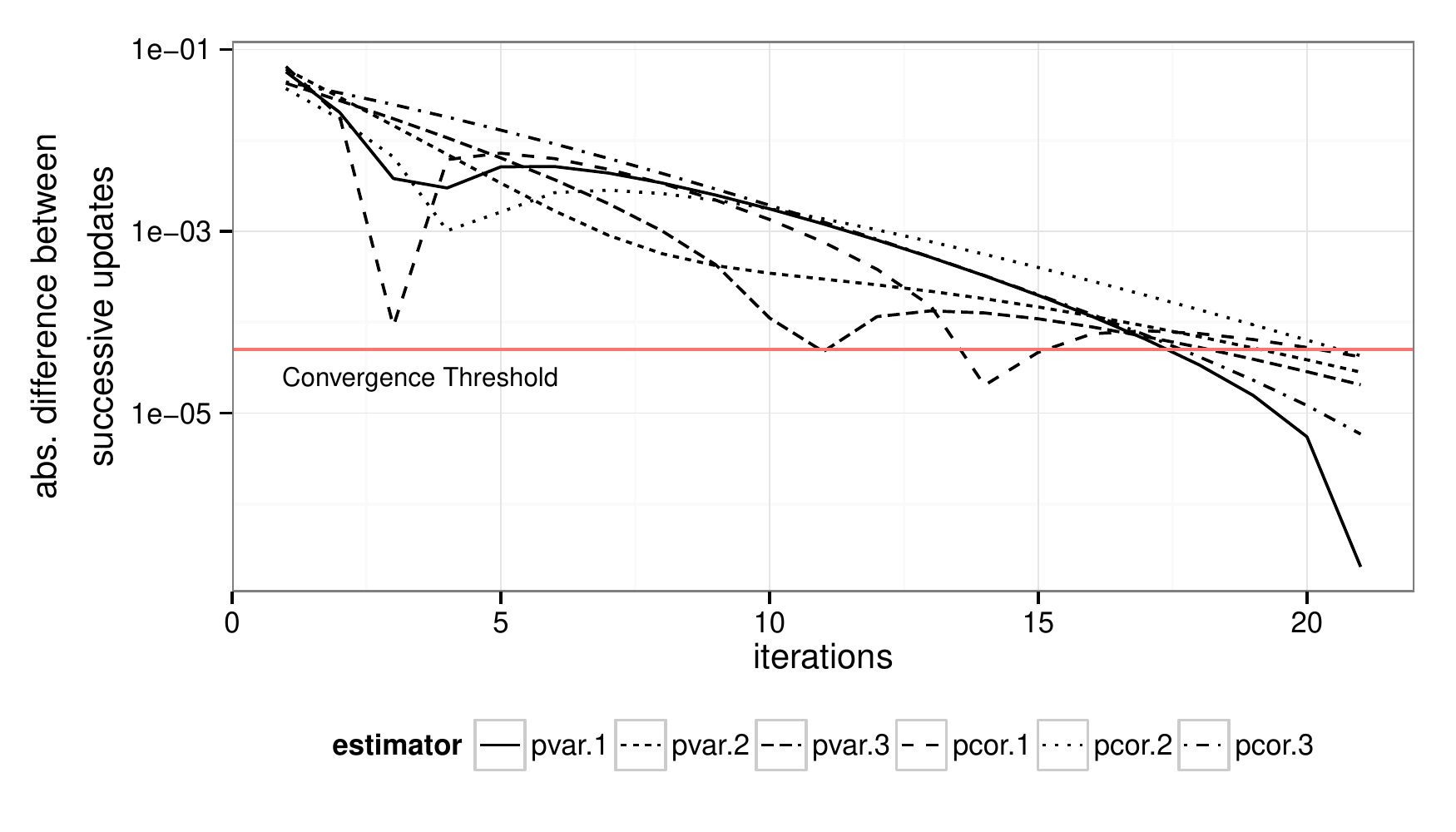}
    \label{fig:weightsWcSpace}
  }  
  \caption[Optional caption for list of figures]{Illustrations of the
    non-convergence of SPACE and convergence of CONCORD. The y-axes
    are log scaled. For SPACE, log absolute difference between entries
    of successive estimates becomes constant (thus indicating
    non-convergence).}
  \label{fig:subfigureExample}
\end{figure}

\noindent
With theory in hand, we now proceed to numerically illustrate the 
convergence properties established above. When CONCORD is 
applied to the dataset in Example 1, convergence is achieved 
(see Figure \ref{fig:weightsWcSpace}), whereas SPACE does not converge (see 
Figure \ref{fig:weightsW}). 

\section{Applications} \label{simulation}

\subsection{Simulated Data} \label{sect:simulated data} 

\subsubsection{Timing Comparison} \label{timing:comparison} 

\noindent
We now proceed to compare the timing performance of CONCORD with
Glasso and the two different versions of SPACE. The acronyms SPACE1
and SPACE2 denote SPACE estimates using uniform weights and partial
variance weights, respectively. We first consider the setting $p =
1000, n = 200$. For the purposes of this simulation study, a
$p \times p$ positive definite matrix $\Omega$ (with $p = 1000$) with
condition number 10 was used.  Thereafter, 50 independent datasets were
generated, each consisting of $n=200$ {\it i.i.d.} samples from a
$\mathcal{N}_p(0,\Sigma=\Omega^{-1})$ distribution. For each dataset,
the four algorithms were run until convergence for a range of penalty
parameter values. We note that the default number of iterations for
SPACE in the R function by \cite{Peng2009} is $3$. However, given the
convergence issues for SPACE, we ran SPACE until convergence or until
50 iterations (whichever is smaller). The timing results (averaged
over the 100 datasets) in the top part of Table \ref{tabletmcrn} below
show wall clock times until convergence (in seconds) for Glasso,
CONCORD, SPACE1 and SPACE2.

One can see that in the $p = 1000, n= 200$ setting, CONCORD is uniformly faster than 
its competitors. Note the low penalty parameter cases correspond to high dimensional settings 
where the estimated covariance matrix is typically poorly conditioned and the log-likelihood surface 
is very flat. The results in Table \ref{tabletmcrn} indicate that in such settings CONCORD is faster than 
its competitors by orders of magnitude (even though Glasso is implemented in Fortran). Both SPACE1 
and SPACE2 are much slower than CONCORD and 
Glasso in this setting. The wall clock time for an iterative algorithm can be 
thought of as a function of the number of iterations until convergence, the 
order of computations for a single iteration, and also the implementation 
details (such as choice of software, efficiency of the code etc.). Note that the 
order of computations for a single iteration is same for SPACE and 
CONCORD, and lower than that of Glasso when $n < p$. It is likely that the 
significant increase in the wall clock time for SPACE is due to implementation 
details and the larger number of iterations required for convergence (or non-convergence, since 
we are stopping SPACE if the algorithm does not satisfy the convergence 
criterion by 50 iterations). 

We further compare the timing performance of CONCORD and Glasso for 
$p = 3000$ with $n = 600$ and $n = 900$ (SPACE is not considered here 
because of the timing issues mentioned above. These issues are amplified in this 
more demanding setting). A $p \times p$ positive definite matrix 
$\Omega$ (with $p = 3000$) with $3 \%$ sparsity is used. Thereafter, 
50 independent datasets were generated, each consisting of $n=600$ 
{\it i.i.d.} samples from a $\mathcal{N}_p(0,\Sigma=\Omega^{-1})$ 
distribution. The same exercise was repeated with $n=900$. The timing 
results (averaged over the 100 datasets) in the bottom part of 
Table \ref{tabletmcrn} below show wall clock times until convergence 
(in seconds) for Glasso, CONCORD, SPACE1 and SPACE2 for various 
penalty parameter values. It can be seen that in both the $n=600$ and 
$n=900$ cases, CONCORD was around ten times faster than Glasso. 

In conclusion, these simulation results in this subsection illustrate that CONCORD is 
much faster as compared to SPACE and Glasso, especially in very high dimensional 
settings. We also note that a downloadable version of the CONCORD algorithm has 
been developed in R, and is freely available at http://cran.r-project.org/web/packages/gconcord. 
\begin{table}
\centering
\footnotesize
\begin{tabular}{||crr|crr|crr|crr||}
  \hline
  \multicolumn{12}{||c||}{$\bf p=1000$, $\bf n=200$}\\\hline
  \multicolumn{3}{||c|}{Glasso} &
  \multicolumn{3}{c|}{CONCORD} &
  \multicolumn{3}{c|}{SPACE1 ($w_i=1$)} & \multicolumn{3}{c||}{SPACE2 ($w_i=\omega_{ii}$)}\\
  $\lambda$ & \multicolumn{1}{c}{NZ} & Time & $\lambda^*$ & \multicolumn{1}{c}{NZ} & Time & $\lambda^*$ & \multicolumn{1}{c}{NZ} & Time & $\lambda^*$ & \multicolumn{1}{c}{NZ} & Time \\\hline
  0.14 & 4.77\% & 87.60 & 0.12 & 4.23\% & 6.12 & 0.10 & 4.49\% & 101.78 & 0.16 & 100.00\% & 19206.55 \\ 
  0.19 & 0.87\% & 71.47 & 0.17 & 0.98\% & 5.10 & 0.17 & 0.64\% &  99.20 & 0.21 &   1.76\% &   222.00 \\ 
  0.28 & 0.17\% &  5.41 & 0.28 & 0.15\% & 5.37 & 0.28 & 0.14\% & 138.01 & 0.30 &   0.17\% &    94.59 \\ 
  0.39 & 0.08\% &  5.30 & 0.39 & 0.07\% & 4.00 & 0.39 & 0.07\% &  75.55 & 0.40 &   0.08\% &   108.61 \\ 
  0.51 & 0.04\% &  6.38 & 0.51 & 0.04\% & 4.76 & 0.51 & 0.04\% &  49.59 & 0.51 &   0.04\% &   132.34 \\ 
   \hline\multicolumn{12}{c}{}\\\hline
  \multicolumn{6}{||c||}{$\bf p=3000$, $\bf n=600$} &
  \multicolumn{6}{c||}{$\bf p=3000$, $\bf n=900$} \\\hline
  \multicolumn{3}{||c|}{Glasso} &
  \multicolumn{3}{c||}{CONCORD} &
  \multicolumn{3}{c|}{Glasso} & \multicolumn{3}{c||}{CONCORD}\\
  $\lambda$ & \multicolumn{1}{c}{NZ} & Time & $\lambda^*$ & \multicolumn{1}{c}{NZ} & \multicolumn{1}{c||}{Time} & $\lambda$ & \multicolumn{1}{c}{NZ} & Time & $\lambda^*$ & \multicolumn{1}{c}{NZ} & Time \\\hline   
  0.09 & 2.71\% & 1842.74 & 0.09 & 2.10\% & \multicolumn{1}{c||}{266.69} & 0.09 & 0.70\% & 1389.96 & 0.09 & 0.64\% & 298.21 \\ 
  0.10 & 1.97\% & 1835.32 & 0.10 & 1.59\% & \multicolumn{1}{c||}{235.49} & 0.10 & 0.44\% & 1395.42 & 0.10 & 0.41\% & 298.00 \\
  0.10 & 1.43\% & 1419.41 & 0.10 & 1.19\% & \multicolumn{1}{c||}{232.67} & 0.10 & 0.27\% & 1334.78 & 0.10 & 0.26\% & 302.15 \\\hline
\end{tabular}
\caption{Timing comparison (in seconds) for $p=1000,\ 3000$ and varying $n$. SPACE is run 
until convergence or 50 iterations (whichever is smaller). Note that SPACE1 and SPACE2 are 
much slower compared than CONCORD and Glasso in wall time, for the $p=1000$ simulation. Hence, 
for $p=3000$, only Glasso and CONCORD are compared. Here, $\lambda$ denotes the value of the 
penalty parameter for the respective algorithms, with $\lambda^* = \lambda/n$ for CONCORD and 
SPACE. $NZ$ denotes the percentage of non-zero entries in the corresponding estimator.}
\label{tabletmcrn}
\end{table}

\subsubsection{Model selection comparison} \label{model:selection} 

\noindent
In this section, we perform a simulation study in which we compare the model
selection performance of CONCORD and Glasso when the underlying data is 
drawn from a multivariate-$t$ distribution (the reasons for not considering 
SPACE are provided in a remark at the end of this section). The data is drawn 
from a multivariate-$t$ distribution to illustrate the potential benefit of using 
penalized regression methods (CONCORD) outside the Gaussian setting. 


For the purposes of this study, using a similar approach as in \cite{Peng2009}, a $p 
\times p$ sparse positive definite matrix $\Omega$ 
(with $p = 1000$) with condition number $13.6$ is chosen. Using this $\Omega$ 
for each sample size $n = 200$, $n = 400$ and $n = 800$, 50 datasets, each having {\it i.i.d.} 
multivariate-$t$ distribution with mean zero and covariance matrix 
$\Sigma=\Omega^{-1}$, are generated. We compare the model selection 
performance of Glasso and CONCORD in this heavy tailed setting with 
receiver operating characteristic (ROC) curves, which compare false positive 
rates (FPR) and true positive rates (TPR). Each ROC curve is traced out by 
varying the penalty parameter $\lambda$ over $50$ possible values. 

We use the Area-under-the-curve (AUC) as a means to compare model 
selection performance. This measure is frequently used to compare ROC 
curves \citep{Fawcett2006,Friedman2010}. The AUC of a full ROC curve 
resulting from perfect recovery of zero/non-zero structure in $\Omega$ 
would be 1. In typical real applications, FPR is controlled to be sufficiently 
low. We therefore compare model selection performance when FPR is less 
than 15\% (or 0.15). When controlling FPR to be less than 0.15, a perfect 
method will yield AUC of 0.15. Table \ref{tbl:auc} provides the median 
of the AUCs (divided by 0.15 to normalize to 1), 
as well as the interquartile ranges (IQR) over the 50 datasets for $n = 
200$, $n = 400$ and $n = 800$. 
\begin{table}[H]
    \centering
    \begin{tabular}{||c||cc||cc||cc||}
      \hline
      & \multicolumn{2}{c||}{$\bf n=200$} 
      & \multicolumn{2}{c||}{$\bf n=400$} 
      & \multicolumn{2}{c||}{$\bf n=800$}\\\hline
      Solver  & Median & IQR   & Median & IQR & Median & IQR \\
      \hline                                 
      Glasso  & 0.745  & 0.032 & 0.819 &0.030 & 0.885 &0.029 \\ 
      CONCORD & 0.811  & 0.011 & 0.887 &0.012 & 0.933 &0.013 \\ 
      \hline
    \end{tabular}
    \caption{Median and IQR of area-under-the-curve (AUC) for 50 
    simulations. Each simulation yields a ROC   curve from which the AUC is 
    computed for FPR in the interval [0, 0.15] and normalized to 1.}
    \label{tbl:auc}
\end{table}

\noindent
Table \ref{tbl:auc} above shows that CONCORD has a much better model
selection performance as compared to Glasso. Moreover, it turns out that 
CONCORD has a higher AUC than Glasso for every single one of the $150$ datasets 
($50$ each for $n = 200, 400$ and $800$). We note that CONCORD 
not only recovers the sparsity structure more accurately in general, it also 
has much less variation. 

\vspace{0.1in}

\noindent {\it Remark:} Note that we need to simulate $50$ datasets
for each of the above three sample sizes. For each of these datasets,
an algorithm has to be run for 50 different penalty parameter
values. In totality, this amounts to running the algorithm 7500
times. As we demonstrated in the simulations in Section 
\ref{timing:comparison}, when SPACE is run until convergence (or 
terminated after the number of iterations is 50), then SPACE's intractability
makes it infeasible to run it 7500 times. As an alternative, one could
follow the approach of \cite{Peng2009} and stop SPACE after running
$3$ iterations. However, given the possible non-convergence issues
associated with SPACE, it is not clear if the resulting estimate is
meaningful. Even so, if we follow this approach of stopping SPACE
after three iterations, we find that CONCORD outperforms SPACE1 and
SPACE2. For example, if we consider the $n = 200$ case, then the
median AUC value for SPACE1 is 0.779 (with $IQR = 0.054$) and the
median AUC value for SPACE2 is 0.802 (with $IQR = 0.013$).


\subsection{Application to breast cancer data}

\noindent
We now illustrate the performance of the CONCORD method on a real
dataset. To facilitate comparison, we consider data from a breast
cancer study \citep{Chang2005} on which SPACE was illustrated. This
dataset contains expression levels of 24481 genes on 248 patients with
breast cancer.  The dataset also contains extensive clinical data
including survival times.

Following the approach in \cite{Peng2009} we focus on a smaller subset
of genes. This reduction can be achieved by utilizing clinical
information that is provided together with the microarray expression
dataset. In particular, survival analysis via univariate Cox
regression with patient survival times is used to select a subset of
genes closely associated with breast cancer. A choice of p-value
$<0.0003$ yields a reduced dataset with 1107 genes. This subset of the
data is then mean centered and scaled so that the median absolute
deviation is 1 (as outliers seem to be present). Following a similar
approach to that in \cite{Peng2009}, penalty parameters for each
partial correlation graph estimation method were chosen so that each
partial correlation graph yields 200 edges.
 
Partial correlation graphs can be used to identify genes that are
biologically meaningful and can lead to gene therapeutic targets. In
particular, there is compelling evidence from the biomedical
literature that highly connected nodes are central to biological
networks \citep{Carter2004,Jeong2001,Han2004}. To this end, we focus
on identifying the 10 most highly connected genes (``hub'' genes)
identified by each partial correlation graph estimation method. Table
\ref{tbl:hub genes} in Supplemental Section \ref{sect:biology example}
summarizes the top 10 hub genes obtained by CONCORD, SYMLASSO, SPACE1
and SPACE2. The table also gives references from the biomedical
literature that places these genes in the context of breast cancer.
These references illustrate that most of the identified genes are
indeed quite relevant in the study of breast cancer. It can also be
seen that there is a large level of overlap in the top 10 genes
identified by the four methods. There are also however some notable
differences. For example, \emph{TPX2} has been identified only by
CONCORD. \cite{Bibby2009} suggests that mutation of Aurora A - a known
general cancer related gene - reduces cellular activity and
mislocalization due to loss of interaction with \emph{TPX2}. Moreover,
a recent extensive study by \cite{Maxwell2011}\footnote{{\tt
    http://www.ncbi.nlm.nih.gov/pubmed/22110403}} identifies a gene
regulatory mechanism in which \emph{TPX2}, Aurora A, \emph{RHAMM} and
\emph{BRCA1} play a key role. This finding is especially significant
given that \emph{BRCA1} (breast cancer type 1 susceptibility protein)
is one of the most well known genes linked to breast cancer.  We also
remark that if a higher number of hub genes are targeted (like the top
20 or top 100 vs. the top 10), CONCORD identifies additional genes not
discovered by existing methods. However, identification of even a
single important gene can lead to significant findings and novel gene
therapeutic targets, since many gene silencing experiments often focus
on one or two genes at a time.

We conclude this section by remarking that CONCORD is a useful
addition to the graphical models literature as it is competitive with
other methods in terms of model selection accuracy, timing, relevance
for applications, and also gives provable convergence guarantees.

\subsection{Application to portfolio optimization}

We now consider the efficacy of using CONCORD in a financial portfolio
optimization setting where a stable estimate of the covariance matrix
is often required. We follow closely the exposition to the problem as
given in \cite{Won2012}. A portfolio of financial instruments
constitutes a collection of both risky and risk-free assets held by a
legal entity.  The return on the overall portfolio over a given
holding period is defined as the weighted average of the returns on
the individual assets, where the weights for each asset corresponds to
its proportion in monetary terms. The primary objective of the
portfolio optimization problem is to determine the weights that
maximize the overall return on the portfolio subject to a certain
level of risk (or vice versa). In Markowitz mean-variance portfolio
(MVP) theory, this risk is taken to be the the standard deviation of
the portfolio \citep{Markowitz1952}. As noted in \cite{Luenberger1997}
\& \cite{Merton1980}, the optimal portfolio weights or the optimal
allocation depends critically on the mean and covariance matrix of the
individual asset returns, and hence estimation of these quantities is
central to MVP. As one of the goals in this paper is to illustrate the
efficacy of using CONCORD to obtain a stable covariance matrix
estimate, we shall consider the \emph{minimum variance} portfolio
problem, as compared to the \emph{mean-variance} portfolio
optimization problem. The former requires estimating only the
covariance matrix and thus presents an ideal setting for comparing
covariance estimation methods in the portfolio optimization context
(see \cite{Chan1999} for more details). In particular, we aim to
compare the performance of CONCORD with other covariance estimation 
methods, for the purposes of constructing a minimum variance portfolio. The
performance of each of the different methods and the associated
strategies will be compared over a sustained period of time in order
to assess their respective merits. 

\subsubsection{Minimum variance portfolio rebalancing} \label{mivarporeb} 

\noindent
The minimum variance portfolio selection problem is defined as
follows. Given $p$ risky assets, let $r_{it}$ denote the return of
asset $i$ over period $t$; which in turn is defined as the change in
its price over time period $t$, divided by the price at the beginning
of the period.  As usual, let $\Sigma_t$ denote the covariance matrix
of the daily returns, $r_t^T=(r_{1t},r_{2t},\dots,r_{pt})$. The
portfolio weights $w_k^T=(w_{1k},w_{2k},\ldots,w_{pk})$ denote the
weight of asset $i=1,\dots,p$ in the portfolio for the $k$-th time
period. A long position or a short position for asset $i$ during
period $k$ is given by the sign of $w_{ik}$, i.e., $w_{ik}>0$ for
long, and $w_{ik}<0$ for short positions respectively. The budget
constraint can be written as $\ones^Tw_k=1$, where $\ones$ denotes the
vector of all ones. Note that the risk of a given portfolio as
measured by the standard deviation of its return is simply
$(w_k^T\Sigma w_k)^{1/2}$ .

The minimum variance portfolio selection problem for investment period
$k$ can now be formally defined as follows:
\begin{align}\label{eqn:MVP}
    \mbox{minimize} \; \;  w_k^T\Sigma w_k \; \; \; \; 
    \mbox{subject to} \; \; \ones^Tw_k=1.
\end{align}
As \eqref{eqn:MVP} above is a simple quadratic program, it has an
analytic solution given by $w_k^\star =
(\ones^T\Sigma^{-1}\ones)^{-1}\Sigma^{-1}\ones$. Note that the
solution depends on the theoretical covariance matrix $\Sigma$. In
practice, the parameter $\Sigma$ has to be estimated.

The most basic approach to the portfolio selection problem often makes
the unrealistic assumption that returns are stationary in time. A
standard approach to dealing with the non-stationarity in such
financial time series is to use a periodic rebalancing strategy. In
particular, at the beginning of each investment period
$k=1,2,\dots,K$, portfolio weights $w_k=(w_{1k},\dots,w_{pk})'$ are
computed from the previous $N_{\est}$ days of observed returns
($N_{\est}$ is called the ``estimation horizon''). These portfolio
weights are then held constant for the duration of each investment
period. The process is repeated at the start of the next investment
period and is often referred to as ``rebalancing.'' More details of the 
rebalancing strategy are provided in 
Supplemental section \ref{sec:minvarporeb}. 

\subsubsection{Application to the Dow Jones Industrial Average}

We now consider the problem of investing in the stocks that feature in
the Dow Jones Industrial Average (DJIA) index. The DJIA is a composite
blue chip index consisting of 30 stocks (note that Kraft Foods (KFT)
data was removed in our analysis due to its limited data
span\footnote{KFT was a component stock of the DJIA form 9/22/2008 to
  9/13/2012. From 9/14/2012, KFT was replaced with United Health Group
  (UNH).}. Table \ref{tbl:DJIA component} in Supplemental Section
\ref{sect:dow jones} lists the 29 component stocks used in our
analysis.

Rebalancing time points were chosen to be every four weeks starting
from 1995/02/18 to 2012/10/26 (approximately 17 years), and are shown
in Table \ref{tab:investment periods} in Supplemental Section
\ref{sect:investment periods}. Start and end dates of each period are
selected to be calendar weeks, and need not coincide with a trading
day. The total number of investment periods is 231, and the number of 
trading days in each investment period varies between 15 and 20 days. 
We shall compare the following five methods for estimating the
covariance matrix: sample covariance, graphical lasso (Glasso) of 
\cite{Friedman2008}, CONCORD, condition number regularized estimator
(CondReg) of \cite{Won2012}, and the Ledoit-Wolf estimator of 
\cite{Ledoit2004}. We consider various choices of 
$N_{\est}$, in particular, $N_{\est}\in\{35,40,45,50,75,150,225,300\}$ 
in our analysis. Note that once a choice for $N_{\est}$ is made, it is 
kept constant for all the 231 investment periods. 

Note that for $\ell_1$-penalized regression methods such as the Glasso
and CONCORD methods, a value for the penalty parameter has to be
chosen. For the purposes of this study, cross-validation was performed
within each estimation horizon so as to minimize the residual sum of
squares from out-of-sample prediction averaged over all
stocks. Further details are given in Supplemental Section
\ref{sect:cross validation}. The condition number regularized
(CondReg) and Ledoit-Wolf estimators each use different criteria to
perform cross-validation. The readers is referred to \cite{Won2012}
and \cite{Ledoit2004} for details on the cross-validation procedure
for these methods.

For comparison purposes with \cite{Won2012}, we use the following
quantities to assess the performance of the five MVR strategies: 
{\it Realized return, Realized risk, Realized Sharpe ratio (SR), Turnover, 
Size of the short side} and {\it Normalized wealth growth}. 
Precise definitions of these quantities are given in Supplemental Section 
\ref{sect:perf metrics}.

%
%
%
%
%

Table \ref{tab:realized sharpe ratio} gives the realized Sharpe ratios
of all MVR strategies for the different choices of estimation horizon
$N_{\est}$. The column DJIA stands for the passive index tracking
strategy that tracks the Dow Jones industrial average index. It is
clear from Table \ref{tab:realized sharpe ratio} that the CONCORD
method performs uniformly well across different choices of estimation
horizons.

\begin{table}
  \centering
  \begin{tabular}{rrrrrrr}
    \hline
    $N_{\est}$ & Sample & Glasso & CONCORD & CondReg & Ledoit-Wolf & DJIA \\ 
    \hline
    35 & 0.357 & \textbf{0.489} & \textbf{0.487} & \textbf{0.486} & 0.470 & 0.185 \\ 
    40 & 0.440 & \textbf{0.491} & \textbf{0.490} & 0.473 & 0.439 & 0.185 \\ 
    45 & 0.265 & 0.468 & \textbf{0.473} & 0.453 & 0.388 & 0.185 \\ 
    50 & 0.234 & \textbf{0.481} & \textbf{0.482} & 0.458 & 0.407 & 0.185 \\ 
    75 & 0.379 & 0.403 & \textbf{0.475} & 0.453 & 0.368 & 0.185 \\ 
    150 & 0.286 & 0.353 & \textbf{0.480} & 0.476 & 0.384 & 0.185 \\ 
    225 & 0.367 & 0.361 & \textbf{0.502} & 0.494 & 0.416 & 0.185 \\ 
    300 & 0.362 & 0.359 & \textbf{0.505} & 0.488 & 0.409 & 0.185 \\ 
    \hline
  \end{tabular}

  \caption[Application: realized Sharpe ratio comparison]{
    Realized Sharpe ratio of 
    different investment strategies corresponding to 
    different estimators with various $N_{\est}$.  The maximum annualized Sharpe ratios
    for each row, and others within 1\% of this maximum, are highlighted in bold.}
  \label{tab:realized sharpe ratio}
\end{table}

\begin{figure}
  \centering{
    \includegraphics[width=\textwidth]{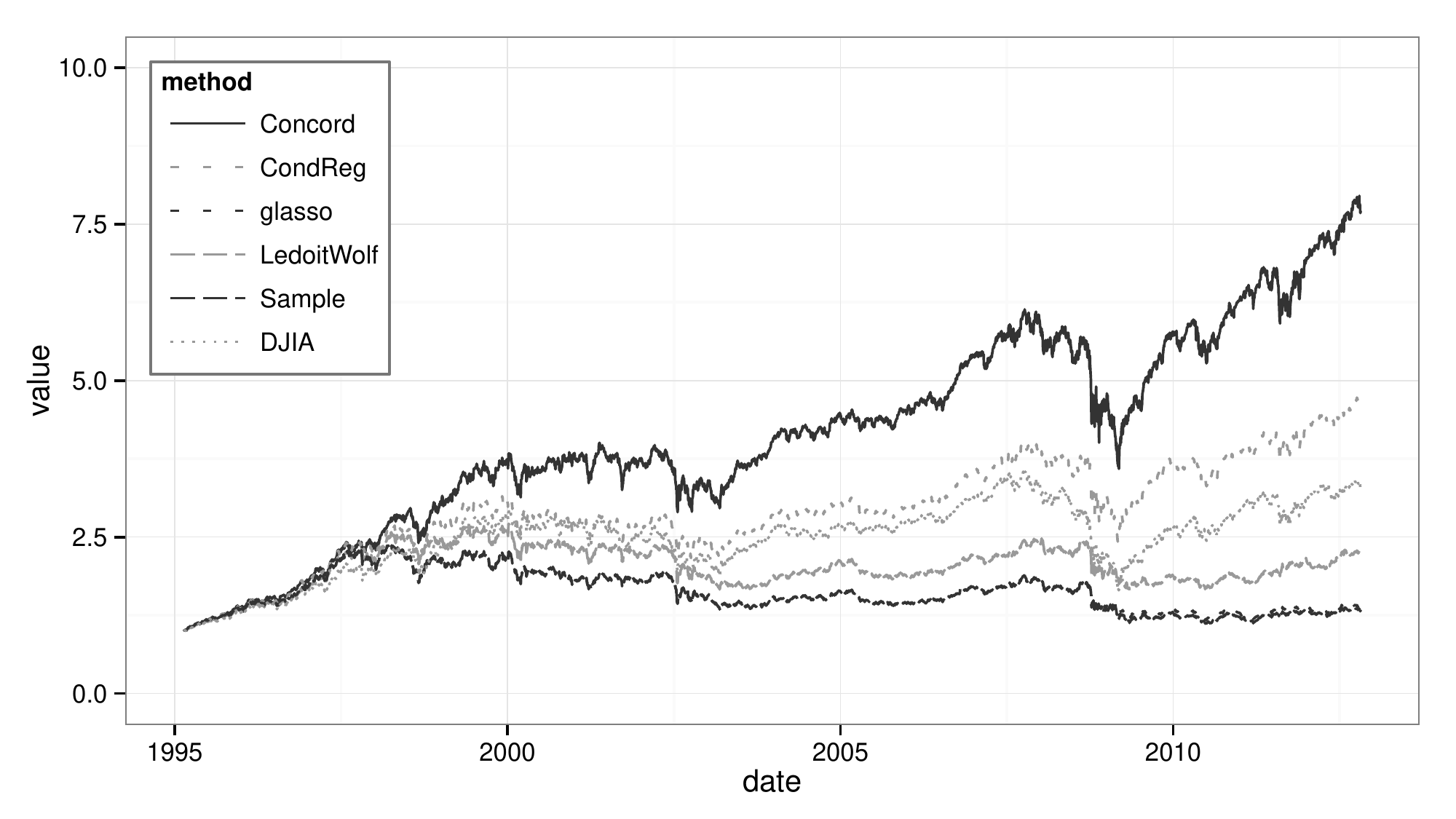} 
  }
  \caption[Normalized wealth growth comparison with trading
  costs]{Normalized wealth growth after adjusting for transaction
    costs (0.5\% of principal) and borrowing costs (interest rate of
    7\% APR) with $N_{\est} = 225$.}
  \label{fig:wealth growth tc and bc}
\end{figure}

Figure \ref{fig:wealth growth tc and bc} shows normalized wealth
growth over the trading horizon for the choice $N_{\est} = 225$. Normalized 
wealth growth curve for another choice $N_{\est} = 75$ is provided in Supplemental 
section \ref{sect:perf metrics}. These plots demonstrate that CONCORD is either
very competitive or better than leading covariance estimation methods.

We also note that trading costs associated with CONCORD are the lowest
for most choices of estimation horizons, and are very comparable with
CondReg for $N_{\est}=\{35,40\}$ (See Table \ref{tab:mean sd trading
  costs} in Supplemental Section \ref{sect:perf metrics}).  Moreover,
CONCORD also has by far the lowest short side for most choices of
estimation horizons. This property reduces the dependence on borrowed
capital for shorting stocks and is also reflected in the higher
normalized wealth growth. 

\section{Large sample properties}\label{sect:largesample}

\noindent
In this section, large sample properties of the CONCORD algorithm,
estimation consistency and oracle properties under suitable regularity
conditions are investigated. We adapt the approach in \cite{Peng2009}
with suitable modifications. Now let the dimension $p = p_n$ vary with
$n$ so that our treatment is relevant to high dimensional
settings. Let $\{\bar{\Omega}_n\}_{n \geq 1}$ denote the sequence of
true inverse covariance matrices. As in \cite{Peng2009}, for
consistency purposes, we assume the existence of suitably accurate
estimates of the diagonal entries, and consider the accuracy of the
estimates of the off-diagonal entries obtained after running the
CONCORD algorithm with diagonal entries fixed. In particular, the
following assumption is made:
\begin{itemize}
\item (A0 - Accurate diagonal estimates) There exist estimates $\{\widehat{\alpha}_{n,ii}\}_{1 
\leq i \leq p_n}$ such that for any $\eta > 0$, there exists a constant $C > 0$ such that 
$$
\max_{1 \leq i \leq p_n} \left| \widehat{\alpha}_{n,ii} - \bar{\omega}_{ii} \right| \leq C \left( 
\sqrt{\frac{\log n}{n}} \right), 
$$

\noindent
holds with probability larger than $1 - O(n^{-\eta})$. 
\end{itemize}

Note that the theory that follows is valid when the estimates
$\{\widehat\alpha_{n,ii}\}_{1\leq i\leq p_n}$ and the estimates of the
off-diagonal entries are obtained from the same dataset. When $\lim
\sup_{n \rightarrow \infty} \frac{p_n}{n} < 1$, \cite{Peng2009} show
that the diagonal entries of $S^{-1}$ can be used as estimates of the
diagonal entries of $\Omega$. However, no such general recipe is
provided in \cite{Peng2009} for the case $p_n > n$. Nevertheless,
establishing consistency in the above framework is useful, as it
indicates that the estimators obtained are statistically well-behaved
when $n$ and $p$ both increase to infinity.

For vectors $\omega^o \in \mathbb{R}^{\frac{p_n(p_n-1)}{2}}$ and
$\omega^d \in \mathbb{R}^{p_n}_+$, the notation ${\mathcal{L}}_n
(\omega^o, \omega^d)$ stands for $\frac{{\mathcal{L}}_{\con}}{n}$
($\mathcal{L}_{\con}$ is defined in \eqref{eq2}) evaluated at a matrix
with off-diagonal entries $\omega^o$ and diagonal entries $\omega^d$.
Let $\bar{\omega}_n^o =((\bar{\omega}_{n,ij}))_{1 \leq i < j \leq
  p_n}$ denote the vector of off-diagonal entries of $\bar{\Omega}_n$,
and $\widehat{\boldsymbol \alpha}_{p_n} \in \mathbb{R}^{p_n}_+$
denotes the vector with entries $\{\widehat{\alpha}_{n,ii}\}_{1 \leq i
  \leq p_n}$.  Let$\mathcal{A}_n$ denote the set of non-zero entries
in the vector $\bar{\omega}_n^o$, and let $q_n = |\mathcal{A}_n|$. Let
$\bar{\Sigma}_n = \bar{\Omega}_n^{-1}$ denote the true covariance
matrix for every $n \geq 1$. The following standard assumptions are
required.
\begin{itemize}
\item (A1 - Bounded eigenvalues) The eigenvalues of $\bar{\Omega}_n$
  are bounded below by $\lambda_{min} > 0$, and bounded above by
  $\lambda_{max} < \infty$ uniformly for all $n$.
\item (A2 - Sub Gaussianity) The random vectors ${\bf Y}^1,\dots,{\bf
    Y}^n$ are \emph{i.i.d.}  sub-Gaussian for every $n \geq 1$, i.e.,
  there exists a constant $c > 0$ such that for every ${\bf x} \in
  \mathbb{R}^{p_n}$, $E \left[ e^{{\bf x}' {\bf Y}^i} \right] \leq
  e^{c {\bf x}' \bar{\Sigma}_n {\bf x}}$, and for every $i, j > 0$,
  there exists $\eta_j > 0$ such that $E \left[ e^{t (Y^i_j)^2} \right]
  < K$ whenever $|t| < \eta_j$. Here $K$ is independent of $i$ and
  $j$.
\item (A3 - Incoherence condition) There exists $\delta < 1$ such that
  for all $(i,j) \notin \mathcal{A}_n$,
$$
\left| \bar{\mathcal{L}}^{''}_{ij,\mathcal{A}_n}(\bar{\Omega}_n) \left[ 
\bar{\mathcal{L}}^{''}_{\mathcal{A}_n, \mathcal{A}_n} (\bar{\Omega}_n) \right]^{-1} 
\sign(\bar{\omega}^o_{\mathcal{A}_n}) \right| \leq \delta, 
$$

\noindent
where for $1 \leq i,j,t,s \leq p_n$ satisfying $i < j$ and $t < s$, 
$$
\bar{\mathcal{L}}^{''}_{ij,ts}(\bar{\Omega}_n) := E_{\bar{\Omega}_n}
\left( (\mathcal{L}^{''}_n (\bar{\Omega}_n))_{ij,ts} \right) = \bar{\Sigma}_{n,js} 
1_{\{i=t\}} + \bar{\Sigma}_{n,it} 1_{\{j=s\}} + \bar{\Sigma}_{n,is} 1_{\{j=t\}} + 
\bar{\Sigma}_{n,jt} 1_{\{i=s\}}. 
$$

\noindent
Conditions analogous to (A3) have been used in
\cite{Zhao2006, Peng2009, Meinshausen2006} to establish high-dimensional 
model selection consistency. In the context of lasso regression, 
\cite{Zhao2006} show that such a condition (which they refer to as an 
irrepresentable condition) is almost necessary and sufficient for model selection 
consistency, and provide some examples when this condition is satisfied. We provide 
some examples of situations where
the condition (A3) is satisfied, along the lines of Zhao and Yu (2006), 
in Supplemental section M. 
\end{itemize}

Define $\bar{\theta}_n^o = ((\bar{\theta}_{n,ij}))_{1 \leq i < j \leq
  p_n} \in \mathbb{R}^{{p_n}({p_n}-1)/2}$ by $\bar\theta_{n,ij} =
\frac{\bar\omega_{n,ij}}{\sqrt{\widehat{\alpha}_{n,ii}
    \widehat{\alpha}_{n,jj}}}$ for $1 \leq i < j \leq p_n$. Let $s_n =
\min_{(i,j) \in \mathcal{A}_n} \bar{\omega}_{n,ij}$.  The assumptions
above can be used to establish the following theorem. 
\begin{thm} \label{thmconcord}
  Suppose that assumptions (A0)-(A3) are satisfied. Suppose $p_n =
  O(n^\kappa)$ for some $\kappa > 0$, $q_n = o \left( \sqrt{n/\log n}
  \right)$, $\sqrt{\frac{q_n \log n}{n}} = o(\lambda_n)$, $\lambda_n
  \sqrt{n/\log n} \rightarrow \infty$, $\frac{s_n}{\sqrt{q_n} \lambda_n} 
  \rightarrow \infty$ and $\sqrt{q_n} \lambda_n\rightarrow 0$, as $n 
  \rightarrow \infty$.  Then there exists a constant $C$ such that for 
  any $\eta > 0$, the following events hold with probability at least 
  $1 - O(n^{-\eta})$.
  \begin{itemize}
  \item There exists a minimizer $\widehat{\omega}^o_n =
    ((\widehat{\omega}_{n,ij}))_{1 \leq i < j \leq p_n}$ of $Q_{\con}
    (\omega^o, \widehat{\boldsymbol \alpha}_n)$.
  \item Any minimizer $\widehat{\omega}^o_n$ of $Q_{\con} (\omega^o,
    \widehat{\boldsymbol \alpha}_n)$ satisfies $ \| \widehat{\omega}^o_n
    - \bar{\omega}^o_n \|_2 \leq C \sqrt{q_n} \lambda_n $ and $
    \sign(\widehat{\omega}_{n,ij}) = \sign(\bar{\omega}_{n,ij}), \;
    \forall\ 1 \leq i < j \leq p_n.$
  \end{itemize}
\end{thm}

\noindent
The proof of the above theorem is provided in 
Supplemental section \ref{sect:asymptotics}.

\section{Conclusion} \label{sect:conclusion}

\noindent
This paper proposes a novel regression based graphical model selection
method that aims to overcome some of the shortcomings of current
methods, but at the same time retain their respective strengths. We
first place the highly useful SPACE method in an optimization
framework, which in turn allows us to identify SPACE with a specific
objective function. These and other insights lead to the formulation
of the CONCORD objective function. It is then shown that the CONCORD
objective function is comprised of quadratic forms, is convex, and can
be regarded as a penalized pseudo-likelihood. A coordinate-wise
descent algorithm that minimizes this objective, via closed form
iterates, is proposed, and subsequently analyzed. The convergence of
this coordinate-wise descent algorithm is established rigorously, thus
ensuring that CONCORD leads to well defined symmetric partial
correlation estimates that are always computable - a guarantee that is
not available with popular regression based methods. Large sample
properties of CONCORD establish consistency of the method as both the
sample size and dimension tend to infinity. The performance of CONCORD
is also illustrated via simulations and is shown to be competitive in
terms of graphical model selection accuracy and timing. CONCORD is
then applied to a biomedical dataset and to a finance dataset, leading
to novel findings. Last but not least, a framework that unifies all
pseudo-likelihood methods is established, yielding important insights. 

Given the attractive properties of CONCORD, a natural question that arises is whether 
one should move away from penalized likelihood estimation (such as Glasso) and rather use only 
pseudo-likelihood methods. We note that CONCORD is attractive over Glasso for several reasons: 
Firstly, it does not assume Gaussianity and is hence more flexible. Secondly, the computational 
complexity per iteration of CONCORD is lower than that of Glasso. Thirdly, CONCORD is faster (in 
terms of wall clock time) than Glasso by an entire order of magnitude in higher dimensions. 
Fourthly, CONCORD delivers better model selection performance. It is however 
important to note that if there is a compelling reason to assume multivariate Gaussianity (which some 
applications may warrant), then using both Glasso and CONCORD can potentially be useful for 
affirming multivariate associations of interest. In this sense, the two classes of methods could 
be complementary in many practical applications. 

\appendix

\newpage
\pagenumbering{gobble}

\setlength{\bibhang}{0.5em}

\bibliographystyle{apalike}
\bibliography{library}

\begin{thebibliography}{}

\bibitem[Banerjee et~al., 2008]{Banerjee2008}
Banerjee, O., {El Ghaoui}, L., and D'Aspremont, A. (2008).
\newblock {Model Selection Through Sparse Maximum Likelihood Estimation for
  Multivariate Gaussian or Binary Data}.
\newblock {\em The Journal of Machine Learning Research}, 9:485--516.

\bibitem[Besag, 1975]{Besag1975}
Besag, J. (1975).
\newblock {Statistical Analysis of Non-Lattice Data}.
\newblock {\em Journal of the Royal Statistical Society. Series D (The
  Statistician)}, 24(3):179--195.

\bibitem[Bibby et~al., 2009]{Bibby2009}
Bibby, R.~A., Tang, C., Faisal, A., Drosopoulos, K., Lubbe, S., Houlston, R.,
  Bayliss, R., and Linardopoulos, S. (2009).
\newblock {A cancer-associated aurora A mutant is mislocalized and misregulated
  due to loss of interaction with TPX2.}
\newblock {\em The Journal of Biological Chemistry}, 284(48):33177--84.

\bibitem[Carter et~al., 2004]{Carter2004}
Carter, S.~L., Brechb\"{u}hler, C.~M., Griffin, M., and Bond, A.~T. (2004).
\newblock {Gene co-expression network topology provides a framework for
  molecular characterization of cellular state.}
\newblock {\em Bioinformatics (Oxford, England)}, 20(14):2242--50.

\bibitem[Chan et~al., 1999]{Chan1999}
Chan, L.~K., Karceski, J., and Lakonishok, J. (1999).
\newblock On portfolio optimization: Forecasting covariances and choosing the
  risk model.
\newblock Working Paper 7039, National Bureau of Economic Research.

\bibitem[Chang et~al., 2005]{Chang2005}
Chang, H.~Y. et~al. (2005).
\newblock {Robustness, scalability, and integration of a wound-response gene
  expression signature in predicting breast cancer survival}.
\newblock {\em Proceedings of the National Academy of Sciences of the United
  States of America}, 102(10):3738--3743.

\bibitem[Fawcett, 2006]{Fawcett2006}
Fawcett, T. (2006).
\newblock {An introduction to ROC analysis}.
\newblock {\em Pattern Recognition Letters}, 27(8):861--874.

\bibitem[Friedman et~al., 2008]{Friedman2008}
Friedman, J., Hastie, T., and Tibshirani, R. (2008).
\newblock {Sparse inverse covariance estimation with the graphical lasso}.
\newblock {\em Biostatistics}, 9(3):432--441.

\bibitem[Friedman et~al., 2010]{Friedman2010}
Friedman, J., Hastie, T., and Tibshirani, R. (2010).
\newblock {Applications of the lasso and grouped lasso to the estimation of
  sparse graphical models}.
\newblock Technical report, Stanford University.

\bibitem[Han et~al., 2004]{Han2004}
Han, J.-D.~J. et~al. (2004).
\newblock {Evidence for dynamically organized modularity in the yeast
  protein-protein interaction network.}
\newblock {\em Nature}, 430(6995):88--93.

\bibitem[Hastie et~al., 2009]{HastieESL}
Hastie, T., Tibshirani, R., and Friedman, J.~H. (2009).
\newblock {\em {The Elements of Statistical Learning}}.
\newblock Springer.

\bibitem[Jensen et~al., 1991]{Jensen1991}
Jensen, S. r.~T., Johansen, S.~r., and Lauritzen, S.~L. (1991).
\newblock {Globally Convergent Algorithms for Maximizing Likelihood Function}.
\newblock {\em Biometrika}, 78(4):867--877.

\bibitem[Jeong et~al., 2001]{Jeong2001}
Jeong, H., Mason, S.~P., Barabasi, A.-L., and Oltvai, Z.~N. (2001).
\newblock {Lethality and centrality in protein networks}.
\newblock {\em Nature}, 411(6833):41--42.

\bibitem[Lauritzen, 1996]{Lauritzen1996}
Lauritzen, S.~L. (1996).
\newblock {\em {Graphical Models}}.
\newblock Oxford University Press, USA.

\bibitem[Ledoit and Wolf, 2004]{Ledoit2004}
Ledoit, O. and Wolf, M. (2004).
\newblock {A well-conditioned estimator for large-dimensional covariance
  matrices}.
\newblock {\em Journal of Multivariate Analysis}, 88(2):365--411.

\bibitem[Lee and Hastie, 2014]{leehstmdml}
Lee, J.~D. and Hastie, T.~J. (2014).
\newblock Learning the structure of mixed graphical models.
\newblock {\em to appear in Journal of Computational and Graphical Statistics}.

\bibitem[Luenberger, 1997]{Luenberger1997}
Luenberger, D.~G. (1997).
\newblock {\em {Investment Science}}.
\newblock Oxford University Press, USA.

\bibitem[Markowitz, 1952]{Markowitz1952}
Markowitz, H. (1952).
\newblock {Portfolio Selection}.
\newblock {\em The Journal of Finance}, 7(1):77--91.

\bibitem[Maxwell et~al., 2011]{Maxwell2011}
Maxwell, C.~A., Benítez, J., Gómez-Baldó, L., Osorio, A., Bonifaci, N.,
  Fernández-Ramires, R., Costes, S.~V., Guinó, E., Chen, H., Evans, G. J.~R.,
  Mohan, P., Català, I., Petit, A., Aguilar, H., Villanueva, A., Aytes, A.,
  Serra-Musach, J., Rennert, G., Lejbkowicz, F., Peterlongo, P., Manoukian, S.,
  Peissel, B., Ripamonti, C.~B., Bonanni, B., Viel, A., Allavena, A., Bernard,
  L., Radice, P., Friedman, E., Kaufman, B., Laitman, Y., Dubrovsky, M.,
  Milgrom, R., Jakubowska, A., Cybulski, C., Gorski, B., Jaworska, K., Durda,
  K., Sukiennicki, G., Lubiński, J., Shugart, Y.~Y., Domchek, S.~M., Letrero,
  R., Weber, B.~L., Hogervorst, F. B.~L., Rookus, M.~A., Collee, J.~M.,
  Devilee, P., Ligtenberg, M.~J., van~der Luijt, R.~B., Aalfs, C.~M., Waisfisz,
  Q., Wijnen, J., van Roozendaal, C. E.~P., Easton, D.~F., Peock, S., Cook, M.,
  Oliver, C., Frost, D., Harrington, P., Evans, D.~G., Lalloo, F., Eeles, R.,
  Izatt, L., Chu, C., Eccles, D., Douglas, F., Brewer, C., Nevanlinna, H.,
  Heikkinen, T., Couch, F.~J., Lindor, N.~M., Wang, X., Godwin, A.~K., Caligo,
  M.~A., Lombardi, G., Loman, N., Karlsson, P., Ehrencrona, H., von
  Wachenfeldt, A., Bjork~Barkardottir, R., Hamann, U., Rashid, M.~U., Lasa, A.,
  Caldés, T., Andrés, R., Schmitt, M., Assmann, V., Stevens, K., Offit, K.,
  Curado, J., Tilgner, H., Guigó, R., Aiza, G., Brunet, J., Castellsagué, J.,
  Martrat, G., Urruticoechea, A., Blanco, I., Tihomirova, L., Goldgar, D.~E.,
  Buys, S., John, E.~M., Miron, A., Southey, M., Daly, M.~B., Schmutzler,
  R.~K., Wappenschmidt, B., Meindl, A., Arnold, N., Deissler, H.,
  Varon-Mateeva, R., Sutter, C., Niederacher, D., Imyamitov, E., Sinilnikova,
  O.~M., Stoppa-Lyonne, D., Mazoyer, S., Verny-Pierre, C., Castera, L.,
  de~Pauw, A., Bignon, Y.-J., Uhrhammer, N., Peyrat, J.-P., Vennin, P.,
  Fert~Ferrer, S., Collonge-Rame, M.-A., Mortemousque, I., Spurdle, A.~B.,
  Beesley, J., Chen, X., Healey, S., Barcellos-Hoff, M.~H., Vidal, M., Gruber,
  S.~B., Lázaro, C., Capellá, G., McGuffog, L., Nathanson, K.~L., Antoniou,
  A.~C., Chenevix-Trench, G., Fleisch, M.~C., Moreno, V., Pujana, M.~A., HEBON,
  EMBRACE, SWE-BRCA, BCFR, {GEMO Study Collaborators}, and kConFab (2011).
\newblock Interplay between brca1 and rhamm regulates epithelial apicobasal
  polarization and may influence risk of breast cancer.
\newblock {\em PLoS Biol}, 9(11):e1001199.

\bibitem[Mazumder and Hastie, 2012]{Mazumder2012}
Mazumder, R. and Hastie, T. (2012).
\newblock {Exact Covariance Thresholding into Connected Components for
  Large-Scale Graphical Lasso}.
\newblock {\em The Journal of Machine Learning Research}, 13:781--794.

\bibitem[Meinshausen and B\"{u}hlmann, 2006]{Meinshausen2006}
Meinshausen, N. and B\"{u}hlmann, P. (2006).
\newblock {High-dimensional graphs and variable selection with the Lasso}.
\newblock {\em The Annals of Statistics}, 34(3):1436--1462.

\bibitem[Merton, 1980]{Merton1980}
Merton, R.~C. (1980).
\newblock On estimating the expected return on the market: An exploratory
  investigation.
\newblock Working Paper 444, National Bureau of Economic Research.

\bibitem[Newman, 2003]{Newman2003}
Newman, M. (2003).
\newblock The structure and function of complex networks.
\newblock {\em SIAM Review}, 45(2):167--256.

\bibitem[Peng et~al., 2009]{Peng2009}
Peng, J., Wang, P., Zhou, N., and Zhu, J. (2009).
\newblock {Partial Correlation Estimation by Joint Sparse Regression Models}.
\newblock {\em Journal of the American Statistical Association},
  104(486):735--746.

\bibitem[Rocha et~al., 2008]{Rocha2008}
Rocha, G., Zhao, P., and Yu, B. (2008).
\newblock {A path following algorithm for Sparse Pseudo-Likelihood Inverse
  Covariance Estimation (SPLICE)}.
\newblock Technical report, Statistics Department, UC Berkeley, Berkeley, CA.

\bibitem[Speed and Kiiveri, 1986]{Speed1986}
Speed, T.~P. and Kiiveri, H.~T. (1986).
\newblock {Gaussian Markov Distributions over Finite Graphs}.
\newblock {\em The Annals of Statistics}, 14(1):138--150.

\bibitem[Tseng, 1988]{Tseng1988}
Tseng, P. (1988).
\newblock {Coordinate ascent for maximizing nondifferentiable concave
  functions}.
\newblock Technical report, Massachusetts Institute of Technology.

\bibitem[Tseng, 2001]{Tseng2001}
Tseng, P. (2001).
\newblock {Convergence of a block coordinate descent method for
  nondifferentiable minimization}.
\newblock {\em Journal of Optimization Theory and Applications},
  109(3):475--494.

\bibitem[Won et~al., 2012]{Won2012}
Won, J.-H., Lim, J., Kim, S.-J., and Rajaratnam, B. (2012).
\newblock {Condition Number Regularized Covariance Estimation}.
\newblock {\em Journal of the Royal Statistical Society: Series B}.

\bibitem[Xu et~al., 2011]{Xu2011}
Xu, P.-F., Guo, J., and He, X. (2011).
\newblock {An Improved Iterative Proportional Scaling Procedure for Gaussian
  Graphical Models}.
\newblock {\em Journal of Computational and Graphical Statistics},
  20(2):417--431.

\bibitem[Zangwill, 1969]{Zangwill1969}
Zangwill, W. (1969).
\newblock {\em Nonlinear programming: a unified approach}.
\newblock Prentice-Hall international series in management. Prentice-Hall,
  Englewood Cliffs, NJ.

\bibitem[Zhao and Yu, 2006]{Zhao2006}
Zhao, P. and Yu, B. (2006).
\newblock {On Model Selection Consistency of Lasso}.
\newblock {\em Journal of Machine Learning Research}, 7:2541--2563.

\end{thebibliography}


\begin{thebibliography}{}

\bibitem[Albergaria et~al., 2009]{Albergaria2009}
Albergaria, A., Paredes, J., Sousa, B., Milanezi, F., Carneiro, V., Bastos, J.,
  Costa, S., Vieira, D., Lopes, N., Lam, E.~W., Lunet, N., and Schmitt, F.
  (2009).
\newblock {Expression of FOXA1 and GATA-3 in breast cancer: the prognostic
  significance in hormone receptor-negative tumours}.
\newblock {\em Breast Cancer Research}, 11(3):R40.

\bibitem[Bibby et~al., 2009]{Bibby2009}
Bibby, R.~A., Tang, C., Faisal, A., Drosopoulos, K., Lubbe, S., Houlston, R.,
  Bayliss, R., and Linardopoulos, S. (2009).
\newblock {A cancer-associated aurora A mutant is mislocalized and misregulated
  due to loss of interaction with TPX2.}
\newblock {\em The Journal of Biological Chemistry}, 284(48):33177--84.

\bibitem[Davidson et~al., 2011]{Davidson2011}
Davidson, B., Stavnes, H.~T., Holth, A., Chen, X., Yang, Y., Shih, I.-M., and
  Wang, T.-L. (2011).
\newblock {Gene expression signatures differentiate ovarian/peritoneal serous
  carcinoma from breast carcinoma in effusions.}
\newblock {\em Journal of Cellular and Molecular Medicine}, 15(3):535--44.

\bibitem[Du et~al., 2012]{Du2012}
Du, J., Li, L., Ou, Z., Kong, C., Zhang, Y., Dong, Z., Zhu, S., Jiang, H.,
  Shao, Z., Huang, B., and Lu, J. (2012).
\newblock {FOXC1, a target of polycomb, inhibits metastasis of breast cancer
  cells.}
\newblock {\em Breast Cancer Research and Treatment}, 131(1):65--73.

\bibitem[Eeckhoute et~al., 2007]{Eeckhoute2007}
Eeckhoute, J., Keeton, E.~K., Lupien, M., Krum, S.~A., Carroll, J.~S., and
  Brown, M. (2007).
\newblock {Positive cross-regulatory loop ties GATA-3 to estrogen receptor
  alpha expression in breast cancer.}
\newblock {\em Cancer Research}, 67(13):6477--83.

\bibitem[Eschenbrenner et~al., 2011]{Eschenbrenner2011}
Eschenbrenner, J., Winsel, S., Hammer, S., Sommer, A., Mittelstaedt, K.,
  Drosch, M., Klar, U., Sachse, C., Hannus, M., Seidel, M., Weiss, B., Merz,
  C., Siemeister, G., and Hoffmann, J. (2011).
\newblock {Evaluation of activity and combination strategies with the
  microtubule-targeting drug sagopilone in breast cancer cell lines.}
\newblock {\em Frontiers in Oncology}, 1:44.

\bibitem[Glinsky et~al., 2005]{Glinsky2005}
Glinsky, G.~V., Berezovska, O., and Glinskii, A.~B. (2005).
\newblock {Microarray analysis identifies a death-from-cancer signature
  predicting therapy failure in patients with multiple types of cancer.}
\newblock {\em The Journal of clinical investigation}, 115(6):1503--21.

\bibitem[Jiang et~al., 2010]{Jiang2010}
Jiang, S., Katayama, H., Wang, J., Li, S.~A., Hong, Y., Radvanyi, L., Li,
  J.~J., and Sen, S. (2010).
\newblock {Estrogen-induced aurora kinase-A (AURKA) gene expression is
  activated by GATA-3 in estrogen receptor-positive breast cancer cells.}
\newblock {\em Hormones \& Cancer}, 1(1):11--20.

\bibitem[Joosse et~al., 2012]{Joosse2012}
Joosse, S.~A., Hannemann, J., Sp\"{o}tter, J., Bauche, A., Andreas, A.,
  M\"{u}ller, V., and Pantel, K. (2012).
\newblock {Changes in Keratin Expression during Metastatic Progression of
  Breast Cancer: Impact on the Detection of Circulating Tumor Cells.}
\newblock {\em Clinical cancer research : an official journal of the American
  Association for Cancer Research}, 18(4):993--1003.

\bibitem[Katoh, 2008]{Katoh2008}
Katoh, M. (2008).
\newblock {WNT signaling in stem cell biology and regenerative medicine.}
\newblock {\em Current Drug Targets}, 9(7):565--70.

\bibitem[Khare and Rajaratnam, 2014]{kharerajl1}
Khare, K. and Rajaratnam, B. (2014).
\newblock Convergence of cyclic coordinate l1 minimization.
\newblock {\em {Preprint, Department of Statistics, Stanford University (soon
  to be available on arxiv)}}.

\bibitem[Koboldt and Others, 2012]{Koboldt2012}
Koboldt, D.~C. and Others (2012).
\newblock {Comprehensive molecular portraits of human breast tumours.}
\newblock {\em Nature}, 490(7418):61--70.

\bibitem[Kraus et~al., 2010]{Kraus2010}
Kraus, T.~S., Cohen, C., and Siddiqui, M.~T. (2010).
\newblock {Prostate-specific antigen and hormone receptor expression in male
  and female breast carcinoma.}
\newblock {\em Diagnostic Pathology}, 5:63.

\bibitem[Lacroix and Leclercq, 2004]{Lacroix2004}
Lacroix, M. and Leclercq, G. (2004).
\newblock {About GATA3, HNF3A, and XBP1, three genes co-expressed with the
  oestrogen receptor-alpha gene (ESR1) in breast cancer.}
\newblock {\em Molecular and Cellular Endocrinology}, 219(1-2):1--7.

\bibitem[Lee and Hastie, 2014]{leehstmdml}
Lee, J.~D. and Hastie, T.~J. (2014).
\newblock Learning the structure of mixed graphical models.
\newblock {\em to appear in Journal of Computational and Graphical Statistics}.

\bibitem[Licata et~al., 2010]{Licata2010}
Licata, L.~A., Hostetter, C.~L., Crismale, J., Sheth, A., and Keen, J.~C.
  (2010).
\newblock {The RNA-binding protein HuR regulates GATA3 mRNA stability in human
  breast cancer cell lines.}
\newblock {\em Breast Cancer Research and Treatment}, 122(1):55--63.

\bibitem[Maxwell et~al., 2011]{Maxwell2011}
Maxwell, C.~A., Benítez, J., Gómez-Baldó, L., Osorio, A., Bonifaci, N.,
  Fernández-Ramires, R., Costes, S.~V., Guinó, E., Chen, H., Evans, G. J.~R.,
  Mohan, P., Català, I., Petit, A., Aguilar, H., Villanueva, A., Aytes, A.,
  Serra-Musach, J., Rennert, G., Lejbkowicz, F., Peterlongo, P., Manoukian, S.,
  Peissel, B., Ripamonti, C.~B., Bonanni, B., Viel, A., Allavena, A., Bernard,
  L., Radice, P., Friedman, E., Kaufman, B., Laitman, Y., Dubrovsky, M.,
  Milgrom, R., Jakubowska, A., Cybulski, C., Gorski, B., Jaworska, K., Durda,
  K., Sukiennicki, G., Lubiński, J., Shugart, Y.~Y., Domchek, S.~M., Letrero,
  R., Weber, B.~L., Hogervorst, F. B.~L., Rookus, M.~A., Collee, J.~M.,
  Devilee, P., Ligtenberg, M.~J., van~der Luijt, R.~B., Aalfs, C.~M., Waisfisz,
  Q., Wijnen, J., van Roozendaal, C. E.~P., Easton, D.~F., Peock, S., Cook, M.,
  Oliver, C., Frost, D., Harrington, P., Evans, D.~G., Lalloo, F., Eeles, R.,
  Izatt, L., Chu, C., Eccles, D., Douglas, F., Brewer, C., Nevanlinna, H.,
  Heikkinen, T., Couch, F.~J., Lindor, N.~M., Wang, X., Godwin, A.~K., Caligo,
  M.~A., Lombardi, G., Loman, N., Karlsson, P., Ehrencrona, H., von
  Wachenfeldt, A., Bjork~Barkardottir, R., Hamann, U., Rashid, M.~U., Lasa, A.,
  Caldés, T., Andrés, R., Schmitt, M., Assmann, V., Stevens, K., Offit, K.,
  Curado, J., Tilgner, H., Guigó, R., Aiza, G., Brunet, J., Castellsagué, J.,
  Martrat, G., Urruticoechea, A., Blanco, I., Tihomirova, L., Goldgar, D.~E.,
  Buys, S., John, E.~M., Miron, A., Southey, M., Daly, M.~B., Schmutzler,
  R.~K., Wappenschmidt, B., Meindl, A., Arnold, N., Deissler, H.,
  Varon-Mateeva, R., Sutter, C., Niederacher, D., Imyamitov, E., Sinilnikova,
  O.~M., Stoppa-Lyonne, D., Mazoyer, S., Verny-Pierre, C., Castera, L.,
  de~Pauw, A., Bignon, Y.-J., Uhrhammer, N., Peyrat, J.-P., Vennin, P.,
  Fert~Ferrer, S., Collonge-Rame, M.-A., Mortemousque, I., Spurdle, A.~B.,
  Beesley, J., Chen, X., Healey, S., Barcellos-Hoff, M.~H., Vidal, M., Gruber,
  S.~B., Lázaro, C., Capellá, G., McGuffog, L., Nathanson, K.~L., Antoniou,
  A.~C., Chenevix-Trench, G., Fleisch, M.~C., Moreno, V., Pujana, M.~A., HEBON,
  EMBRACE, SWE-BRCA, BCFR, {GEMO Study Collaborators}, and kConFab (2011).
\newblock Interplay between brca1 and rhamm regulates epithelial apicobasal
  polarization and may influence risk of breast cancer.
\newblock {\em PLoS Biol}, 9(11):e1001199.

\bibitem[Mohajeri et~al., 2011]{Mohajeri2011}
Mohajeri, A., Zarghami, N., {Pourhasan Moghadam}, M., Alani, B., Montazeri, V.,
  Baiat, A., and Fekhrjou, A. (2011).
\newblock {Prostate-specific antigen gene expression and telomerase activity in
  breast cancer patients: possible relationship to steroid hormone receptors.}
\newblock {\em Oncology Research}, 19(8-9):375--80.

\bibitem[Pellegrino et~al., 1988]{Pellegrino1988}
Pellegrino, M.~B., Asch, B.~B., Connolly, J.~L., and Asch, H.~L. (1988).
\newblock {Differential expression of keratins 13 and 16 in normal epithelium,
  benign lesions, and ductal carcinomas of the human breast determined by the
  monoclonal antibody Ks8.12.}
\newblock {\em Cancer Research}, 48(20):5831--6.

\bibitem[Ray et~al., 2011]{Ray2011}
Ray, P.~S., Bagaria, S.~P., Wang, J., Shamonki, J.~M., Ye, X., Sim, M.-S.,
  Steen, S., Qu, Y., Cui, X., and Giuliano, A.~E. (2011).
\newblock {Basal-like breast cancer defined by FOXC1 expression offers superior
  prognostic value: a retrospective immunohistochemical study.}
\newblock {\em Annals of Surgical Oncology}, 18(13):3839--47.

\bibitem[R\o~nneberg et~al., 2011]{Ronneberg2011}
R\o~nneberg, J.~A., Fleischer, T., Solvang, H.~K., Nordgard, S.~H., Edvardsen,
  H., Potapenko, I., Nebdal, D., Daviaud, C., Gut, I., Bukholm, I., Naume,
  B.~r., B\o~rresen Dale, A.-L., Tost, J., and Kristensen, V. (2011).
\newblock {Methylation profiling with a panel of cancer related genes:
  association with estrogen receptor, TP53 mutation status and expression
  subtypes in sporadic breast cancer.}
\newblock {\em Molecular Oncology}, 5(1):61--76.

\bibitem[Robinson et~al., 2011]{Robinson2011}
Robinson, J. L.~L., Macarthur, S., Ross-Innes, C.~S., Tilley, W.~D., Neal,
  D.~E., Mills, I.~G., and Carroll, J.~S. (2011).
\newblock {Androgen receptor driven transcription in molecular apocrine breast
  cancer is mediated by FoxA1.}
\newblock {\em The EMBO Journal}, 30(15):3019--27.

\bibitem[Sauter et~al., 2004]{Sauter2004}
Sauter, E.~R., Lininger, J., Magklara, A., Hewett, J.~E., and Diamandis, E.~P.
  (2004).
\newblock {Association of kallikrein expression in nipple aspirate fluid with
  breast cancer risk.}
\newblock {\em International Journal of Cancer}, 108(4):588--91.

\bibitem[Shimo et~al., 2007]{Shimo2007}
Shimo, A., Nishidate, T., Ohta, T., Fukuda, M., Nakamura, Y., and Katagiri, T.
  (2007).
\newblock {Elevated expression of protein regulator of cytokinesis 1, involved
  in the growth of breast cancer cells.}
\newblock {\em Cancer Science}, 98(2):174--81.

\bibitem[Shimo et~al., 2008]{Shimo2008}
Shimo, A., Tanikawa, C., Nishidate, T., Lin, M.-L., Matsuda, K., Park, J.-H.,
  Ueki, T., Ohta, T., Hirata, K., Fukuda, M., Nakamura, Y., and Katagiri, T.
  (2008).
\newblock {Involvement of kinesin family member 2C/mitotic
  centromere-associated kinesin overexpression in mammary carcinogenesis.}
\newblock {\em Cancer Science}, 99(1):62--70.

\bibitem[Sizemore and Keri, 2012]{Sizemore2012}
Sizemore, S.~T. and Keri, R.~A. (2012).
\newblock {The Forkhead Box Transcription Factor FOXC1 Promotes Breast Cancer
  Invasion by Inducing Matrix Metalloprotease 7 (MMP7) Expression.}
\newblock {\em The Journal of Biological Chemistry}, 287(29):24631--40.

\bibitem[Tkocz et~al., 2012]{Tkocz2011}
Tkocz, D., Crawford, N.~T., Buckley, N.~E., Berry, F.~B., Kennedy, R.~D.,
  Gorski, J.~J., Harkin, D.~P., and Mullan, P.~B. (2012).
\newblock {BRCA1 and GATA3 corepress FOXC1 to inhibit the pathogenesis of
  basal-like breast cancers.}
\newblock {\em Oncogene}, 31(32):3667--3678.

\bibitem[Wang et~al., 2012]{Wang2012}
Wang, J., Ray, P.~S., Sim, M.-S., Zhou, X.~Z., Lu, K.~P., Lee, A.~V., Lin, X.,
  Bagaria, S.~P., Giuliano, A.~E., and Cui, X. (2012).
\newblock {FOXC1 regulates the functions of human basal-like breast cancer
  cells by activating NF-$\kappa$B signaling.}
\newblock {\em Oncogene}.

\bibitem[Yan et~al., 2010]{Yan2010}
Yan, W., Cao, Q.~J., Arenas, R.~B., Bentley, B., and Shao, R. (2010).
\newblock {GATA3 inhibits breast cancer metastasis through the reversal of
  epithelial-mesenchymal transition}.
\newblock {\em The Journal of Biological Chemistry}, 285(18):14042--14051.

\bibitem[Yang et~al., 2002]{Yang2002}
Yang, Q., Nakamura, M., Nakamura, Y., Yoshimura, G., Suzuma, T., Umemura, T.,
  Tamaki, T., Mori, I., Sakurai, T., and Kakudo, K. (2002).
\newblock {Correlation of prostate-specific antigen promoter polymorphisms with
  clinicopathological characteristics in breast cancer.}
\newblock {\em Anticancer Research}, 22(3):1825--8.

\bibitem[Zheng et~al., 2012]{Zheng2012}
Zheng, Y., Huo, D., Zhang, J., Yoshimatsu, T.~F., Niu, Q., and Olopade, O.~I.
  (2012).
\newblock {Microsatellites in the Estrogen Receptor (ESR1, ESR2) and Androgen
  Receptor (AR) Genes and Breast Cancer Risk in African American and Nigerian
  Women.}
\newblock {\em PLoS ONE}, 7(7):e40494.

\end{thebibliography}

\newpage
\pagenumbering{arabic}
\section*{Supplemental Section}

\section{Proof of Lemma \ref{convexity of symlasso}}
\label{sect:proof convexity}
Let ${\bf Y}$ denote the $n \times p$ matrix with $j^{th}$ column
given by ${\bf Y}_j$ for $j = 1,2, \ldots, p$. Define
$Q_{\sym}({\boldsymbol\alpha},\breve\Omega) = \frac{1}{2}
\left(\sum_{j=1}^p
  \mathcal{L}_{\sym,j}(\alpha_{jj},\breve\Omega_j)\right) + \lambda
\left(\sum_{1\leq i<j\leq p}
  |\omega_{ij}|\right) \label{symmlasso.objective} $ so that
\begin{align}
  \mathcal{L}_{\sym,j}(\alpha_{jj}, \breve\Omega_j) &= n\log\alpha_{jj} +
  \frac{1}{\alpha_{jj}}\| {\bf Y}_j + {\bf
    Y}\breve\Omega_j\alpha_{jj}\|_2^2
\end{align}
where ${\boldsymbol\alpha}=(\alpha_{11}\ \alpha_{22}\ \cdots\
\alpha_{pp})'$, $\alpha_{jj}=1/\omega_{jj}$ and $\breve\Omega_j$ is
the $j^{th}$ column of $\breve\Omega$. Recall that $\breve\Omega$ is
the matrix $\Omega$ with zeros in place of the diagonal entries. If follows
that 
\begin{align}
  \frac{\partial Q_{\sym}({\boldsymbol\alpha},\breve\Omega)}{\partial
    \alpha_{jj}} = \frac{n}{\alpha_{jj}} - \frac{{\bf Y}_j'{\bf
      Y}_j}{\alpha^2_{jj}}+\breve\Omega_j'{\bf Y}'{\bf
    Y}\breve\Omega_j, \mbox{\quad and \quad}
  \frac{\partial^2
    Q_{\sym}({\boldsymbol\alpha},\breve\Omega)}{\partial
    \alpha_{jj}^2} &= -\frac{n}{\alpha_{jj}^2}+2\frac{{\bf Y}_j'{\bf
      Y}_j}{\alpha_{jj}^3}
\end{align}
It is clear that in general $\partial^2
Q_{\sym}({\boldsymbol\alpha},\breve\Omega)/\partial \alpha_{jj}^2
\not\geq 0$.  Hence, $Q_{\sym}({\boldsymbol\alpha},\breve\Omega)$ is
not convex.

\section{Proof of Lemma \ref{lemma:splice}}
\label{sect:proof splice}
\begin{proof} 
  i) Rewrite the SPLICE objective function
  $Q_{\spl}(\mathbf{B},\mathbf{D})=\mathcal{L}_{\spl}(\mathbf{B},\mathbf{D})
  + \lambda\sum_{i<j}|\beta_{ij}|$ where
  \begin{align*}
    \mathcal{L}_{\spl}(\mathbf{B},\mathbf{D})=\frac{1}{2}\left[n\log\det(\mathbf{D}^2)
      + \tr(\mathbf{D}^{-2}\mathbf{A})\right],
  \end{align*} 
  and
  $\mathbf{A}=[a_{ij}]=(\mathbf{I}-\mathbf{B})\mathbf{Y}'\mathbf{Y}(\mathbf{I}-\mathbf{B}')$. The
  function $\mathcal{L}_{\spl}(\mathbf{B},\mathbf{D})$ with all
  variables fixed except $d_{jj}$ is given by
  \begin{align*}
    \mathcal{L}_{\spl,j}(\mathbf{B},d_{jj})=\frac{1}{2}\left[n\log d_{jj}^2 +
    \frac{a_{jj}}{d_{jj}^2}\right] + \mbox{constants}.
  \end{align*} Now,
  \begin{align*} 
    \frac{\partial Q_{\spl}(\mathbf{B},\mathbf{D})}{\partial d_{jj}}
    &=
    \frac{n}{d_{jj}} - \frac{a_{jj}}{d_{jj}^3}\\
    \frac{\partial ^2 Q_{\spl} (\mathbf{B},\mathbf{D})}{\partial
      d_{jj}^2} &= -\frac{n}{d_{jj}^2} + 3\frac{a_{jj}}{d_{jj}^4}
  \end{align*}  
  It is clear in general $\partial
  Q_{\spl}^2(\mathbf{B},\mathbf{D})/\partial d_{jj}^2 \not\geq 0$. Hence
  $Q_{\spl}(\mathbf{B},\mathbf{D})$ is not convex.
  
  ii) Similarly, define
  $Q_{\spl}^*(\mathbf{B},\mathbf{C})=\mathcal{L}_{\spl}^*(\mathbf{B},\mathbf{C})
  + \lambda\sum_{i<j}|\beta_{ij}|$ where
  $$\mathcal{L}_{\spl}^*(\mathbf{B},\mathbf{C}) = \frac{1}{2}\left[n \log
  \mathbf{C}^{-2} + \tr(\mathbf{C}^2\mathbf{A})\right].$$
  It is clear that for
  a fixed $\mathbf{C}$, $\mathcal{L}_{\spl}^*(\mathbf{B},\mathbf{C})$ is a
  convex function in $\mathbf{B}$ \citep{Rocha2008}. Now for a fixed
  $\mathbf{B}$ let
  \begin{align*}
    \mathcal{L}_{\spl,j}^*(\mathbf{B},c_{jj}) &=
    \frac{1}{2}\left[-2n\log c_{jj} + c_{jj}^2 a_{jj}\right] + \mbox{constants}\\
    \frac{\partial Q_{\spl}^*(\mathbf{B},\mathbf{C})}{\partial c_{jj}}
    &=
    -\frac{n}{c_{jj}} + c_{jj}a_{jj}\\
    \frac{\partial ^2Q_{\spl}^*(\mathbf{B},\mathbf{C})}{\partial
      c_{jj}^2} &= \frac{n}{c_{jj}^2} + a_{jj}
  \end{align*}
  Now, note that $\partial (Q_{\spl}^*)^2(\mathbf{B},\mathbf{C})/\partial
  c_{jj}^2 \geq 0$ since $a_{jj}\geq 0$.
  
  To see that $a_{jj}\geq 0$ note that $\mathbf{A} =
  (\mathbf{I}-\mathbf{B})\mathbf{Y}'\mathbf{Y}(\mathbf{I}-\mathbf{B}')
  = {\bf G}'{\bf G}$, where
  $\mathbf{G}=\mathbf{Y}(\mathbf{I}-\mathbf{B}')$ Now,
  $a_{jj}=\mathbf{G}_{\bullet j}'\mathbf{G}_{\bullet
    j}=\|\mathbf{G}_{\bullet j}\|^2 \geq 0$
\end{proof}

\section{Proof of Lemma \ref{ipm}}
\label{sect:proof ipm}
Note that for $1 \leq i \leq p$, 
\begin{equation} \label{eq3}
Q_{\con} (\Omega) = -n \log \omega_{ii} + \frac{n}{2}\left(\omega_{ii}^2 s_{ii} + 2 
\omega_{ii} \sum_{j \neq i} \omega_{ij} s_{ij}\right) + \mbox{ terms independent of } \omega_{ii}. 
\end{equation}
where $s_{ij}={\bf Y}_i'{\bf Y}_j/n$. Hence,
\begin{eqnarray*}
  \frac{\partial}{\partial \omega_{ii}} Q_{\con} (\Omega) = 0 
  &\Leftrightarrow& -\frac{1}{\omega_{ii}} + \omega_{ii} s_{ii} + \sum_{j \neq i} \omega_{ij} 
  s_{ij} = 0\\
  &\Leftrightarrow& \omega_{ii} = \frac{- \sum_{j \neq i} \omega_{ij} s_{ij} + \sqrt{\left( 
        \sum_{j \neq i} \omega_{ij} s_{ij} \right)^2 + 4 s_{ii}}}{2 s_{ii}}, 
\end{eqnarray*}

\noindent
Note that since $\omega_{ii}>0$ the positive root has been retained as the solution. 

Also, for $1 \leq i < j \leq p$, 
\begin{equation} \label{eq4}
Q_{\con} (\Omega) = n\frac{s_{ii} + s_{jj}}{2} \omega_{ij}^2 + n\left( 
\sum_{j' \neq j} \omega_{ij'} s_{jj'} + \sum_{i' \neq i} \omega_{i'j} s_{ii'} \right) \omega_{ij} 
+ \lambda |\omega_{ij}| + \mbox{ terms independent of } \omega_{ij}. 
\end{equation}
 
\noindent
It follows that 
$$
\left( T_{ij} (\Omega) \right)_{ij} = \frac{S_{\lambda\over n}\left(-\left(\sum_{j' \neq j} 
\omega_{ij'} s_{jj'} + \sum_{i' \neq i} \omega_{i'j} s_{ii'}\right)\right)}{s_{ii} + s_{jj}},
$$
where $S_{\eta}$ is the soft-thresholding operator given by
$S_{\eta}(x) = \sign(x)(|x|-\eta)_+ $.

\section{Proof of Lemma \ref{eq:residuals}}
\label{sect:proof residual}
Let ${\bf Y}_j$ denote $j^{th}$ column of the data matrix ${\bf
  Y}$. Then, using the identity $\sum_{k=1}^p \omega_{ik}s_{jk} =
\omega_{ij}s_{jj}+\sum_{k\neq j} \omega_{ik}s_{jk} = \omega_{ii}s_{ij}
+ \sum_{k\neq i}\omega_{ik}s_{jk}$,
\begin{align*}
  \sum_{k \neq j} \omega_{ik} s_{jk} &= -\omega_{ij} s_{jj} +
  \omega_{ii}\left( s_{ij} + \sum_{k\neq i}
    \frac{\omega_{ik}}{\omega_{ii}} s_{jk}
  \right)\nonumber\\
  &= -\omega_{ij} s_{jj} + \omega_{ii}{\bf Y}_j'\left( {\bf Y}_i +
    \sum_{k\neq i} \frac{\omega_{ik}}{\omega_{ii}} {\bf Y}_k
  \right)\nonumber\\
  &= -\omega_{ij} s_{jj} + \omega_{ii}{\bf Y}_j'{\bf r}_i,
\end{align*}
where ${\bf r}_i={\bf Y}_i + \sum_{k\neq i}
\frac{\omega_{ik}}{\omega_{ii}} {\bf Y}_k$ is an $n$-vector of
residuals after regressing the $i^{th}$ variable on the
rest. \hfill$\Box$

\section{Proof of Lemma \ref{eq:residual updates}}
\label{sect:proof residual updates}

\begin{enumerate}
\item Result follows easily from inspecting ${\bf r}_k$ and ${\bf r}_l$.
\item If $\omega_{kl}$ is updated to $\omega_{kl}^*$, it 
follows from part 1 that among all the residual vectors, 
only ${\bf r}_k$ and ${\bf r}_l$ change values. The residual 
vector ${\bf r}_k$ can be updated as follows: 
$${\bf r}_k^* = {\bf r}_k + \frac{(\omega_{kl}^*- 
\omega_{kl})}{\omega_{kk}}{\bf Y}_l\,.$$

\noindent
Clearly, this update requires $O(n)$ operations. The 
vector ${\bf r}_l$ can be updated similarly. 
\item Result follows easily from inspecting ${\bf r}_k$.
\item If $\omega_{kk}$ is updated to $\omega_{kk}^*$, it follows 
from part 3 that among all the residual vectors, only ${\bf r}_k$ 
changes value. The residual vector ${\bf r}_k$ can be updated as 
follows: 
$$ {\bf r}_k^* = ({\bf r}_k - {\bf Y}_k) 
\frac{\omega_{kk}}{\omega_{kk}^*} + {\bf Y}_k\,.$$ 
Clearly, this update requires $O(n)$ operations. \hfill$\Box$
\end{enumerate}

\section{Proof of Lemma \ref{lemma:unify}}
\label{sect:proof unify}

\begin{proof}
  \noindent{\bf (CONCORD)} Let $A=nS$ Expanding the $\ell_2$-norm of
  the residual, we have
  \begin{align*}
    \|\omega_{ii}{\bf Y}_i + \sum_{j \neq i} \omega_{ij} {\bf
      Y}_j\|_2^2
    = \|\sum_{j=1}^p \omega_{ij} {\bf Y}_j\|_2^2
    = \|{\bf Y}\omega_{i\bullet}\|_2^2
    = \omega_{i\bullet}'{\bf Y}'{\bf Y}\omega_{i\bullet}
    = \omega_{i\bullet}'{\bf A}\omega_{i\bullet}
  \end{align*}
  Hence, \eqref{PL:concord} is equivalent to
  \begin{align*}
    \mathcal{L}_{\con}(\Omega) = \frac{1}{2} \sum_{i=1}^p \left( - 2n
      \log \omega_{ii} + \omega_{i\bullet}'\mathbf{A}\omega_{i\bullet}
    \right) &=
    - n \sum_{i=1}^p \log \omega_{ii} + \frac{1}{2}\sum_{i=1}^p\omega_{i\bullet}'A\omega_{i\bullet}\\
    &= -n \log \left(\prod_{i=1}^p \omega_{ii}\right) +
    \frac{n}{2}\tr(\Omega \mathbf{S} \Omega)\\
    &= \frac{n}{2}\left(-\log\det\Omega_D^2 + \tr(\mathbf{S}\Omega^2)\right).
  \end{align*}
   Hence, $G_{\con}(\Omega)=\Omega_D$ and
  $H_{\con}(\Omega)=\Omega^2$

  \vspace{1cm}\noindent{\bf (SPACE with unit weights)}
  Reparameterizing \eqref{PL:space1} using the identity
  $-\rho^{ij}\sqrt{\omega_{jj}/\omega_{ii}}=\omega_{ij}/\omega_{ii}$,
  the $\ell_2$-norm of the residual can be expressed as follows.
  \begin{align*}
    \|{\bf Y}_i + \sum_{j \neq i} \frac{\omega_{ij}}{\omega_{ii}}
    {\bf Y}_j\|_2^2
    = \|\frac{1}{\omega_{ii}}(\omega_{ii}{\bf Y}_i + \sum_{j \neq i}
    \omega_{ij} {\bf Y}_j)\|_2^2
    = \frac{1}{\omega_{ii}^2}
    \omega_{i\bullet}'\mathbf{A}\omega_{i\bullet}
  \end{align*}
  Hence, \eqref{PL:space1} is equivalent to 
  \begin{align*}
    \mathcal{L}_{\spc,1}(\Omega) &= - \frac{n}{2} \log\det\Omega_D +
    \frac{1}{2}\sum_{i=1}^p \frac{1}{\omega_{ii}^2}
    \omega_{i\bullet}'\mathbf{A}\omega_{i\bullet}\\
    &= - \frac{n}{2} \log\det\Omega_D + \frac{n}{2}\sum_{i=1}^p
    \frac{\omega_{i\bullet}'}{\omega_{ii}}\mathbf{S}\frac{\omega_{i\bullet}}{\omega_{ii}}\\
    &=-\frac{n}{2}\log\det\Omega_D +
    \frac{1}{2}\tr(\Omega_D^{-1}\Omega \mathbf{A}\Omega\Omega_D^{-1})\\
    &= \frac{n}{2}\left(-\log\det\Omega_D +
      \tr(\mathbf{S}\Omega\Omega_D^{-2}\Omega)\right).
  \end{align*}
  Therefore, $G_{\spc,1}(\Omega)=\Omega_D$ and
  $H_{\spc,1}(\Omega)=\Omega\Omega_D^{-2}\Omega$.

  \vspace{1cm}\noindent{\bf (SPACE with $\omega_{ii}$ weights)}
  Similar to the analysis for SPACE1 with unit weights, the
  $\ell_2$-norm of the residual for the SPACE2 formulation (i.e., with
  weights $\omega_{ii}$) can be expressed as follows.
  \begin{align*}
    \omega_{ii}\|{\bf Y}_i - \sum_{j \neq i}
    \rho^{ij}\sqrt{\frac{\omega_{jj}}{\omega_{ii}}} {\bf Y}_j\|_2^2 &=
    \omega_{ii}\left(\frac{1}{\omega_{ii}^2}
      \omega_{i\bullet}'\mathbf{A}\omega_{i\bullet}\right)\\
    &= \frac{1}{\omega_{ii}}
    \omega_{i\bullet}'\mathbf{A}\omega_{i\bullet}
  \end{align*}
  Hence, \eqref{PL:space2} is equivalent to
  \begin{align*}
    \mathcal{L}_{\spc,2}(\Omega) &= - \frac{n}{2} \log\det\Omega_D +
    \frac{1}{2}\sum_{i=1}^p \frac{1}{\omega_{ii}}
    \omega_{i\bullet}'\mathbf{A}\omega_{i\bullet}\\
    &= - \frac{n}{2} \log\det\Omega_D + \frac{n}{2}\sum_{i=1}^p
    \frac{\omega_{i\bullet}'}{\sqrt{\omega_{ii}}}\mathbf{S}\frac{\omega_{i\bullet}}{\sqrt{\omega_{ii}}}\\
    &=-\frac{n}{2}\log\det\Omega_D +
    \frac{n}{2}\tr(\Omega_D^{-1/2}\Omega \mathbf{S} \Omega\Omega_D^{-1/2})\\
    &= \frac{n}{2}\left(-\log\det\Omega_D +
      \tr(\mathbf{S}\Omega\Omega_D^{-1}\Omega)\right)
  \end{align*}
  Therefore, $G_{\spc,2}(\Omega)=\Omega_D$ and
  $H_{\spc,2}(\Omega)=\Omega\Omega_D^{-1}\Omega$.

  \vspace{1cm}\noindent{\bf (SYMLASSO)} Reparameterizing
  \eqref{PL:symlasso} by $\alpha_{ii}=1/\omega_{ii}$ and
  $-\rho^{ij}\sqrt{\omega_{jj}/\omega_{ii}}=\omega_{ij}/\omega_{ii}$
  yields \eqref{PL:space2}. It follows that
  $G_{\sym}(\Omega)=\Omega_D$,
  $H_{\sym}(\Omega)=\Omega\Omega_D^{-1}\Omega$.

  \vspace{1cm}\noindent{\bf (SPLICE)} Reparameterizing
  \eqref{PL:splice} by $d_{ii}^2=1/\omega_{ii}$ and
  $\beta_{ij}=\rho^{ij}\sqrt{\omega_{jj}/\omega_{ii}}$ yields
  \eqref{PL:space2}. It follows that $G_{\spl}(\Omega)=\Omega_D$,
  $H_{\spl}(\Omega)=\Omega\Omega_D^{-1}\Omega$.
\end{proof}

\section{Effect of correction factor}\label{sect:correction}

Following steps similar to proof of Lemma \ref{ipm}, the update
formulas for
$\bar{Q}_{\con}(\Omega)=\mathcal{L}_{\con}(\Omega)+\lambda\sum_{i<j}|\omega_{ij}|$
of \eqref{PL:concord} 
can be shown to be
\begin{align}
  (T_{kk}(\Omega))_{kk} &= \frac{-\sum_{j\neq k}\omega_{kj}s_{kj} +
    \sqrt{\left(\sum_{j\neq k}\omega_{kj}s_{kj}\right)^2 +
      2s_{kk}}}{2s_{kk}} \label{eq:raw diag update}\\
  (T_{kl}(\Omega))_{kl} &= \frac{S_{\frac{\lambda}{n}}\left(-\left(
        \sum_{j\neq l}\omega_{kj}s_{jl} + \sum_{j\neq
          k}\omega_{lj}s_{jk}\right)\right)}{s_{kk}+s_{ll}} \label{eq:raw
    offdiag update}
\end{align}

\subsection{Numerical example}
Analysis on a dataset $(n=1000)$ generated from following $\Omega$ was
used for this example.
\begin{align*}
\Omega = 
  \begin{pmatrix}
    1.0 & 0.3 & 0.0\\
    0.3 & 1.0 & 0.3\\
    0.0 & 0.3 & 1.0    
  \end{pmatrix}
\end{align*}
Without penalty, i.e. $\lambda=0$, computed solutions $\Omega_{\con}$
from using CONCORD and $\Omega_{\mbox{\scriptsize uncorrected}}$ from
using update formulas \eqref{eq:raw diag update} and \eqref{eq:raw
  offdiag update} are
\begin{align*}
  \Omega_{\mbox{\scriptsize uncorrected}} =
  \begin{pmatrix}[r]
     0.675 & 0.089 & -0.015\\
     0.089 & 0.658 &  0.117\\
    -0.015 & 0.117 &  0.668
  \end{pmatrix},\ 
  \Omega_{\con} =
  \begin{pmatrix}
    0.974 & 0.257 & 0.007\\
    0.257 & 0.983 & 0.344\\
    0.007 & 0.344 & 0.978
  \end{pmatrix}
\end{align*}
It is clear that the estimate $\Omega_{\con}$ with the correction
factor performs better parameter estimation.

\section{Proof of Theorem \ref{cnvconcord}} \label{sect:convergence:concord}

\noindent
\citesec{kharerajl1} establish convergence of the cyclic coordinatewise minimization algorithm 
for a general class of objective functions. The proof of convergence for CONCORD relies on 
showing that the corresponding objective function is a special case of the general class of 
objective functions considered in \citesec{kharerajl1}. A more detailed version of the 
following argument can be found in \citesec[Section 4.1]{kharerajl1}. We provide the main 
steps here for convenience and completeness. 

Let ${\bf y} = {\bf y} (\Omega) \in \mathbb{R}^{p^2}$ denote a vectorized version of 
$\Omega$ obtained by shifting the corresponding diagonal entry at the bottom of 
each column of $\Omega$, and then stacking the columns on top of each other. Let 
$P^i$ denote the $p \times p$ permutation matrix such that $P^i {\bf z} = (z_1, 
\cdots, z_{i-1}, z_{i+1}, \cdots, z_p, z_i)$ for every ${\bf z} \in \mathbb{R}^p$. It 
follows by the definition of ${\bf y}$ that 
$$
{\bf y} = {\bf y} (\Omega) = ((P^1 \Omega_{\cdot 1})^T, (P^2 \Omega_{\cdot 2})^T, 
\cdots, (P^p \Omega_{\cdot p})^T )^T. 
$$

\noindent
Let ${\bf x} = {\bf x} ({\Omega}) \in \mathbb{R}^{\frac{p(p+1)}{2}}$ be the symmetric 
version of ${\bf y}$, obtained by removing all $\omega_{ij}$ with $i > j$ from ${\bf y}$. 
More precisely, 
$$
{\bf x} = {\bf x} (\Omega) = (\omega_{11}, \omega_{12}, \omega_{22}, \cdots, 
\omega_{1p}, \omega_{2p}, \cdots, \omega_{pp})^T. 
$$

\noindent
Let $\tilde{P}$ be the $p^2 \times \frac{p(p+1)}{2}$ matrix such that every entry of $\tilde{P}$ 
is either $0$ or $1$, exactly one entry in each row of $\tilde{P}$ is equal to $1$, and ${\bf y} 
= \tilde{P} {\bf x}$. Let $\tilde{S}$ be a $p^2 \times p^2$ block diagonal matrix with $p$ 
diagonal blocks, and the $i^{th}$ diagonal block is equal to $\tilde{S}^i := \frac{1}{2} 
P^i S (P^i)^T$, where $S = \frac{1}{n} {\bf Y}^T {\bf Y}$. It follows that 
\begin{eqnarray}
\frac{1}{2} \sum_{i=1}^p \Omega_{\cdot i}^T S \Omega_{\cdot i} = \frac{1}{2} 
\sum_{i=1}^p \Omega_{\cdot i}^T (P^i)^T P^i S (P^i)^T P^i \Omega_{\cdot i} 
&=& \frac{1}{2} \sum_{i=1}^p (P^i \Omega_{\cdot i})^T (P^i S (P^i)^T) (P^i 
\Omega_{\cdot i}) \nonumber\\
&=& {\bf y}^T \tilde{S} {\bf y} \nonumber\\
&=& {\bf x}^T \tilde{P}^T \tilde{S} \tilde{P} {\bf x}. \label{eq1.1} 
\end{eqnarray}

\noindent
Note that for every $1 \leq i \leq p$, the matrix $\tilde{S}^i =
\frac{1}{2} P^i S (P^i)^T$ is positive semi-definite. Let
$\tilde{S}^{1/2}$ denote the $p^2 \times p^2$ block diagonal matrix
with $p$ diagonal blocks, such that the $i^{th}$ diagonal block is
given by $(\tilde{S}^i)^{1/2}$. Let $E = \tilde{S}^{1/2}
\tilde{P}$. It follows by (\ref{eq1.1}) that
\begin{equation} \label{eq1.2}
\frac{1}{2} \sum_{i=1}^p \Omega_{\cdot i}^T S \Omega_{\cdot i} = (E {\bf x})^T 
(E {\bf x}). 
\end{equation}

\noindent
By the definition of ${\bf x} (\Omega)$, we obtain 
\begin{equation} \label{eq1.3}
\omega_{ii} = x_{\frac{i(i+1)}{2}} 
\end{equation} 

\noindent
for every $1 \leq i \leq p$. Let 
$$
S_0 = \left\{ j: \; 1 \leq j \leq \frac{p(p+1)}{2}, \; j \neq \frac{i(i+1)}{2} \mbox{ for any } 1 
\leq i \leq p \right\}, 
$$ 

\noindent
and 
$$
\mathcal{X} = \{{\bf x} \in \mathbb{R}^{\frac{p(p+1)}{2}}: x_j \geq 0 \mbox{ for every } j 
\in S_0^c\}. 
$$

\noindent
It follows by (\ref{eq2}), (\ref{eq1.2}) and (\ref{eq1.3}) that the CONCORD algorithm can be viewed 
as a cyclic coordinatewise minimization algorithm for minimizing the function 
\begin{equation} \label{eq60}
Q_{con} ({\bf x}) = n \left\{ {\bf x}^T E^T E {\bf x} - \sum_{i \in S_0^c} \log x_i +  \frac{\lambda}{n} 
\sum_{j \in S_0} |x_j| \right\}, 
\end{equation}

\noindent
subject to ${\bf x} \in \mathcal{X}$. For every $1 \leq i \leq p(p+1)/2$, there exist $1 \leq k,l 
\leq p$ such that $x_i = \omega_{kl}$. Note that $\|E_{\cdot i}\|^2 = \frac{S_{kk} + S_{ll}}{2} > 
0$. It also follows from \citesec[Lemma 4.1]{kharerajl1} that for every $\xi \in \mathbb{R}$, the 
set $R_\xi := \{{\bf x} \in \mathcal{X}: Q_{con} ({\bf x}) \leq \xi\}$ is bounded in the sense 
that for every $i \in S_0$, $x_i$ is uniformly bounded above and below, and for every $i \in 
S_0^c$, $x_i$ is uniformly bounded above and below (from zero). It follows by 
\citesec[Theorem 3.1]{kharerajl1} that the sequence of iterates produced by the CONCORD 
algorithm converges. 

\section{Application to breast cancer data}
\label{sect:biology example}

\begin{table}[H]
  \begin{center}
    \small
    \begin{tabularx}{\textwidth}{|r|c|c|c|c|X|}
      \hline
      Gene Symbol  
      & \begin{sideways}CONCORD\end{sideways} 
      & \begin{sideways}SYMLASSO\,\,\end{sideways} 
      & \begin{sideways}SPACE1\end{sideways} 
      & \begin{sideways}SPACE2\end{sideways} 
      & Reference \\\hline
      \emph{HNF3A (FOXA1)} & + & + & + & +                        
      & \citesec{Koboldt2012,Albergaria2009,Davidson2011,Lacroix2004,Robinson2011}\\\hline
      \emph{TONDU}         & + & + & + & +                        & \\\hline
      \emph{FZD9}          & + & + & + & +                        & \citesec{Katoh2008,Ronneberg2011}\\\hline
      \emph{KIAA0481}      & + & + & + & +                        & [Gene record discontinued]\\\hline
      \emph{KRT16}         & + & + & + &                          
      & \citesec{Glinsky2005,Joosse2012,Pellegrino1988}\\\hline
      \emph{KNSL6 (KIF2C)} & + &   &   & +                        & \citesec{Eschenbrenner2011,Shimo2007,Shimo2008}\\\hline
      \emph{FOXC1}         & + & + & + & +                        
      & \citesec{Du2012,Sizemore2012,Wang2012,Ray2011,Tkocz2011} \\\hline
      \emph{PSA}           & + & + &   & +                        
      & \citesec{Kraus2010,Mohajeri2011,Sauter2004,Yang2002}\\\hline
      \emph{GATA3}         & + & + & + & +                        
      & \citesec{Koboldt2012,Davidson2011,Albergaria2009,Eeckhoute2007,Jiang2010,Licata2010,Yan2010}\\\hline
      \emph{C20ORF1 (TPX2)}& + &   &   &                          
      & \citesec{Maxwell2011,Bibby2009}\\\hline
      \emph{E48}           &   & + & + & +                        & \\\hline
      \emph{ESR1}          &   &   &   & +                        & \citesec{Zheng2012}\\\hline
    \end{tabularx}
    \caption{Summary of the top hub genes identified by each of the
      four methods, CONCORD, SYMLASSO, SPACE1 \& SPACE2: Genes
      indicated by `+' denote the 10 most highly connected genes for
      each of the methods. References are provided at the end of this
      supplemental section.}
    \label{tbl:hub genes}
  \end{center}
\end{table}

\section{Application to portfolio optimization}
\label{sect:finance example}
 
\subsection{Constituents of Dow Jones Industrial Average}
\label{sect:dow jones}

\begin{table}[H]
  \small
  \centering
  \begin{tabular}{clrrr}\hline
    Symbol& Description                                 & Return (\%) & Risk (\%) & SR \\\hline
    AA    & Alcoa Inc.                                  &  9.593 & 41.970 & 0.109\\
    AXP   & American Express Company                    & 18.706 & 38.913 & 0.352\\
    BA    & The Boeing Company                          & 13.417 & 32.685 & 0.258\\
    BAC   & Bank of America Corporation                 & 13.182 & 48.588 & 0.168\\
    CAT   & Caterpillar Inc.                            & 19.042 & 35.050 & 0.401\\
    CSCO  & Cisco Systems, Inc.                         & 22.650 & 44.565 & 0.396\\
    CVX   & Chevron Corporation                         & 15.486 & 26.716 & 0.392\\
    DD    & E. I. du Pont de Nemours and Company        & 10.591 & 30.537 & 0.183\\
    DIS   & The Walt Disney Company                     & 12.312 & 32.800 & 0.223\\
    GE    & General Electric Company                    & 12.449 & 31.667 & 0.235\\
    HD    & The Home Depot, Inc.                        & 17.266 & 34.422 & 0.356\\
    HPQ   & Hewlett-Packard Company                     & 10.769 & 40.727 & 0.142\\
    IBM   & International Business Machines Corporation & 18.715 & 29.944 & 0.458\\
    INTC  & Intel Corporation                           & 18.325 & 41.543 & 0.321\\
    JNJ   & Johnson \& Johnson                          & 13.664 & 22.087 & 0.392\\
    JPM   & JPMorgan Chase \& Co.                       & 18.292 & 42.729 & 0.311\\
    KO    & The Coca-Cola Company                       & 10.617 & 24.092 & 0.233\\
    MCD   & McDonald's Corp.                            & 14.457 & 26.114 & 0.362\\
    MMM   & 3M Company                                  & 12.596 & 25.353 & 0.300\\
    MRK   & Merck \& Co. Inc.                           & 12.385 & 29.616 & 0.249\\
    MSFT  & Microsoft Corporation                       & 18.612 & 33.904 & 0.401\\
    PFE   & Pfizer Inc.                                 & 14.376 & 29.060 & 0.323\\
    PG    & Procter \& Gamble Co.                       & 13.262 & 24.241 & 0.341\\
    T     & AT\&T, Inc.                                 & 11.231 & 28.781 & 0.217\\
    TRV   & The Travelers Companies, Inc.               & 14.726 & 31.706 & 0.307\\
    UTX   & United Technologies Corp.                   & 18.618 & 28.760 & 0.474\\
    VZ    & Verizon Communications Inc.                 & 11.403 & 27.728 & 0.231\\
    WMT   & Wal-Mart Stores Inc.                        & 15.495 & 27.955 & 0.375\\
    XOM   & Exxon Mobil Corporation                     & 15.466 & 25.764 & 0.406\\\hline
  \end{tabular}
  \caption[Dow Jones Industrial Average component stocks]{Dow Jones 
    Industrial Average component stocks and their respective realized returns, 
    realized risk and Sharpe ratios. The risk-free rate is set at 5\%.}
  \label{tbl:DJIA component}
\end{table}

\subsection{Investment periods}
\label{sect:investment periods}

\begin{table}[H]
  \centering
  \scriptsize
  \begin{tabular}{|r|c||r|c||r|c||r|c|}\hline
    $k$ & Date Range & $k$ & Date Range & $k$ & Date Range & $k$ & Date Range\\\hline
     1& 95/02/18-95/03/17& 59&99/07/31-99/08/27&117&04/01/10-04/02/06&175&08/06/21-08/07/18\\
     2& 95/03/18-95/04/14& 60&99/08/28-99/09/24&118&04/02/07-04/03/05&176&08/07/19-08/08/15\\
     3& 95/04/15-95/05/12& 61&99/09/25-99/10/22&119&04/03/06-04/04/02&177&08/08/16-08/09/12\\
     4& 95/05/13-95/06/09& 62&99/10/23-99/11/19&120&04/04/03-04/04/30&178&08/09/13-08/10/10\\
     5& 95/06/10-95/07/07& 63&99/11/20-99/12/17&121&04/05/01-04/05/28&179&08/10/11-08/11/07\\
     6& 95/07/08-95/08/04& 64&99/12/18-00/01/14&122&04/05/29-04/06/25&180&08/11/08-08/12/05\\
     7& 95/08/05-95/09/01& 65&00/01/15-00/02/11&123&04/06/26-04/07/23&181&08/12/06-09/01/02\\
     8& 95/09/02-95/09/29& 66&00/02/12-00/03/10&124&04/07/24-04/08/20&182&09/01/03-09/01/30\\
     9& 95/09/30-95/10/27& 67&00/03/11-00/04/07&125&04/08/21-04/09/17&183&09/01/31-09/02/27\\
    10& 95/10/28-95/11/24& 68&00/04/08-00/05/05&126&04/09/18-04/10/15&184&09/02/28-09/03/27\\
    11& 95/11/25-95/12/22& 69&00/05/06-00/06/02&127&04/10/16-04/11/12&185&09/03/28-09/04/24\\
    12& 95/12/23-96/01/19& 70&00/06/03-00/06/30&128&04/11/13-04/12/10&186&09/04/25-09/05/22\\
    13& 96/01/20-96/02/16& 71&00/07/01-00/07/28&129&04/12/11-05/01/07&187&09/05/23-09/06/19\\
    14& 96/02/17-96/03/15& 72&00/07/29-00/08/25&130&05/01/08-05/02/04&188&09/06/20-09/07/17\\
    15& 96/03/16-96/04/12& 73&00/08/26-00/09/22&131&05/02/05-05/03/04&189&09/07/18-09/08/14\\
    16& 96/04/13-96/05/10& 74&00/09/23-00/10/20&132&05/03/05-05/04/01&190&09/08/15-09/09/11\\
    17& 96/05/11-96/06/07& 75&00/10/21-00/11/17&133&05/04/02-05/04/29&191&09/09/12-09/10/09\\
    18& 96/06/08-96/07/05& 76&00/11/18-00/12/15&134&05/04/30-05/05/27&192&09/10/10-09/11/06\\
    19& 96/07/06-96/08/02& 77&00/12/16-01/01/12&135&05/05/28-05/06/24&193&09/11/07-09/12/04\\
    20& 96/08/03-96/08/30& 78&01/01/13-01/02/09&136&05/06/25-05/07/22&194&09/12/05-10/01/01\\
    21& 96/08/31-96/09/27& 79&01/02/10-01/03/09&137&05/07/23-05/08/19&195&10/01/02-10/01/29\\
    22& 96/09/28-96/10/25& 80&01/03/10-01/04/06&138&05/08/20-05/09/16&196&10/01/30-10/02/26\\
    23& 96/10/26-96/11/22& 81&01/04/07-01/05/04&139&05/09/17-05/10/14&197&10/02/27-10/03/26\\
    24& 96/11/23-96/12/20& 82&01/05/05-01/06/01&140&05/10/15-05/11/11&198&10/03/27-10/04/23\\
    25& 96/12/21-97/01/17& 83&01/06/02-01/06/29&141&05/11/12-05/12/09&199&10/04/24-10/05/21\\
    26& 97/01/18-97/02/14& 84&01/06/30-01/07/27&142&05/12/10-06/01/06&200&10/05/22-10/06/18\\
    27& 97/02/15-97/03/14& 85&01/07/28-01/08/24&143&06/01/07-06/02/03&201&10/06/19-10/07/16\\
    28& 97/03/15-97/04/11& 86&01/08/25-01/09/21&144&06/02/04-06/03/03&202&10/07/17-10/08/13\\
    29& 97/04/12-97/05/09& 87&01/09/22-01/10/19&145&06/03/04-06/03/31&203&10/08/14-10/09/10\\
    30& 97/05/10-97/06/06& 88&01/10/20-01/11/16&146&06/04/01-06/04/28&204&10/09/11-10/10/08\\
    31& 97/06/07-97/07/04& 89&01/11/17-01/12/14&147&06/04/29-06/05/26&205&10/10/09-10/11/05\\
    32& 97/07/05-97/08/01& 90&01/12/15-02/01/11&148&06/05/27-06/06/23&206&10/11/06-10/12/03\\
    33& 97/08/02-97/08/29& 91&02/01/12-02/02/08&149&06/06/24-06/07/21&207&10/12/04-10/12/31\\
    34& 97/08/30-97/09/26& 92&02/02/09-02/03/08&150&06/07/22-06/08/18&208&11/01/01-11/01/28\\
    35& 97/09/27-97/10/24& 93&02/03/09-02/04/05&151&06/08/19-06/09/15&209&11/01/29-11/02/25\\
    36& 97/10/25-97/11/21& 94&02/04/06-02/05/03&152&06/09/16-06/10/13&210&11/02/26-11/03/25\\
    37& 97/11/22-97/12/19& 95&02/05/04-02/05/31&153&06/10/14-06/11/10&211&11/03/26-11/04/22\\
    38& 97/12/20-98/01/16& 96&02/06/01-02/06/28&154&06/11/11-06/12/08&212&11/04/23-11/05/20\\
    39& 98/01/17-98/02/13& 97&02/06/29-02/07/26&155&06/12/09-07/01/05&213&11/05/21-11/06/17\\
    40& 98/02/14-98/03/13& 98&02/07/27-02/08/23&156&07/01/06-07/02/02&214&11/06/18-11/07/15\\
    41& 98/03/14-98/04/10& 99&02/08/24-02/09/20&157&07/02/03-07/03/02&215&11/07/16-11/08/12\\
    42& 98/04/11-98/05/08&100&02/09/21-02/10/18&158&07/03/03-07/03/30&216&11/08/13-11/09/09\\
    43& 98/05/09-98/06/05&101&02/10/19-02/11/15&159&07/03/31-07/04/27&217&11/09/10-11/10/07\\
    44& 98/06/06-98/07/03&102&02/11/16-02/12/13&160&07/04/28-07/05/25&218&11/10/08-11/11/04\\
    45& 98/07/04-98/07/31&103&02/12/14-03/01/10&161&07/05/26-07/06/22&219&11/11/05-11/12/02\\
    46& 98/08/01-98/08/28&104&03/01/11-03/02/07&162&07/06/23-07/07/20&220&11/12/03-11/12/30\\
    47& 98/08/29-98/09/25&105&03/02/08-03/03/07&163&07/07/21-07/08/17&221&11/12/31-12/01/27\\
    48& 98/09/26-98/10/23&106&03/03/08-03/04/04&164&07/08/18-07/09/14&222&12/01/28-12/02/24\\
    49& 98/10/24-98/11/20&107&03/04/05-03/05/02&165&07/09/15-07/10/12&223&12/02/25-12/03/23\\
    50& 98/11/21-98/12/18&108&03/05/03-03/05/30&166&07/10/13-07/11/09&224&12/03/24-12/04/20\\
    51& 98/12/19-99/01/15&109&03/05/31-03/06/27&167&07/11/10-07/12/07&225&12/04/21-12/05/18\\
    52& 99/01/16-99/02/12&110&03/06/28-03/07/25&168&07/12/08-08/01/04&226&12/05/19-12/06/15\\
    53& 99/02/13-99/03/12&111&03/07/26-03/08/22&169&08/01/05-08/02/01&227&12/06/16-12/07/13\\
    54& 99/03/13-99/04/09&112&03/08/23-03/09/19&170&08/02/02-08/02/29&228&12/07/14-12/08/10\\
    55& 99/04/10-99/05/07&113&03/09/20-03/10/17&171&08/03/01-08/03/28&229&12/08/11-12/09/07\\
    56& 99/05/08-99/06/04&114&03/10/18-03/11/14&172&08/03/29-08/04/25&230&12/09/08-12/10/05\\
    57& 99/06/05-99/07/02&115&03/11/15-03/12/12&173&08/04/26-08/05/23&231&12/10/06-12/10/26\\
    58& 99/07/03-99/07/30&116&03/12/13-04/01/09&174&08/05/24-08/06/20& &                     \\\hline
  \end{tabular}
  \caption{Investment periods in YY/MM/DD format}
  \label{tab:investment periods}
\end{table}

\subsection{Details of minimum variance portfolio rebalancing} 
\label{sec:minvarporeb}

\noindent
The investment
period during which a set of portfolio weights are held constant is
also referred to as the ``holding period''. The number of trading days
in the $k$-th investment period, $L_k$, may vary if rebalancing time
points are chosen to coincide with either calendar months, weeks or
fiscal quarters. Let $t$ index the number of an arbitrary day during the
entire investment horizon. The number of trading days $T_j$ in the
first $j$ investment periods is given by
\begin{align}
  T_j = \sum_{k=1}^j L_k,
\end{align}
where $j=1,2,\dots, K$ with $T_0=0$. We consider holding $N_{\est}$
constant for all investment periods, $k=1,2,\dots$. For convenience,
denote by $k_t$ the investment period that trading day $t$ belongs to:
i.e., $k_t=k(t):=\{k: t \in [T_{k-1},T_{k}]\}$.

The algorithm for the minimum variance portfolio rebalancing strategy
(MVR) can now be described as follows: At the beginning of time period
$k$, that is after $T_{k-1}$ days, compute an estimate of the
covariance matrix $\widehat\Sigma_k$ for period $k$ from $N_{\est}$
past returns: i.e., $\{r_t: t\in
[T_{k-1}-N_{\est}+1,T_{k-1}]\}$. Then, compute a new set of portfolio
weights
$w_k=(\ones^T\widehat\Sigma_k^{-1}\ones)^{-1}\widehat\Sigma_k^{-1}\ones$,
and hold this portfolio constant until the $T_k$-th trading day. The
process is then repeated for the next holding period. 

\subsection{Details of cross-validation} \label{sect:cross validation}

Consider the matrix of returns ${\bf R}$ for all the stocks in the
portfolio in the estimation horizon preceding the start of the
investment period $(k-1)$.
\begin{align*}
  {\bf R} = ((r_{ti})),\mbox{ where } i\in \{1,\dots,p\},\
  t\in\{T_{k-1}-N_{\est}+1,\dots,T_{k-1}\}.
\end{align*}

Hence, ${\bf R}$ is an $N_{\est}$-by-$p$ matrix, and the column vector
${\bf R}_j$ is an $N_{\est}$-vector of returns for the $j$-th stock.

Now denote by $\Omega(\lambda)=((\omega_{ij}(\lambda)))_{1\leq i,j\leq
  p}$ an estimate of $\Omega$ obtained by $\ell_1$-regularization
methods such as Glasso or CONCORD. The use of $\lambda$ makes explicit
the dependence of these estimation methods on the penalty parameter
$\lambda$. The data are the over the estimation horizon is divided
into $m$-folds. The penalty parameter is chosen so as to minimize the
out of sample predictive risk (PR) given by
\begin{align*}
  PR(\lambda) = \sum_{m=1}^M \left\{ \frac{1}{N_m}\sum_{i=1}^p \|{\bf
      R}^{(m)}_{i}-\sum_{j\neq i}\beta^{(\backslash m)}_{ij}(\lambda)
    {\bf R}^{(m)}_{j}\|_2^2\right\},
\end{align*}
where ${\bf R}_i^{(m)}$ is the vector of returns for stock $i$ in fold
$m$, and where $N_m$ is the number of observations in the $m$-th
fold. The regression coefficient $\beta^{(\backslash
  m)}_{ij}(\lambda)$ is determined as follows: $\beta^{(\backslash
  m)}_{ij}(\lambda)=-\frac{\omega^{(\backslash
    m)}_{ij}(\lambda)}{\omega^{(\backslash m)}_{ii}(\lambda)}$, with
$\Omega^{(\backslash m)}(\lambda)$ based on using all the available
data within a given estimation horizon except for fold $m$. The
optimal choice of penalty parameter $\lambda^*$ is then determined as
follows:

\begin{align*}
  \lambda^* = \arg\inf_{\lambda\geq 0} PR(\lambda).
\end{align*}

\subsection{Performance metrics} \label{sect:perf metrics}
For comparison purposes with \citep{Won2012}, we use the following
quantities to assess the performance of the five MVR strategies. The
formulas for these metrics are given below.

\begin{itemize}
\item \emph{Realized return}: The average return of the portfolio over
  the entire investment horizon.
\begin{align*}
  r_p = \frac{1}{T} \sum_{t=1}^T r_{t}' w_{k_t}
\end{align*}

\item \emph{Realized risk}: The risk (standard error) of the portfolio
  over the entire investment horizon.
\begin{align*}
  \sigma_p = \left[
      \frac{1}{T} \sum_{t=1}^T (r_{t}' w_{k_t} - r_p)^2
    \right]^{1/2}
\end{align*}

\item \emph{Realized Sharpe ratio (SR)}: The realized excess return of
  the portfolio over the risk-free rate per unit realized risk 
  for the entire investment horizon.
\begin{align}
  SR = \frac{r_p - r_f}{\sigma_p} \label{eq:sharpe ratio}
\end{align}

\item \emph{Turnover}: The amount of new portfolio assets purchased or
  sold over each trading period. The turnover for the $k$-th
  investment period when the portfolio weights $w_k$ are held constant
  is given by
\begin{align}
  TO(k) = \sum_{i=1}^p \left| w_{ik} -
    \left(\prod_{t=T_{k-1}+1}^{T_{k-1}+L_k} (1 + r_{it})\right)\, w_{i(k-1)}
  \right|
  \label{eq:turnover}
\end{align}
with $w_{i0}=0$ for all $i=1,\dots,p$.

\item \emph{Size of the short side} The proportion of the negative
  weights to the sum of the absolute weights of each portfolio. The
  short side for the $k$-th investment period is given by
\begin{align*}
  SS(k) = \frac{\sum_{i=1}^{p}|\min
    (w_{ik},0)|}{\sum_{i=1}^{p}|w_{ik}|}
\end{align*}

The average and standard error of the short sides over the all
investment periods is
\begin{align*}
  \overline{SS} = \frac{1}{K} \sum_{k=1}^{K} SS(k),\quad
  \hat\sigma_{SS} = \left[\frac{1}{K} \sum_{k=1}^{K} (SS(k) -
    \overline{SS})^2\right]^{1/2}
\end{align*}

\item \emph{Normalized wealth growth}: Accumulated wealth derived from
  the portfolio over the trading period when the initial budget is
  normalized to one. Note that both transaction costs and borrowing
  costs are taken into account. Let $W(t-1)$ denote the wealth of the
  portfolio after the $(t-1)$-th trading day. Then, the wealth of the
  portfolio after the $t$-th trading day is given by
\begin{align*}
  W(t) = 
  \begin{cases}
    W(t-1)\left(1+r_{t}'w_{k_t} - TC(k_t) - BC(k_t)\right), 
    & t = T_{k_t-1}+1\\
    W(t-1)\left(1+r_{t}'w_{k_t}\right), 
    & t \neq T_{k_t-1}+1
  \end{cases},
\end{align*}
where $TC(k)$ and $BC(k)$ are transaction costs (of trading stocks)
and borrowing costs (of capital for taking short positions on stocks),
respectively. On the first day of each trading period, we adjust the
return for these trading costs. Denote the transaction cost rate by
$r_c$, then the transaction cost incurred at the beginning of period
$k$ is given by
\begin{align}
  TC(k) = r_c \cdot TO(k).\label{eq:transaction cost}
\end{align}
The borrowing cost rate, $BC(k)$, depends on the short side of the
portfolio weights during the $(k-1)$-th period. Denote the borrowing
daily percentage by $r_b$, then the borrowing cost rate is given by
\begin{align}
  BC(k) = ((1 + r_b)^{L_{k-1}}-1)\sum_{i=1}^p |\min
  (w_{i(k-1)},0)|.\label{eq:borrowing cost}
\end{align}
\end{itemize}

\begin{table}
  \centering

  \footnotesize
  \begin{tabular}{rrrrrrr}
    \hline
    $N_{\est}$ & Sample & Glasso & CONCORD & CondReg & LedoitWolf & DJIA \\ 
    \hline
    35  & \textbf{17.08} (33.86) & 13.10 (16.57) & 13.29          (17.04) & 13.62          (17.74) & 12.33 (\textbf{15.58}) & 8.51 (18.96) \\ 
    40  & \textbf{16.66} (26.52) & 13.13 (16.57) & 13.34          (17.02) & 13.39          (17.74) & 11.78 (\textbf{15.46}) & 8.51 (18.96) \\ 
    45  & 11.13          (23.19) & 12.74 (16.52) & \textbf{13.05} (17.04) & \textbf{13.05} (17.77) & 10.99 (\textbf{15.43}) & 8.51 (18.96) \\ 
    50  &  9.90          (20.95) & 12.89 (16.39) & \textbf{13.21} (17.04) & 13.08          (17.65) & 11.25 (\textbf{15.36}) & 8.51 (18.96) \\ 
    75  & 11.61          (17.45) & 11.28 (15.57) & \textbf{13.10} (17.04) & 12.77          (17.15) & 10.56 (\textbf{15.10}) & 8.51 (18.96) \\ 
    150 &  9.40          (15.41) & 10.28 (14.97) & \textbf{13.20} (17.08) & 12.76          (16.30) & 10.63 (\textbf{14.66}) & 8.51 (18.96) \\ 
    225 & 10.49          (14.98) & 10.38 (14.89) & \textbf{13.58} (17.10) & 12.92          (16.04) & 11.04 (\textbf{14.52}) & 8.51 (18.96) \\ 
    300 & 10.41          (14.95) & 10.37 (14.95) & \textbf{13.66} (17.16) & 12.85          (16.07) & 10.94 (\textbf{14.52}) & 8.51 (18.96) \\ 
    \hline
  \end{tabular}

  \caption[Realized returns and risks comparison]{Realized returns of 
    different investment strategies corresponding to 
    different estimators with various $N_{\est}$ (realized risks are
    given in parentheses). The maximum annualized returns and risks 
    are highlighted in bold.}
  \label{tab:realized risk}

\end{table}

\begin{table}
  \centering
  \begin{tabular}{rrrrrr}
    \hline
    $N_{\est}$ & Sample & Glasso & CONCORD & CondReg & LedoitWolf \\ 
    \hline
    35  & 8.42 (3.19) & 0.45 (0.12) & \textbf{0.38} (\textbf{0.10}) & 0.39 (0.27) & 1.40 (0.38) \\ 
    40  & 5.81 (2.28) & 0.41 (0.12) & \textbf{0.34} (\textbf{0.10}) & 0.37 (0.26) & 1.29 (0.36) \\ 
    45  & 4.58 (1.65) & 0.39 (0.12) & \textbf{0.31} (\textbf{0.10}) & 0.36 (0.23) & 1.20 (0.35) \\ 
    50  & 3.74 (1.19) & 0.39 (0.13) & \textbf{0.28} (\textbf{0.09}) & 0.36 (0.25) & 1.11 (0.33) \\ 
    75  & 2.03 (0.67) & 0.50 (0.19) & \textbf{0.21} (\textbf{0.08}) & 0.43 (0.29) & 0.86 (0.29) \\ 
    150 & 0.87 (0.32) & 0.73 (0.27) & \textbf{0.14} (\textbf{0.07}) & 0.40 (0.22) & 0.54 (0.23) \\ 
    225 & 0.57 (0.24) & 0.56 (0.22) & \textbf{0.11} (\textbf{0.07}) & 0.31 (0.13) & 0.41 (0.18) \\ 
    300 & 0.44 (0.21) & 0.44 (0.23) & \textbf{0.09} (\textbf{0.07}) & 0.24 (0.11) & 0.33 (0.17) \\ 
    \hline
  \end{tabular}

  \caption[Asset turnover comparison]{Average turnovers for 
    various estimation horizons, $N_{\est}$ (standard errors are
    given in parentheses).
    The minimum average and standard error values for 
    each row are highlighted in bold.}
  \label{tab:turnover mean}
\end{table}

\begin{table}
  \centering
  \begin{tabular}{rrrrrr}
    \hline
    $N_{\est}$ & Sample & Glasso & CONCORD & CondReg & LedoitWolf \\ 
    \hline
    35  & 41.13 (3.18) & 0.66  (0.84) & \textbf{0.05} (\textbf{0.14}) & 1.75  (5.00) & 20.50 (6.64) \\ 
    40  & 38.64 (3.47) & 0.64  (0.75) & \textbf{0.05} (\textbf{0.14}) & 1.78  (5.04) & 20.45 (6.63) \\ 
    45  & 36.89 (4.26) & 0.90  (0.85) & \textbf{0.05} (\textbf{0.14}) & 1.84  (4.95) & 20.31 (6.61) \\ 
    50  & 35.46 (4.38) & 1.35  (1.19) & \textbf{0.04} (\textbf{0.11}) & 2.17  (5.44) & 20.33 (6.66) \\ 
    75  & 30.89 (5.37) & 8.67  (3.76) & \textbf{0.04} (\textbf{0.11}) & 4.91  (7.38) & 20.13 (6.83) \\ 
    150 & 25.65 (6.25) & 23.48 (4.68) & \textbf{0.02} (\textbf{0.07}) & 9.07  (6.31) & 19.60 (6.82) \\ 
    225 & 23.68 (6.69) & 23.36 (6.27) & \textbf{0.01} (\textbf{0.05}) & 10.71 (3.22) & 19.26 (6.91) \\ 
    300 & 22.45 (6.90) & 22.42 (6.87) & \textbf{0.00} (\textbf{0.02}) & 9.95  (2.93) & 18.85 (7.10) \\ 
    \hline
  \end{tabular}

  \caption[Short side comparison]{Average short sides for 
    various estimation horizons, $N_{\est}$ (standard
    errors are given in parentheses).
    The minimum average and standard error values for 
    each row are highlighted in bold.}
  \label{tab:shortside}
\end{table}

\begin{table}
  \centering
  \begin{tabular}{rrrrrr}
    \hline
    $N_{\est}$ & Sample & Glasso & CONCORD & CondReg & LedoitWolf \\ 
    \hline
    35  & 567.958 (214.05) & 22.635 (5.62)  & \textbf{18.642} (\textbf{4.53}) & 20.757 (17.46) & 91.316 (25.19) \\ 
    40  & 394.508 (149.90) & 20.660 (5.70)  & \textbf{16.858} (\textbf{4.40}) & 20.013 (16.78) & 85.661 (24.16) \\ 
    45  & 315.340 (108.87) & 19.899 (5.80)  & \textbf{15.470} (\textbf{4.22}) & 19.419 (15.27) & 80.524 (23.39) \\ 
    50  & 260.887 (81.13)  & 20.146 (6.39)  & \textbf{14.081} (\textbf{4.06}) & 19.695 (16.04) & 76.154 (22.43) \\ 
    75  & 150.242 (45.87)  & 30.942 (10.92) & \textbf{10.516} (\textbf{3.17}) & 25.191 (19.19) & 63.481 (20.94) \\ 
    150 & 75.700  (27.88)  & 61.495 (18.40) & \textbf{6.596}  (\textbf{2.24}) & 26.788 (12.83) & 46.680 (17.78) \\ 
    225 & 56.242  (22.09)  & 54.117 (18.82) & \textbf{5.155}  (\textbf{1.80}) & 22.973 (6.08)  & 39.441 (15.72) \\ 
    300 & 46.904  (20.09)  & 47.118 (20.72) & \textbf{4.404}  (\textbf{1.67}) & 18.823 (5.16)  & 35.065 (14.89) \\ 
    \hline
  \end{tabular}

  \caption[Trading cost comparison]{
    Average trading costs in basis points for various estimation
    horizons, $N_{\est}$ (standard errors are given in
    parentheses). Borrowing rate is taken to be 7\% APR and
    transaction cost rate is taken to be 0.5\% of principal for each
    transaction. The minimum transaction cost for each row is
    highlighted in bold.} 
  \label{tab:mean sd trading costs}
\end{table}

\begin{figure}
  \centering{
    \includegraphics[width=\textwidth]{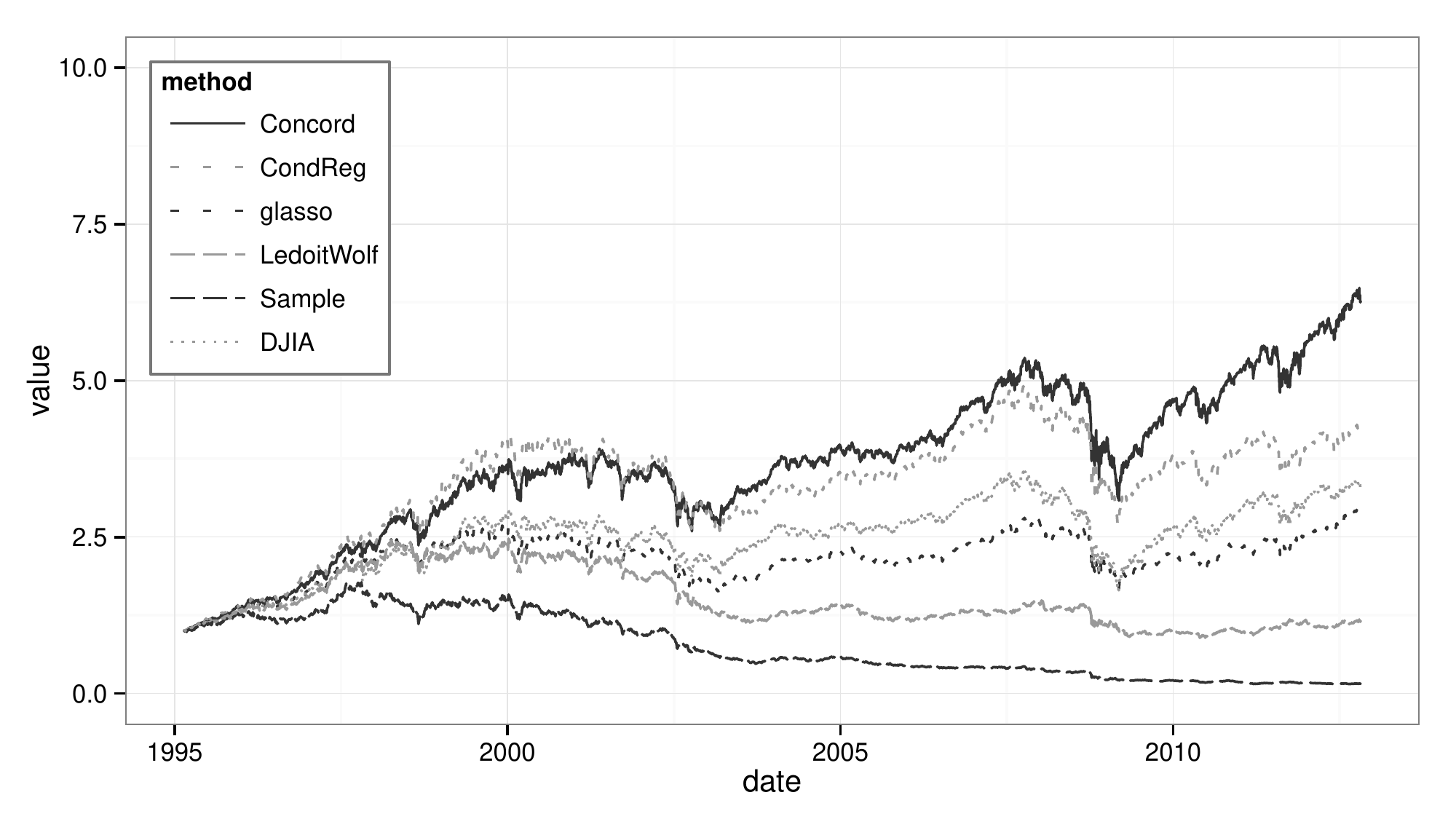} 
  }
  \caption[Normalized wealth growth comparison with trading
  costs]{Normalized wealth growth after adjusting for transaction
    costs (0.5\% of principal) and borrowing costs (interest rate of
    7\% APR) with $N_{\est} = 75$.}
  \label{fig:wealth growth tc and bc:1}
\end{figure}

\begin{figure}
  \centering  
  \subfigure[ $N_{\est}= 35 $ ]{
    \includegraphics[trim = 0mm 3mm 35mm 5mm,clip,width=0.455\textwidth]{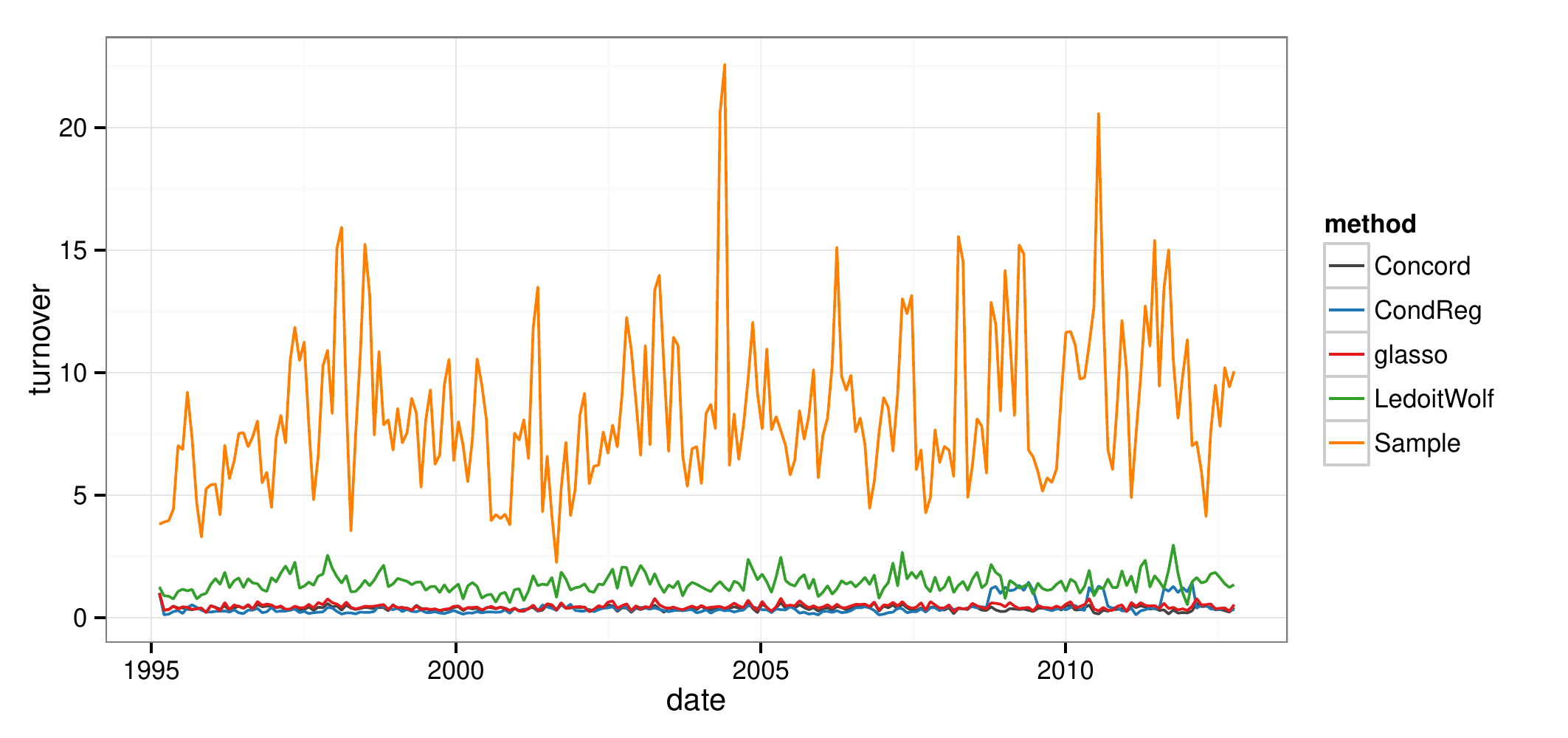} 
  }\subfigure[ $N_{\est}= 40 $ ]{
    \includegraphics[trim = 8mm 3mm 0mm 5mm,clip,width=0.525\textwidth]{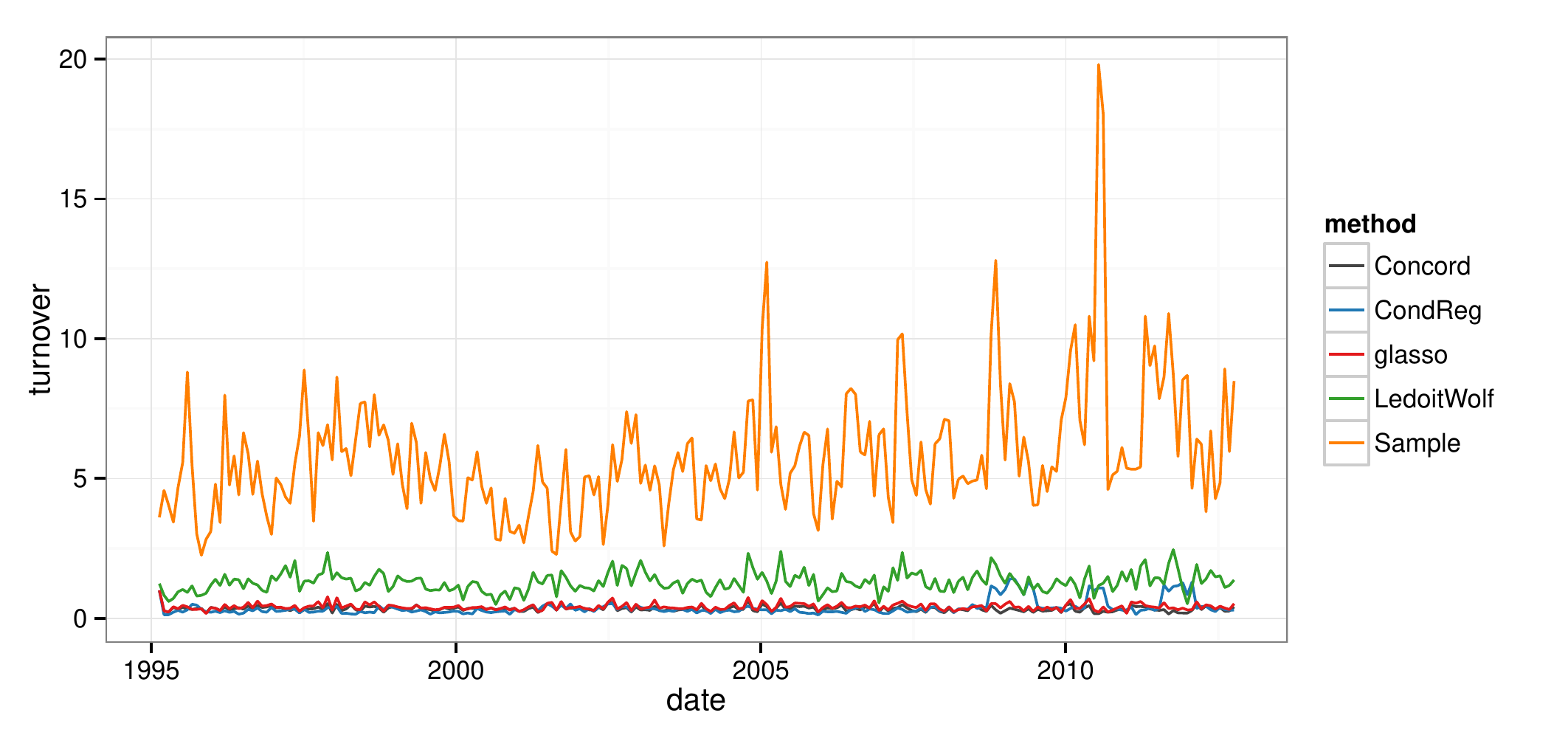} 
  }
  \subfigure[ $N_{\est}= 45 $ ]{
    \includegraphics[trim = 0mm 3mm 35mm 5mm,clip,width=0.455\textwidth]{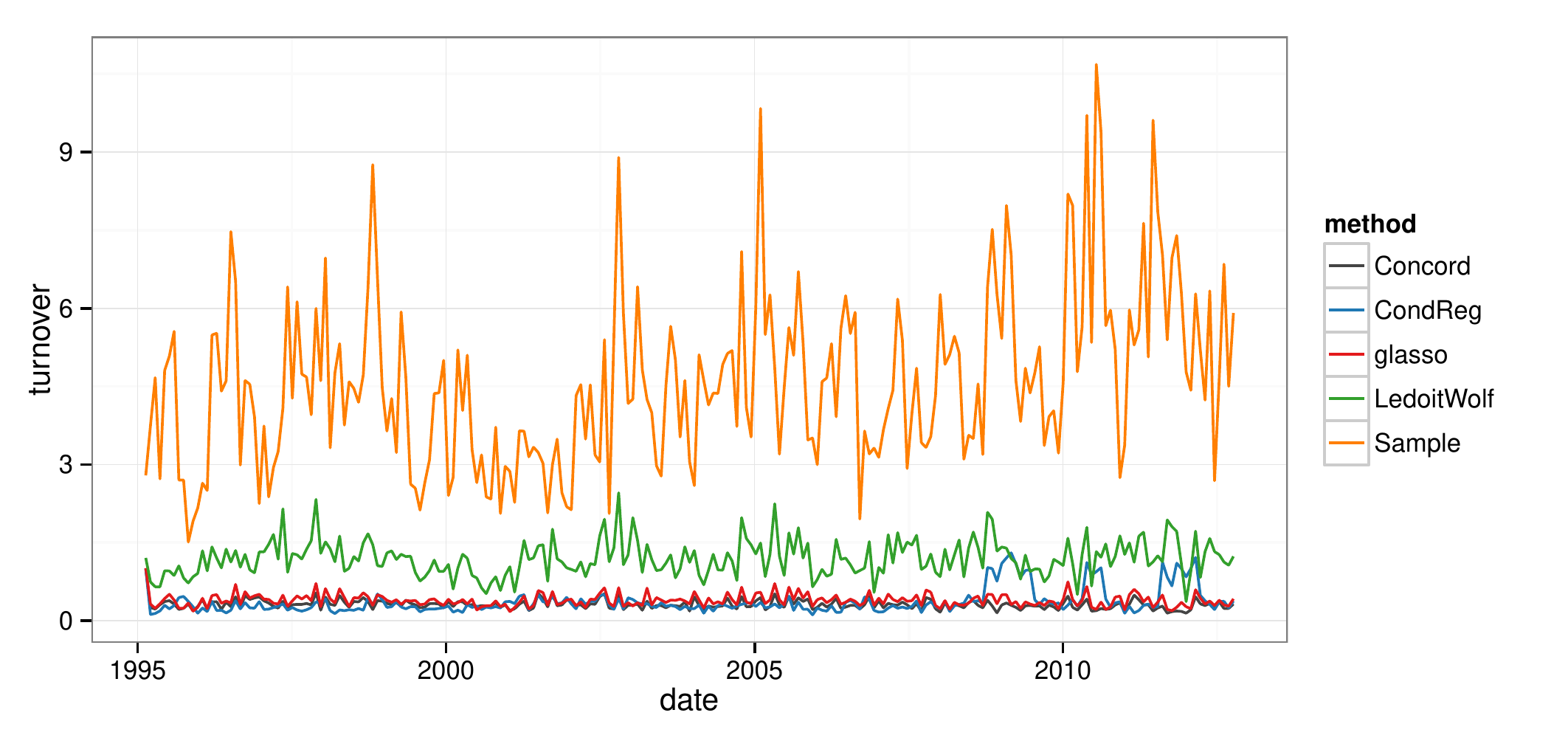} 
  }\subfigure[ $N_{\est}= 50 $ ]{
    \includegraphics[trim = 8mm 3mm 0mm 5mm,clip,width=0.525\textwidth]{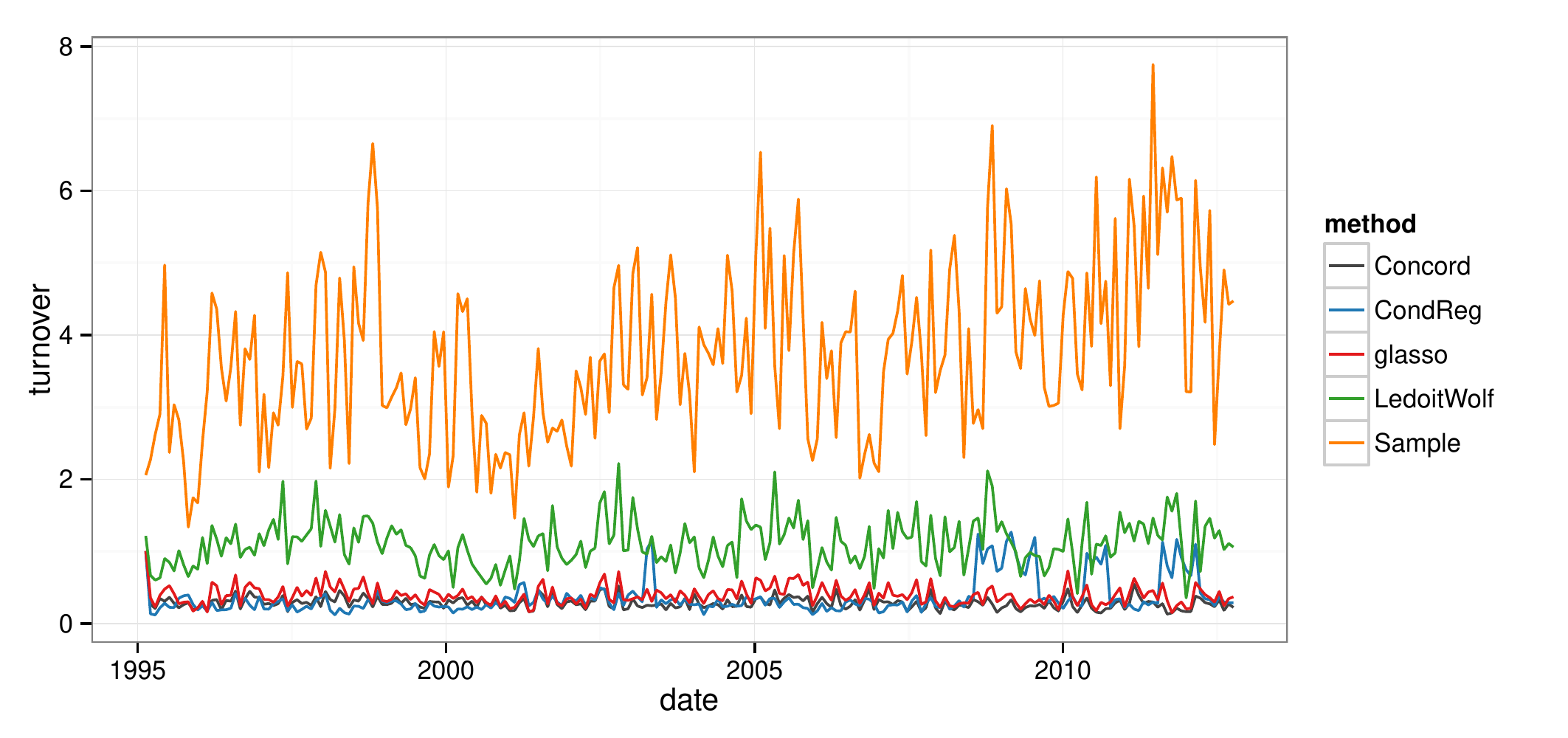} 
  }
  \subfigure[ $N_{\est}= 75 $ ]{
    \includegraphics[trim = 0mm 3mm 35mm 5mm,clip,width=0.455\textwidth]{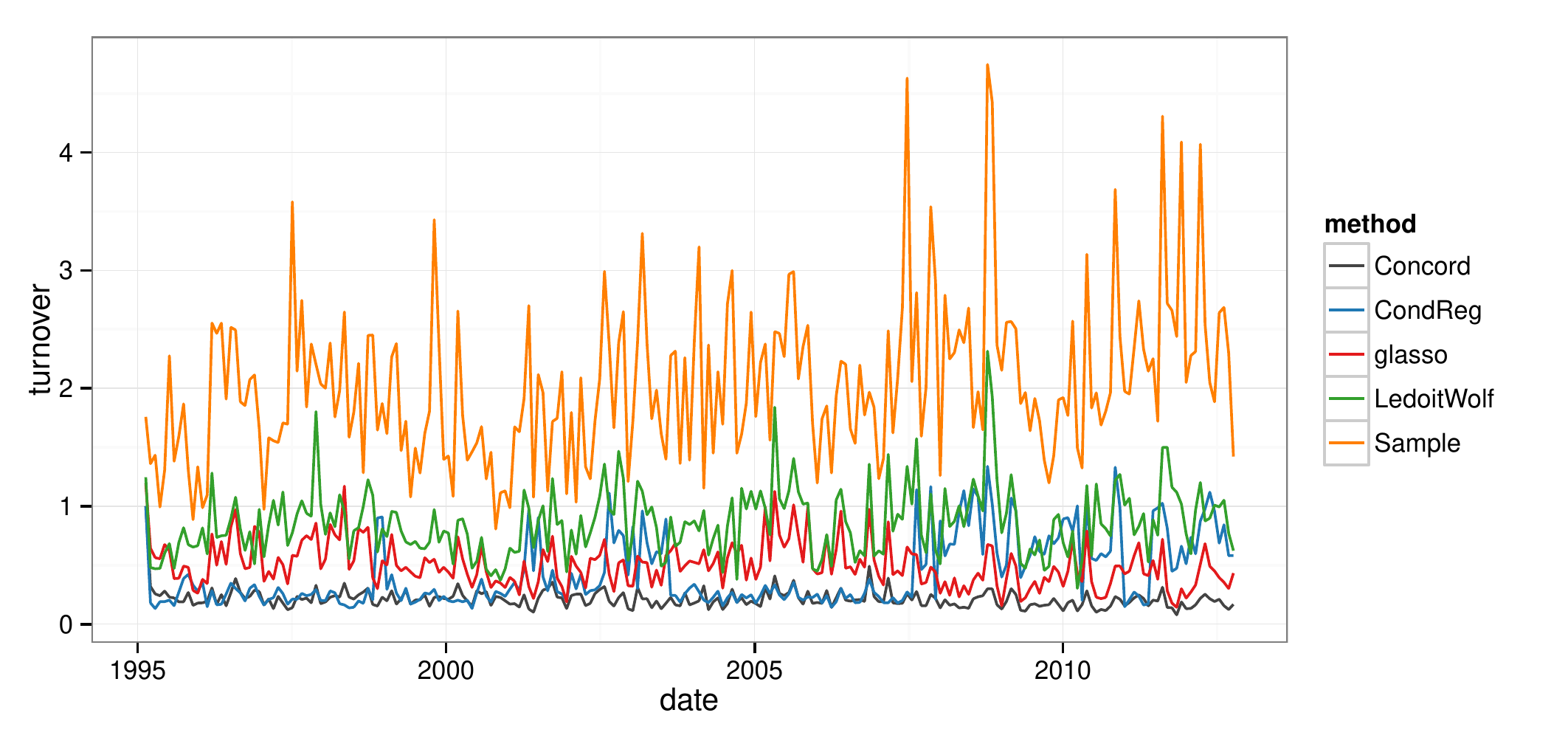} 
  }\subfigure[ $N_{\est}= 150 $ ]{
    \includegraphics[trim = 8mm 3mm 0mm 5mm,clip,width=0.525\textwidth]{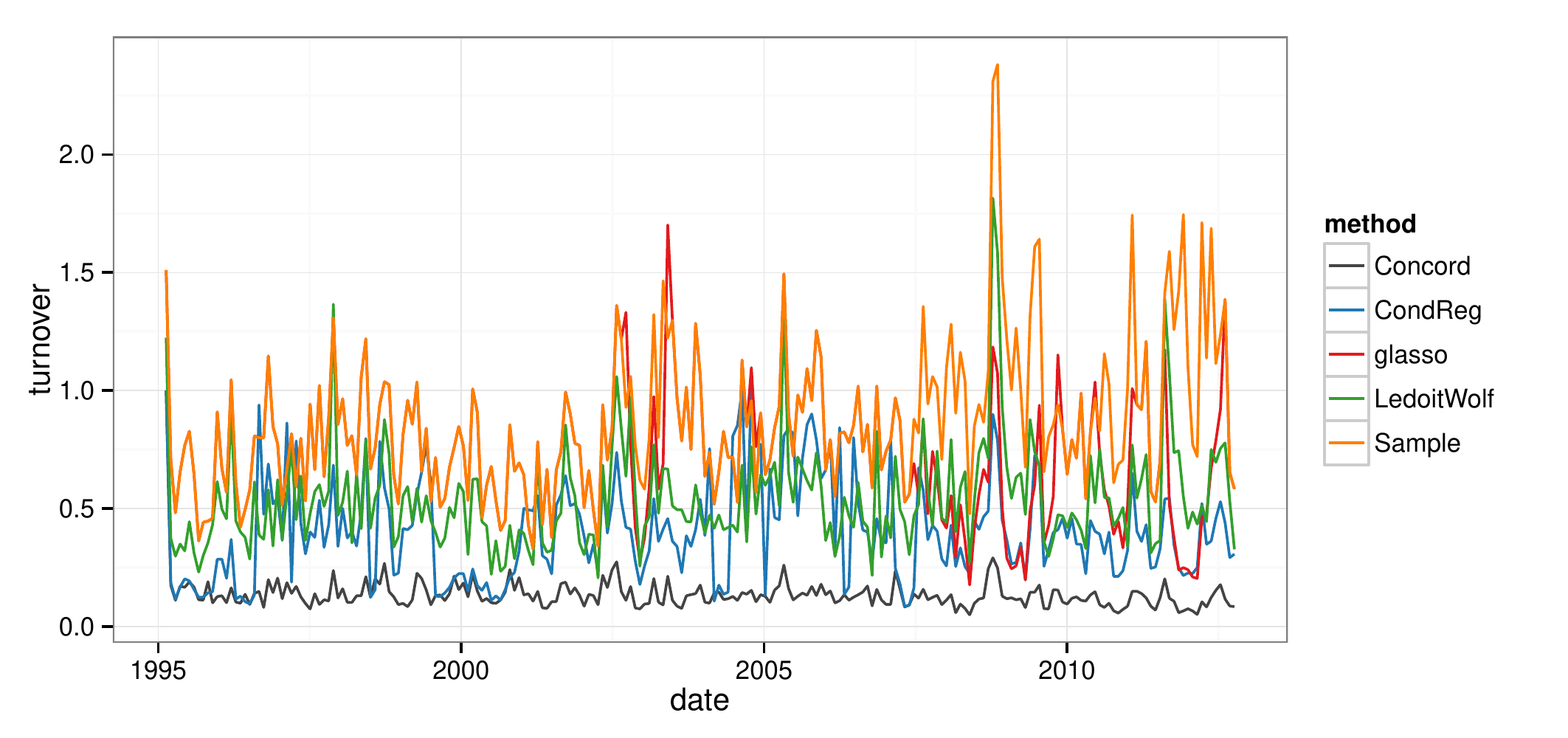} 
  }
  \subfigure[ $N_{\est}= 225 $ ]{
    \includegraphics[trim = 0mm 3mm 35mm 5mm,clip,width=0.455\textwidth]{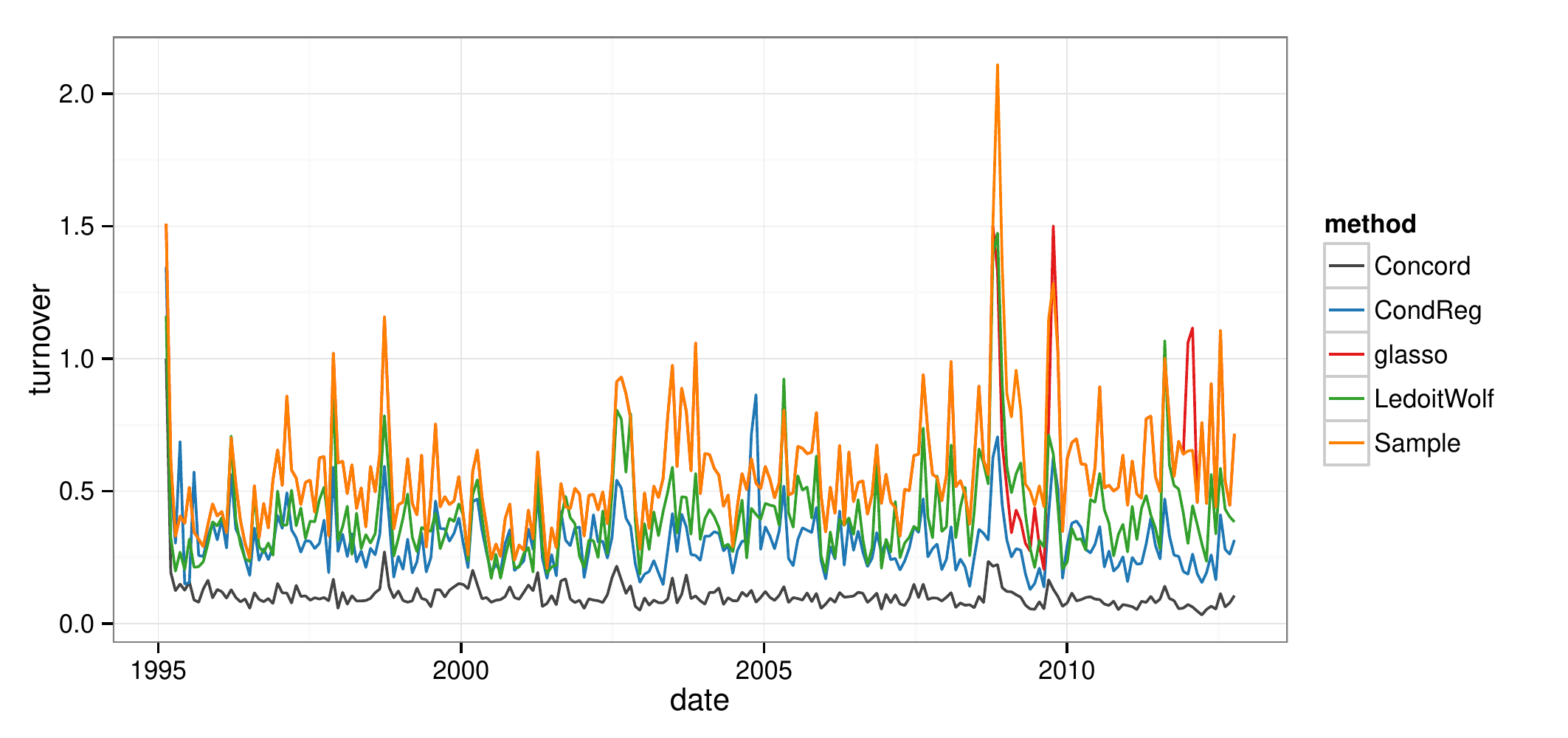} 
  }\subfigure[ $N_{\est}= 300 $ ]{
    \includegraphics[trim = 8mm 3mm 0mm 5mm,clip,width=0.525\textwidth]{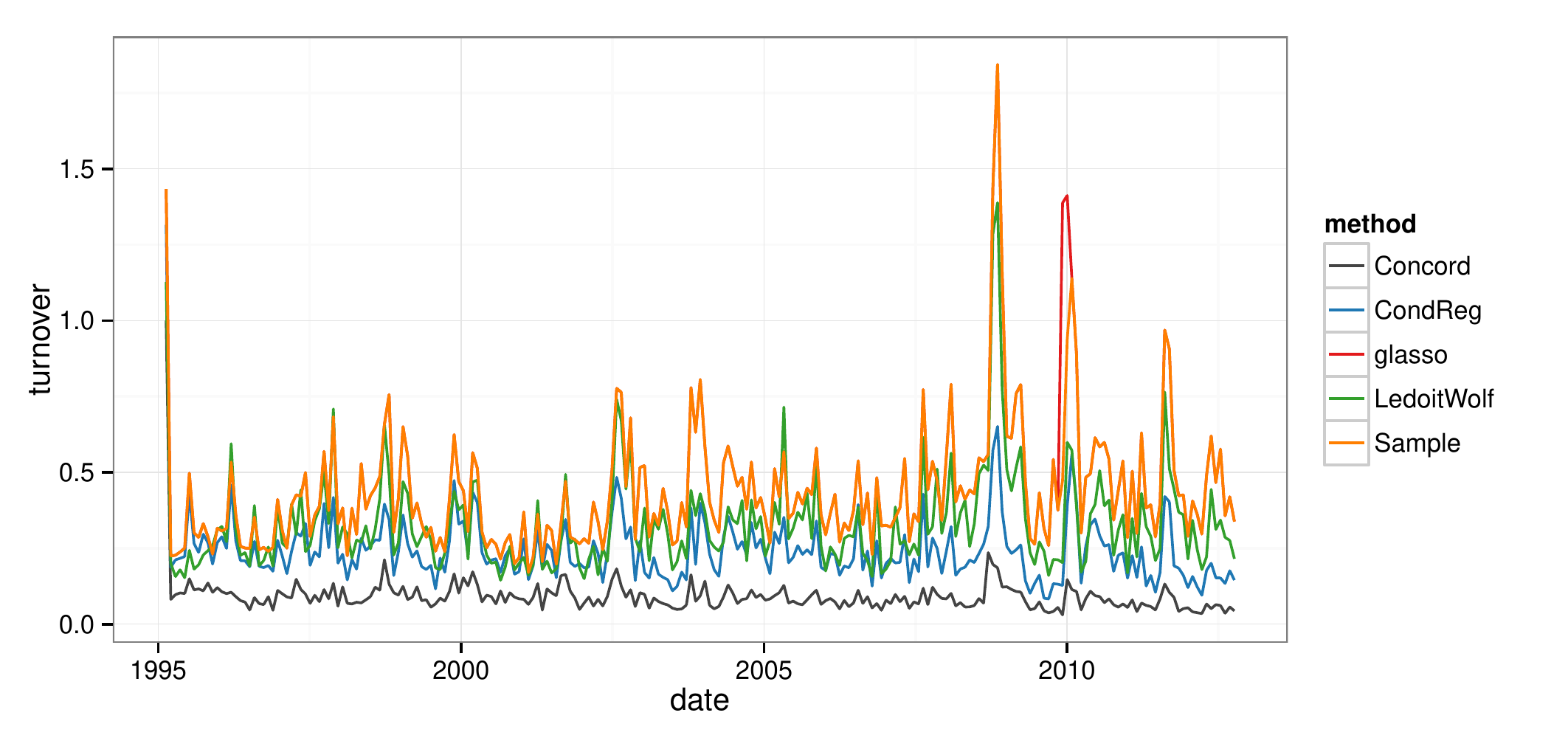} 
  }
  \caption[Turnover comparison]{Turnover in percentage points.}
\end{figure}

\begin{figure}
  \centering
  \subfigure[ $N_{\est}= 35 $ ]{
    \includegraphics[trim = 0mm 3mm 35mm
    5mm,clip,width=0.48\textwidth]{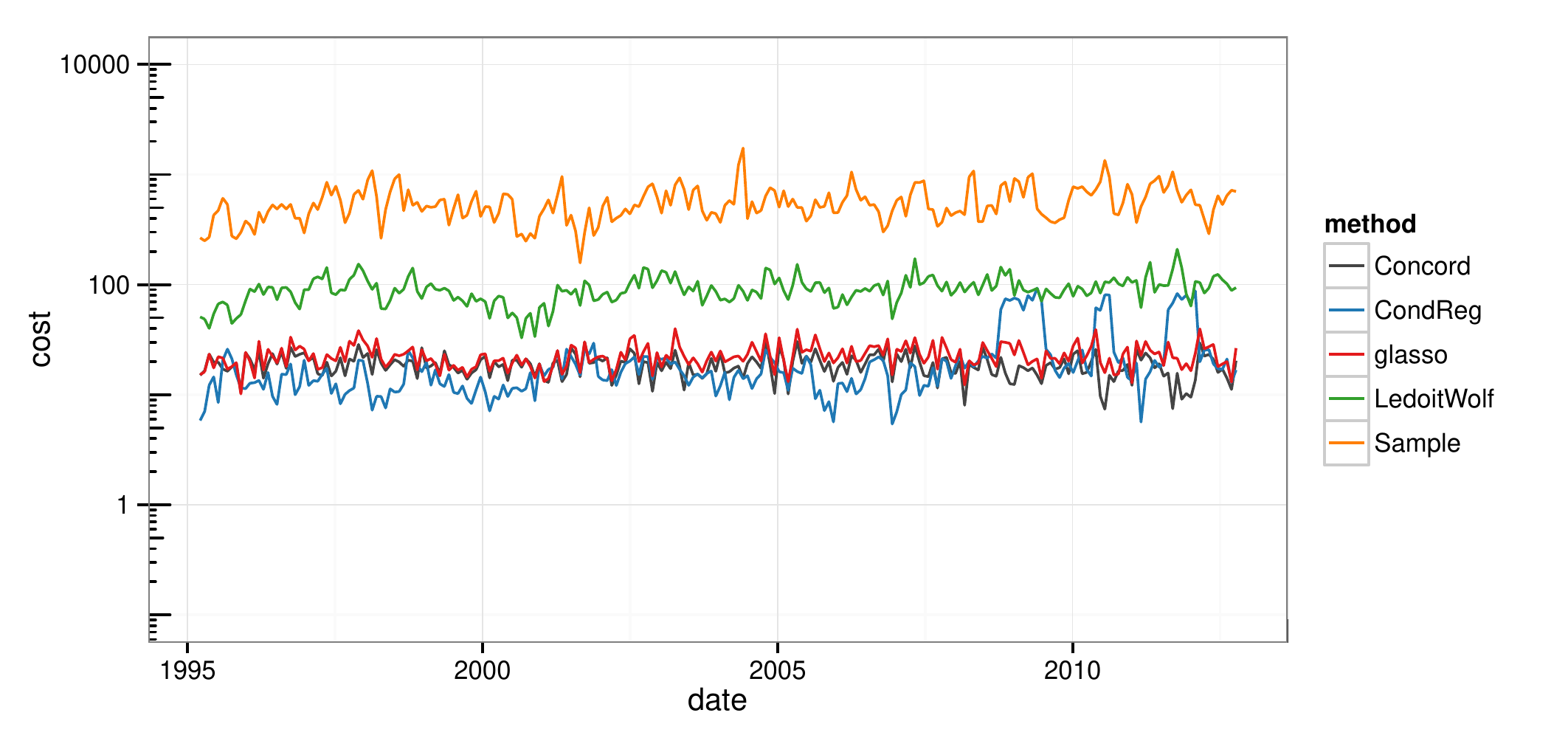}
  }\subfigure[ $N_{\est}= 40 $ ]{
    \includegraphics[trim = 8mm 0mm 10mm
    0mm,clip,width=0.51\textwidth]{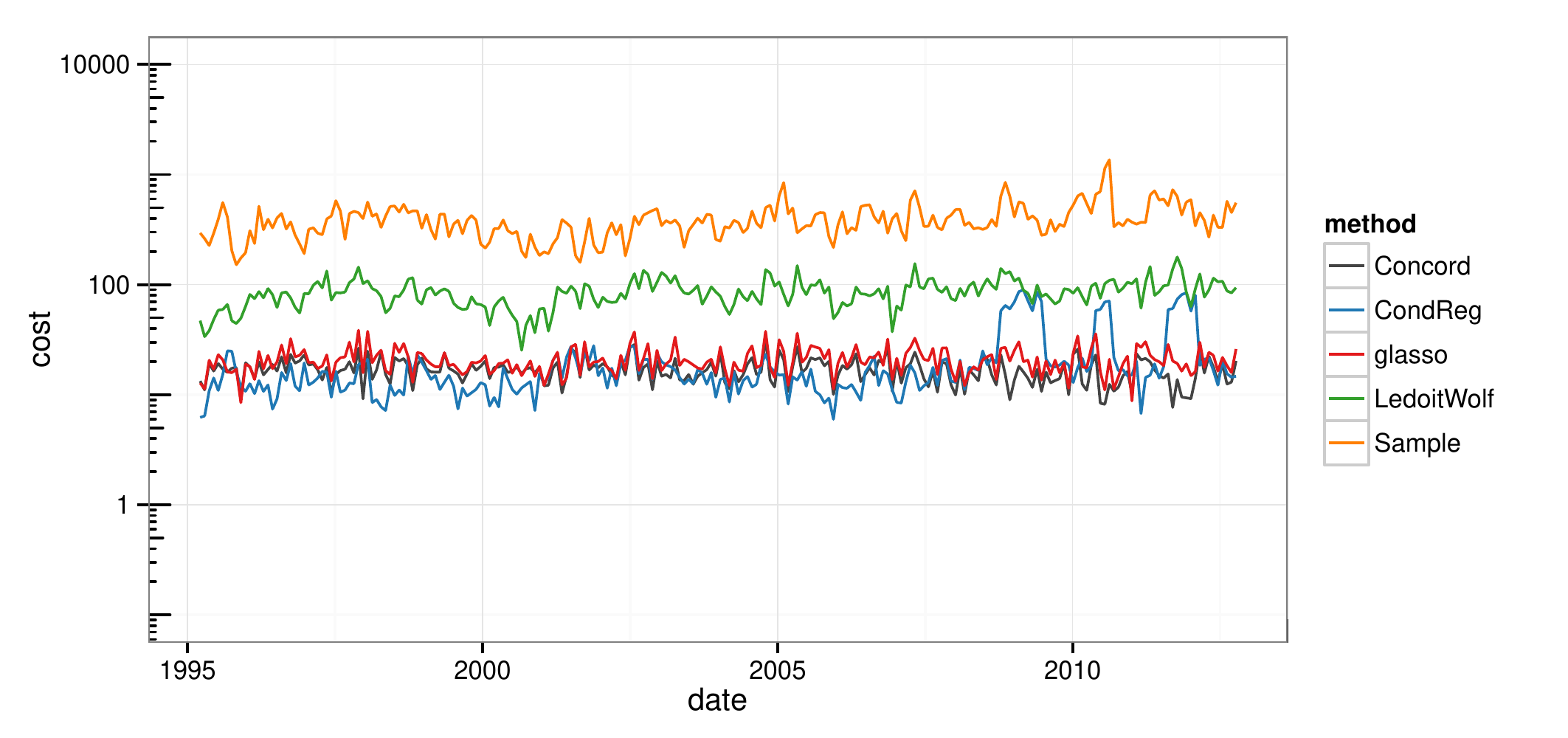}
  }
  \subfigure[ $N_{\est}= 45 $ ]{
    \includegraphics[trim = 0mm 3mm 35mm
    5mm,clip,width=0.48\textwidth]{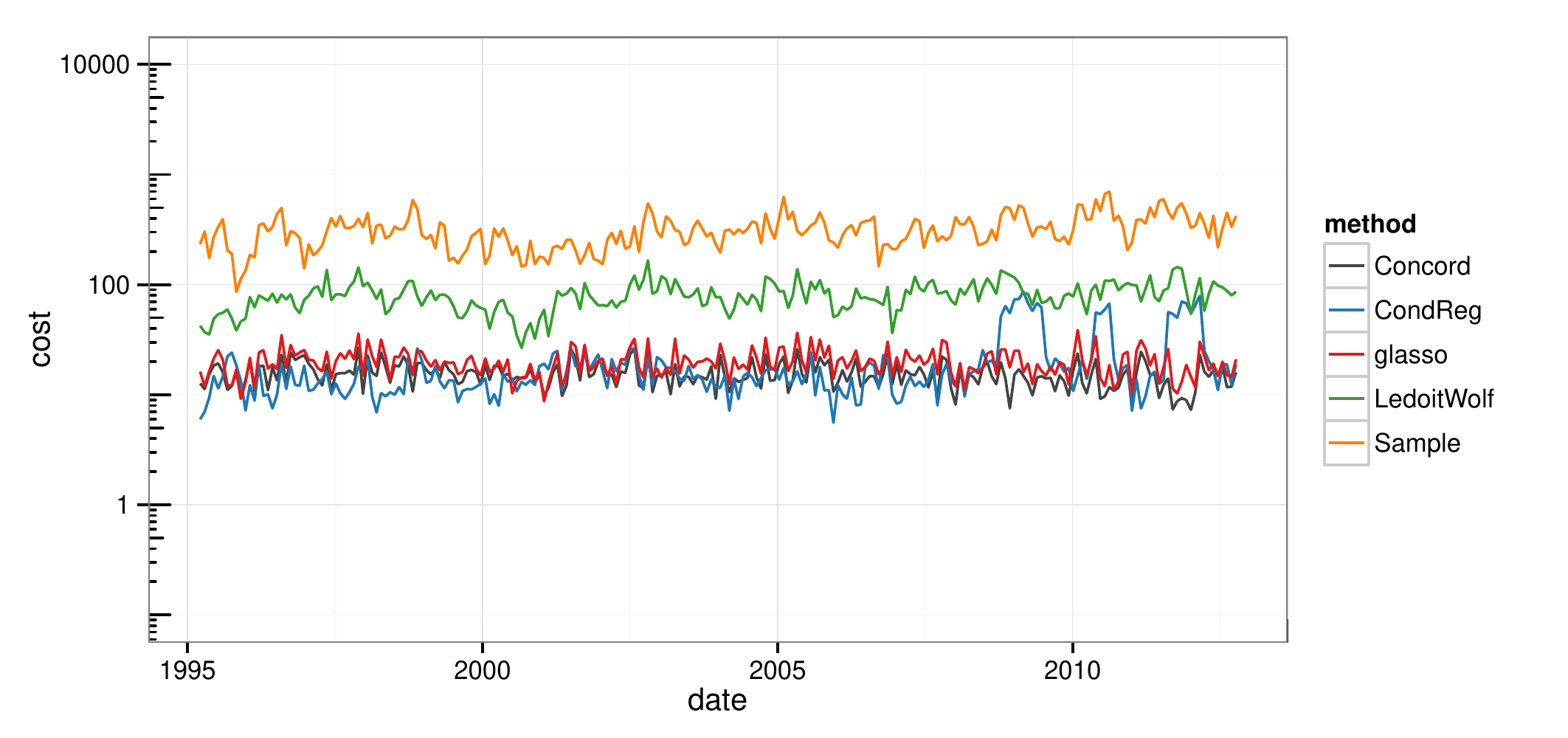}
  }\subfigure[ $N_{\est}= 50 $ ]{
    \includegraphics[trim = 8mm 0mm 10mm
    0mm,clip,width=0.51\textwidth]{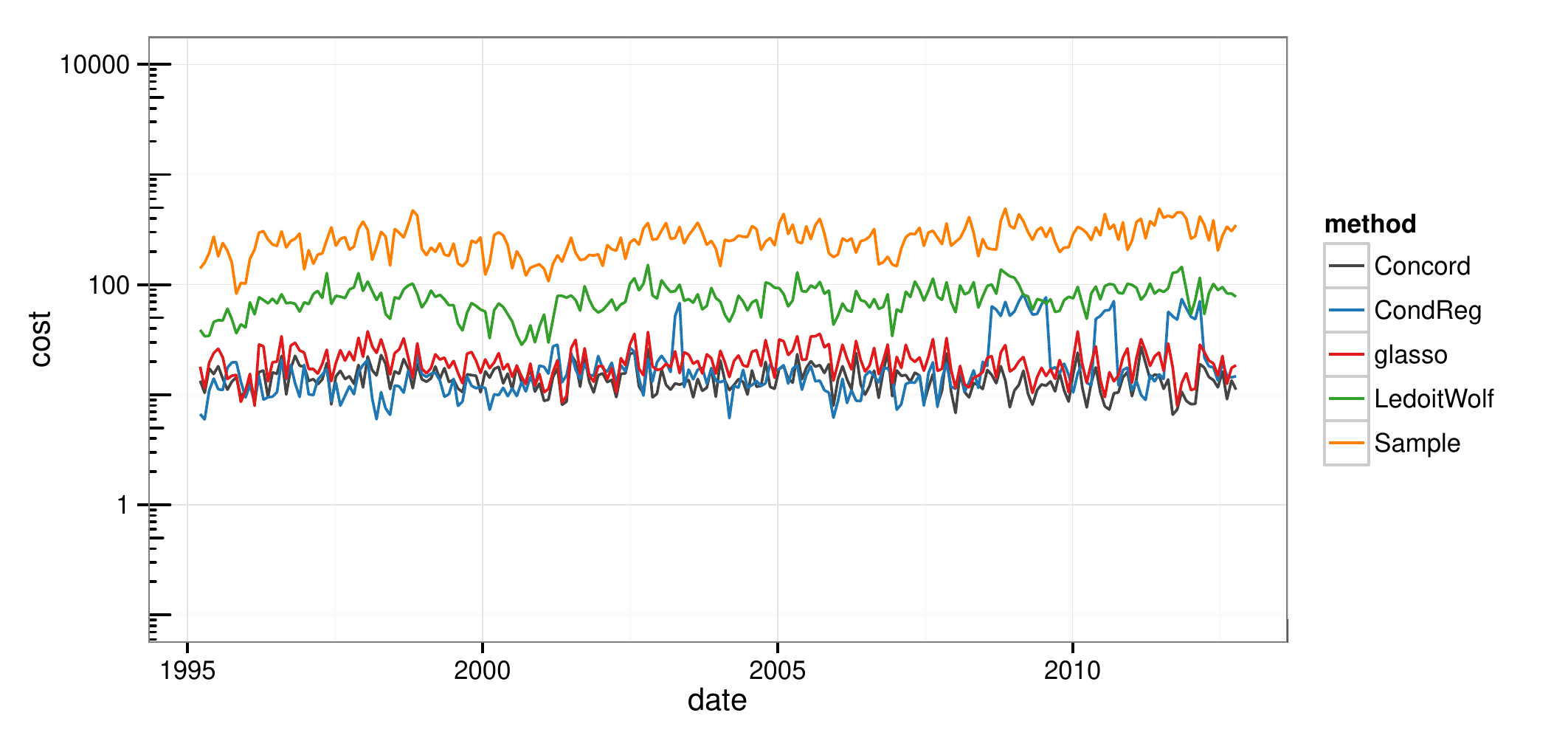}
  }
  \subfigure[ $N_{\est}= 75 $ ]{
    \includegraphics[trim = 0mm 3mm 35mm
    5mm,clip,width=0.48\textwidth]{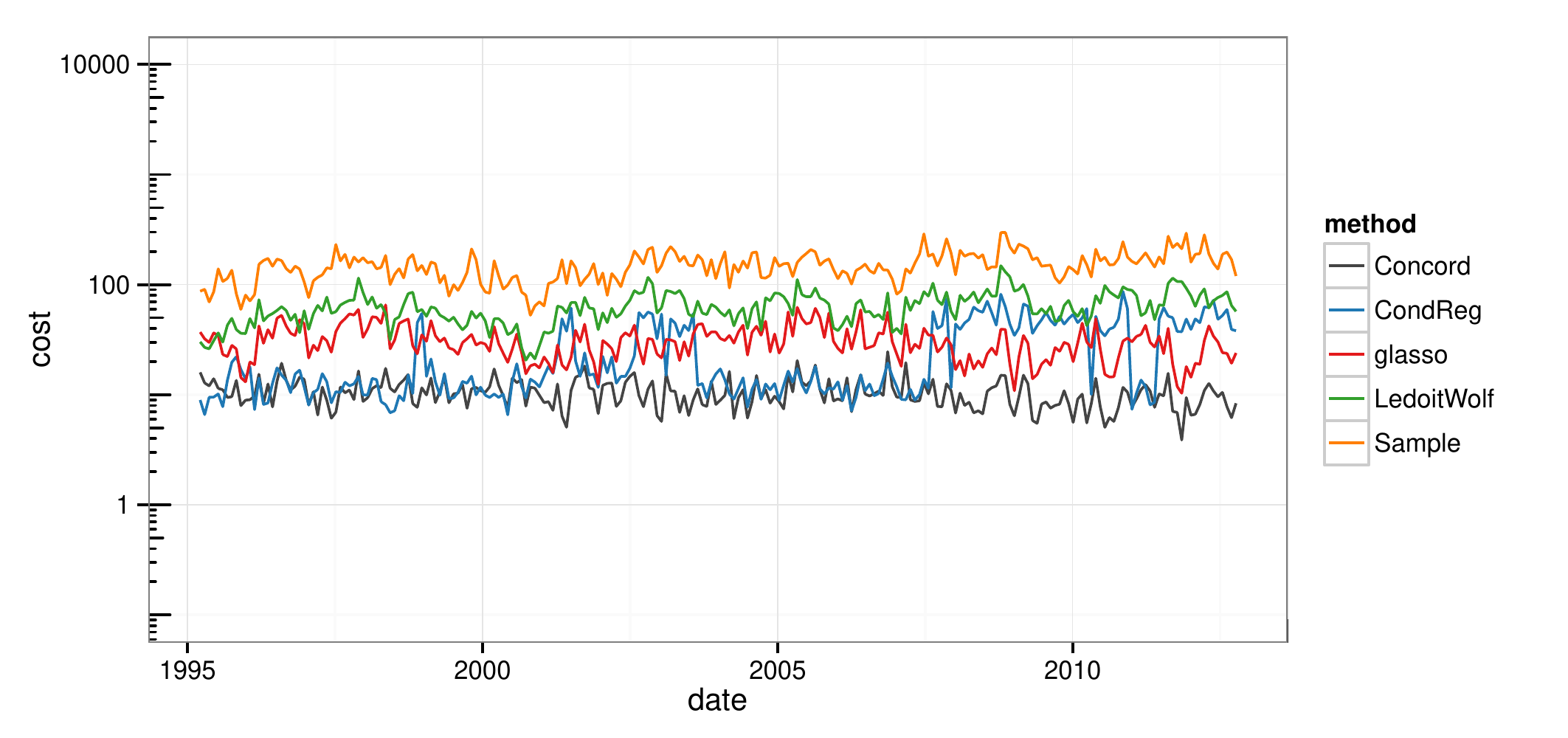}
  }\subfigure[ $N_{\est}= 150 $ ]{
    \includegraphics[trim = 8mm 0mm 10mm
    0mm,clip,width=0.51\textwidth]{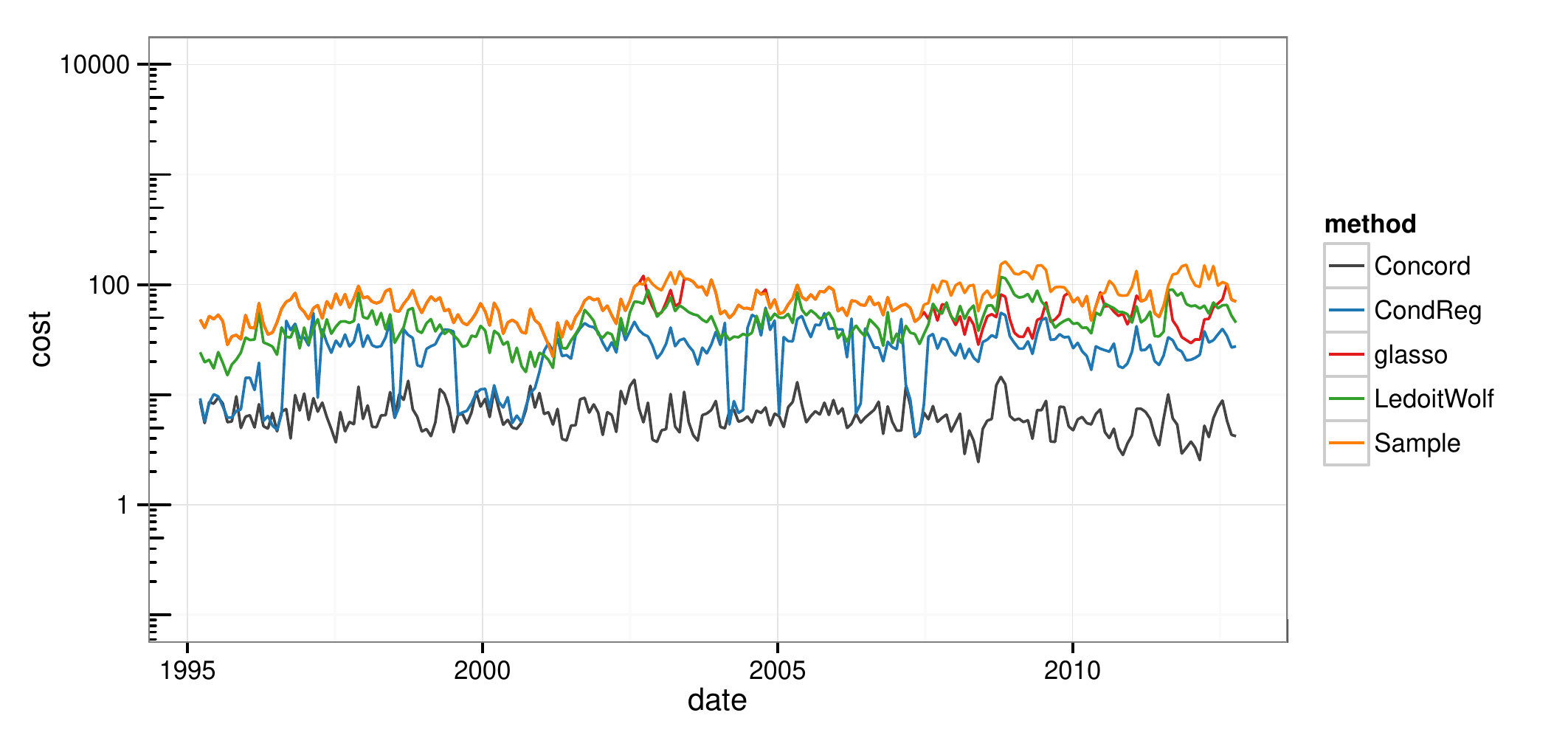}
  }
  \subfigure[ $N_{\est}= 225 $ ]{
    \includegraphics[trim = 0mm 3mm 35mm
    5mm,clip,width=0.48\textwidth]{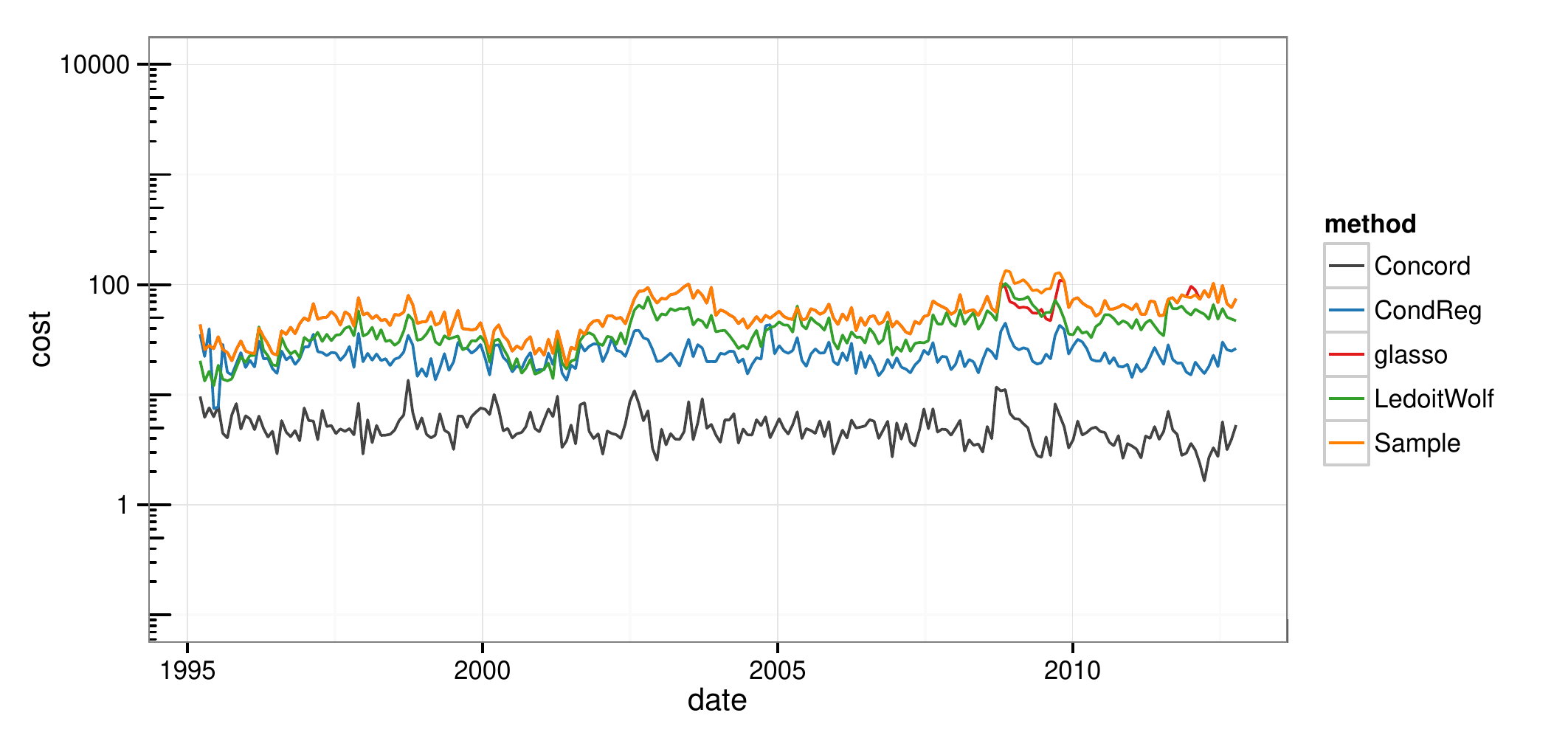}
  }\subfigure[ $N_{\est}= 300 $ ]{
    \includegraphics[trim = 8mm 0mm 10mm
    0mm,clip,width=0.51\textwidth]{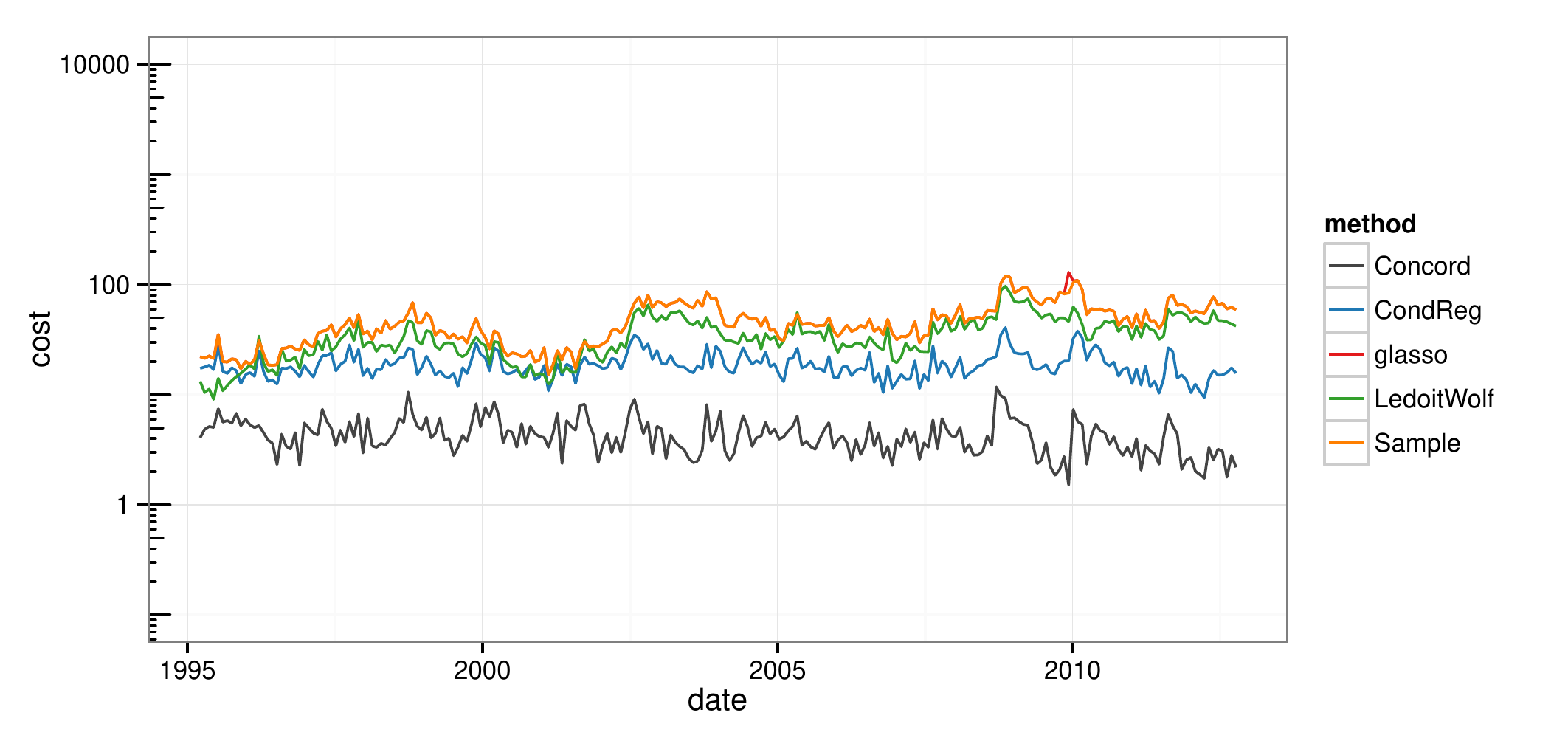}
  }
  \caption[Effect of trading costs on wealth growth]{Trading costs in
    basis points for each trading period. Borrowing rate is taken to
    be 7\% APR and transaction cost rate is taken to be 0.5\% APR. The
    y-axes are log-scaled.}
  \label{fig:trading costs}
\end{figure}

\section{Proof of Theorem \ref{thmconcord}} \label{sect:asymptotics}

\noindent
The result follows by noting the following straightforward facts 
\begin{enumerate}
\item The existence of a minimizer follows by the convexity of
  $Q_{\con}$.
\item By assumptions (A0) and (A1), for any $\eta > 0$,
  $\{\widehat{\alpha}_{n,ii}\}_{1 \leq i \leq p_n}$ are uniformly
  bounded away from zero and infinity with probability larger than $1
  - O(n^{-\eta})$.
\item When the diagonal entries are fixed at
  $\{\widehat{\alpha}_{n,ii}\}_{1 \leq i \leq p_n}$, then the
  objective function $Q_{\con}$ (reparameterized from $\omega^o$ to
  $\theta$) is same as the objective function of SPACE with weights
  $w_i = \widehat{\alpha}_{n,ii}^2$ (which are uniformly bounded),
  except that the penalty term is now \linebreak $\sum_{1 \leq i < j
    \leq p_n} \lambda_n \sqrt{\widehat{\alpha}_{n,ii}
    \widehat{\alpha}_{n,jj}} \theta_{ij}$, instead of $\sum_{1 \leq i
    < j \leq p_n} \lambda_n \theta_{ij}$ as in
  $Q_{\spc}$. 
\item Since $\bar\theta_{n,ij} =
  \frac{\bar\omega_{n,ij}}{\sqrt{\widehat{\alpha}_{n,ii}
      \widehat{\alpha}_{n,jj}}}$, using the uniform boundedness of
  $\{\widehat\alpha_{n,ii}\}_{1\leq i\leq p_n}$, there exists a
  constant $C_1$ such that for any $\eta > 0$,
  $$
  \| \widehat\omega_n^o - \bar{\omega}^o_n \|_2 \leq C_1 \|
  \widehat\theta^o_n - \bar{\theta}^o_n \|_2
  $$
  holds with probability larger than $1 - O(n^{-\eta})$.

\item For $1 \leq i < j \leq p_n$, $\sign(\hat\omega_{n,ij})
  = \sign(\hat\theta_{n,ij})$, since they differ by a positive
  multiplicative constant.

\item When the penalty term in SPACE is replaced by $\sum_{1 \leq i <
    j \leq p_n} \lambda_n \sqrt{\widehat{\alpha}_{n,ii}
    \widehat{\alpha}_{n,jj}} \theta_{ij}$, the uniform boundedness of
  $\{\widehat{\alpha}_{n,ii}\}_{1 \leq i \leq p_n}$ implies that
  Theorems $1$, $2$ and $3$ of \cite{Peng2009} hold with trivial
  modifications at appropriate places. The result now follows
  immediately using these theorems along with the above
  assertions. \hfill$\Box$
\end{enumerate}

\noindent
{\emph Remark:} Note that Theorem 2 on the consistency of
  CONCORD has been formulated as to exactly parallel the result
  given for SPACE by \cite{Peng2009}. An accurate estimator of
  $\bar\omega_{ii}$ when $p_n > n$ can be obtained by using the
  inverse of the sample conditional variance of each variable. In
  practice, however, once can simply use the diagonal estimates given by
  CONCORD, and there is no need for recourse to external
  estimates. Note also that CONCORD estimates themselves always exist,
  regardless of the sample size, and with certainty will lead to
  estimates, even when $p_n > n$. This property follows directly from
  the convergence of the CONCORD algorithm. 

\section{Joint convexity of the SYMLASSO in the $\Omega$ parameterization} 
\label{sect:symlasso:convexity:omega}

\noindent
We will show that the SYMLASSO objective function in
(\ref{eq:symlassoobjective}) is jointly convex if we reparameterize in
terms of $\Omega$ (see also \citesec{leehstmdml}). However, the SYMLASSO
objective function is not in general strictly convex if $n < p$, and
hence the convergence of the coordinatewise descent algorithm is not
guaranteed. It follows from the proof of Lemma \ref{lemma:unify} that
the SYMLASSO objective function (in terms of $\Omega$) is given by
\begin{eqnarray*}
Q_{\sym} (\Omega) 
&=& \frac{n}{2} \left[ -\log |\Omega_D| + tr(S \Omega \Omega_D^{-1} \Omega)
\right] + \lambda \sum_{1 \leq i < j \leq p} |\omega_{ij}|\\
&=& \frac{n}{2} \left[ - \sum_{i=1}^p \log \omega_{ii} + \frac{1}{\omega_{ii}} 
\omega_{i\bullet}^T S \omega_{i \bullet} \right] + \lambda \sum_{1 \leq i < j \leq p} 
|\omega_{ij}|. 
\end{eqnarray*}

\noindent
To prove the convexity of $Q_{\sym} (\Omega)$, we first prove the following lemma. 
\begin{lemma}
Consider the function $f$ on $\mathbb{R}_+ \times \mathbb{R}^k$ defined by $f({\bf a}) = 
\frac{{\bf a}^T A {\bf a}}{a_1}$. If $A$ is positive semi-definite, then $f$ is a convex function. 
\end{lemma}

\noindent
{\it Proof} It follows by straightforward manipulations that 
\begin{equation} \label{symlassoc1}
f({\bf a}) = A_{11} a_1 + 2 \sum_{j=2}^{k+1} A_{1j} a_j + \frac{{\bf a}_{-1}^T A_{-1} 
{\bf a}_{-1}}{a_1}, 
\end{equation}

\noindent
where ${\bf a}_{-1} := (a_j)_{j=2}^{k+1}$ and $A_{-1}$ is the principle submatrix of $A$ obtained by 
excluding the first row and the first column. Since the first two terms above are clearly convex 
functions of ${\bf a}$, it suffices to prove that the third term $\frac{{\bf a}_{-1}^T A_{-1} 
{\bf a}_{-1}}{a_1}$ is a convex function of ${\bf a}$. Again, by straightforward manipulations, 
it follows that the Hessian matrix of this term is given by 
$$
H = \frac{2}{a_1^3} \left( \begin{matrix} 
{\bf a}_{-1}^T A_{-1} {\bf a}_{-1} & - (a_1 A_{-1} {\bf a}_{-1})^T \cr 
- a_1 A_{-1} {\bf a}_{-1} & a_1^2 A_{-1} 
\end{matrix} \right). 
$$

\noindent
Hence, for any ${\bf b} \in \mathbb{R}^{k+1}$ (with ${\bf b}_{-1} := (b_j)_{j=2}^{k+1}$), it follows 
that 
\begin{eqnarray}
& & {\bf b}^T H {\bf b} \nonumber\\
&=& \frac{2}{a_1^3} \left( b_1^2 {\bf a}_{-1}^T A_{-1} {\bf a}_{-1} - 2 b_1 a_1 {\bf b}_{-1}^T 
A_{-1} {\bf a}_{-1} + a_1^2 {\bf b}_{-1}^T A_{-1} {\bf b}_{-1} \right). \label{symlassoc2} 
\end{eqnarray}

\noindent
Since $A_{-1}$ is positive semi-definite, it follows that if ${\bf b}_{-1}^T A_{-1} {\bf b}_{-1} = 0$, 
then $A_{-1} {\bf b}_{-1} = 0$. In this case 
$$
{\bf b}^T H {\bf b} = \frac{2}{a_1^3} \left( b_1^2 {\bf a}_{-1}^T A_{-1} {\bf a}_{-1} \right) 
\geq 0. 
$$

\noindent
If ${\bf b}_{-1}^T A_{-1} {\bf b}_{-1} > 0$, then it follows by (\ref{symlassoc2}) that 
\begin{eqnarray*}
& & {\bf b}^T H {\bf b}\\
&=& \frac{2 b_1^2}{a_1^3} \left( {\bf a}_{-1}^T A_{-1} {\bf a}_{-1} - \frac{({\bf b}_{-1}^T A_{-1} 
{\bf a}_{-1})^2}{{\bf b}_{-1}^T A_{-1} {\bf b}_{-1}} \right) + \frac{2}{a_1^3} \left( 
a_1 \sqrt{{\bf b}_{-1}^T A_{-1} {\bf b}_{-1}} - b_1 \frac{{\bf b}_{-1}^T A_{-1} 
{\bf a}_{-1}}{\sqrt{{\bf b}_{-1}^T A_{-1} {\bf b}_{-1}}} \right)^2\\
&\geq& 0. 
\end{eqnarray*}

\noindent
The last statement follows by noting that $\left( {\bf a}_{-1}^T A_{-1} {\bf a}_{-1} \right) 
\left( {\bf b}_{-1}^T A_{-1} {\bf b}_{-1} \right) \geq ({\bf b}_{-1}^T A_{-1} 
{\bf a}_{-1})^2$ (using the positive semi-definiteness of $A_{-1}$ and the Cauchy-Schwarz 
inequality). Hence $H$ is a positive semi-definite matrix, which combined with (\ref{symlassoc1}) 
implies that $f$ is a convex function. \hfill$\Box$ 

\medskip

\noindent
It follows by the above lemma that $\frac{1}{\omega_{ii}} \omega_{i\bullet}^T S \omega_{i\bullet}$ is 
a convex function in $\omega_{i \bullet}$ (and hence $\Omega$) for every $1 \leq i \leq p$. Since $-\log x$ 
and $|x|$ are convex functions, it follows that $Q_{\sym} (\Omega)$ is a convex function. 

\section{Examples where the Incoherence condition (A3) is satisfied} \label{sect:incoherence}

\noindent
We now present two lemmas which outline settings where the Incoherence condition (A3) is 
satisfied. The first lemma shows that (A3) is satisfied if the true correlations are sufficiently 
small. This lemma can be regarded as a parallel result to \cite[Corollary 2]{Zhao2006}, 
which shows that the irrepresentable condition for lasso regression is satisfied if the entries 
of $\frac{1}{n} X_n^T X_n$ ($X_n$ being the regression design matrix) are bounded by 
$\frac{c}{2q_n - 1}$ for some $0 \leq c < 1$. 
\begin{lemma}
Let 
$$
d_n := \max_{1 \leq i \leq p_n} |\{j: \bar{\omega}_{n,ij} \neq 0\}|. 
$$

\noindent
The incoherence condition (A3) is satisfied if 
$$
\frac{|\bar{\Sigma}_{n,ij}|}{\sqrt{\bar{\Sigma}_{n,ii} \bar{\Sigma}_{n,jj}}} \leq \frac{\sqrt{2} \delta 
\lambda_{min}}{\sqrt{q_n d_n} \lambda_{max}}, 
$$

\noindent
for every $n \geq 1$ and $1 \leq i \neq j \leq p_n$. 
\end{lemma}

\noindent
{\it Proof:} It can be shown by straightforward algebraic manipulations that 
$$
\bar{\mathcal{L}}^{''}_{\mathcal{A}_n, \mathcal{A}_n} (\bar{\Omega}_n) = U_n^T V_n U_n, 
$$

\noindent
where $V_n$ is a $p_n$-block diagonal matrix with the $i^{th}$ diagonal block given by 
$\bar{\Sigma}_n$ without the $i^{th}$ row and column, and $U_n$ is an appropriate $p_n (p_n - 1) 
\times q_n$ orthogonal matrix with $0$ and $1$ elements. Each column of $U_n$ has exactly two 
$1$'s. Hence for any ${\bf x} \in \mathbb{R}^{q_n}$, it follows that ${\bf x}^T U_n^T U_n {\bf x} = 2 
{\bf x}^T {\bf x}$. It follows that the smallest eigenvalue of $U_n^T V_n U_n$ is bounded below by 
$\frac{2}{\lambda_{max}}$. Consequently, the largest eigenvalue of $(U_n^T V_n U_n)^{-1}$ is bounded 
above by $\frac{\lambda_{max}}{2}$. 

Since the diagonal entries of $\bar{\Sigma}_n$ are uniformly bounded above by $\frac{1}
{\lambda_{min}}$, it follows 
that 
$$
|\bar{\Sigma}_{n,kl}| \leq \frac{\sqrt{2} \delta}{\sqrt{q_n d_n} \lambda_{max}}, 
$$

\noindent
for every $n \geq 1$ and $1 \leq k \neq l \leq p_n$. Note that for every $(i,j) \notin \mathcal{A}_n$,  
$\bar{\mathcal{L}}^{''}_{ij,\mathcal{A}_n}(\bar{\Omega}_n)$ has at most $2 d_n$ non-zero entries. 
Hence, we get that 
$$
\left\| \bar{\mathcal{L}}^{''}_{ij,\mathcal{A}_n}(\bar{\Omega}_n) \right\| \leq \sqrt{2d_n}  \times 
\frac{\sqrt{2} \delta}{\sqrt{q_n d_n} \lambda_{max}} = \frac{2 \delta}{\sqrt{q_n} \lambda_{max}}. 
$$

\noindent
Finally, we note from the discussion above that 
\begin{eqnarray*}
& & \left| \bar{\mathcal{L}}^{''}_{ij,\mathcal{A}_n}(\bar{\Omega}_n) \left[ 
\bar{\mathcal{L}}^{''}_{\mathcal{A}_n, \mathcal{A}_n} (\bar{\Omega}_n) \right]^{-1} 
\sign(\bar{\omega}^o_{\mathcal{A}_n}) \right|\\
&\leq& \left\| \bar{\mathcal{L}}^{''}_{ij,\mathcal{A}_n}(\bar{\Omega}_n) \right\| 
\left\| \left[ \bar{\mathcal{L}}^{''}_{\mathcal{A}_n, \mathcal{A}_n} (\bar{\Omega}_n) 
\right]^{-1} \right\| \left\| \sign(\bar{\omega}^o_{\mathcal{A}_n})  \right\|\\
&\leq& \frac{2 \delta}{\sqrt{q_n} \lambda_{max}} \times \frac{\lambda_{max}}{2} \times 
\sqrt{q_n}\\
&=& \delta. 
\end{eqnarray*}

\noindent
Hence (A3) is satisfied. \hfill$\Box$ 

\medskip 

\noindent
The next lemma shows that the Incoherence condition (A3) holds if the true $\bar{\Omega}_n$'s 
are tridiagonal matrices satisfying some mild conditions. This lemma can be regarded as a parallel 
result to \cite[Corollary 3]{Zhao2006}. 
\begin{lemma}
Suppose that $\bar{\Omega}_n$ is a tridiagonal matrix with all diagonal entries equal to $1$ and the
non-zero off-diagonal entries equal to $\rho_n$, for every $n \geq 1$. If $\rho := \sup_n |\rho_n|$ 
satisfies 
$$
\frac{8 \rho}{(1 - \rho^2)(2 - \rho^4/2)} \leq \delta, 
$$

\noindent
then (A3) is satisfied. 
\end{lemma}

\noindent
{\it Proof:} Using standard results for inverse of tridiagonal matrices, it follows that 
$$
\bar{\Sigma}_{n,ij} = \frac{\rho_n^{|i-j|}}{1 - \rho_n^2}, 
$$

\noindent
for every $1 \leq i,j \leq p_n$. Note that $\mathcal{A}_n = \{(i-1, i): \; 2 \leq i \leq p_n\}$, and 
$|\mathcal{A}_n| = p_n - 1$. Hence, $\bar{\mathcal{L}}^{''}_{\mathcal{A}_n, \mathcal{A}_n} 
(\bar{\Omega}_n)$ is a tridiagonal matrix (with the $i^{th}$ row corresponding to the edge 
$(i,i+1)$), with 
$$
\bar{\mathcal{L}}^{''}_{i(i+1), i(i+1)} (\bar{\Omega}_n) = \bar{\Sigma}_{n,ii} + 
\bar{\Sigma}_{n,(i+1)(i+1)} = \frac{2}{1 - \rho_n^2}, 
$$

\noindent
for every $1 \leq i \leq p_n - 1$, and 
$$
\bar{\mathcal{L}}^{''}_{i(i+1), (i+1)(i+2)} (\bar{\Omega}_n) = \bar{\Sigma}_{n,i(i+2)} = 
\frac{\rho_n^2}{1 - \rho_n^2}, 
$$

\noindent
for every $1 \leq i \leq p_n - 2$. Again, using standard results for inverse of tridiagonal 
matrices, it follows that 
$$
\left( \bar{\mathcal{L}}^{''}_{\mathcal{A}_n, \mathcal{A}_n} (\bar{\Omega}_n) \right)^{-1}_{i(i+1), 
j(j+1)} = \frac{(1 - \rho_n^2) (\rho_n^2/2)^{|i-j|}}{2 - \rho_n^4/2}, 
$$

\noindent
for every $1 \leq i,j \leq p_n - 1$. Using the fact that $\sum_{i=0}^\infty a^i = \frac{1}{1 - a}$ 
for $|a| < 1$, we conclude that each entry in $\left( \bar{\mathcal{L}}^{''}_{\mathcal{A}_n, 
\mathcal{A}_n} \right)^{-1} (\bar{\Omega}_n) \sign(\bar{\omega}^o_{\mathcal{A}_n})$ is bounded 
above in absolute value by $\frac{2}{2 - \rho_n^4/2}$. Moreover, if $i < j$ and 
$(i,j) \notin \mathcal{A}_n$, then $\bar{\mathcal{L}}^{''}_{ij,\mathcal{A}_n}(\bar{\Omega}_n)$ 
has at most four non-zero entries (entries corresponding to the edges $(i-1,i), (i,i+1), (j-1,j)$ and 
$(j,j+1)$, if applicable). All of these non-zero entries are bounded above in absolute value by 
$\frac{|\rho_n|}{1 - \rho_n^2}$. It follows that for every $(i,j) \notin \mathcal{A}_n$, 
\begin{eqnarray*}
& & \left| \bar{\mathcal{L}}^{''}_{ij,\mathcal{A}_n}(\bar{\Omega}_n) \left[ 
\bar{\mathcal{L}}^{''}_{\mathcal{A}_n, \mathcal{A}_n} (\bar{\Omega}_n) \right]^{-1} 
\sign(\bar{\omega}^o_{\mathcal{A}_n}) \right|\\
&\leq& \frac{4 |\rho_n|}{1 - \rho_n^2} \times \frac{2}{2 - \rho_n^4/2}\\
&=& \frac{8 |\rho_n|}{(1 - \rho_n^2)(2 - \rho_n^4/2)}\\
&\leq& \frac{8 |\rho|}{(1 - \rho^2)(2 - \rho^4/2)}\\
&\leq& \delta. 
\end{eqnarray*}

\noindent
Hence (A3) is satisfied. \hfill$\Box$ 

\section{Non-convergence of SPACE}
\label{sect:nonconvergence of space}

\noindent
We provide a simple example where the SPACE algorithm (with uniform 
weights) does not converge, and the iterates alternate between two 
matrices.  A sample of $n = 4$ \emph{i.i.d.} vectors was generated 
from the $\mathcal{N} ({\bf 0}, \Sigma)$ distribution with $\Sigma$ as 
in \eqref{eq:sigmadef}. The standardized data is as follows:
\begin{align}
  \begin{pmatrix}[r]
    0.659253 & -0.635923 & 0.492419 \\ 
    0.994414 & -1.015863 & 1.115863 \\ 
    -1.150266 & 1.141668 & -1.135115 \\ 
    -0.503401 & 0.510117 & -0.473166 \\ 
  \end{pmatrix}.
\end{align}
The SPACE algorithm was implemented with choice of weights $w_i = 1$
and $\lambda = 0.2$. Again, after the first few iterations, it turns out that 
successive SPACE iterates alternate between
$$
\left( \begin{matrix} 
    1.432570 & 1.416740 & -2.132500 \\ 
    1.416740 & 3552.598070 & 0.000000 \\ 
    -2.132500 & 0.000000 & 89.163310 \\ 
  \end{matrix} 
\right) \mbox{ and } 
\left( \begin{matrix} 
    3552.565950 & 1.416720 & 0.000000 \\ 
    1.416720 & 1.404240 & 2.100770 \\ 
    0.000000 & 2.100770 & 123.137260 \\ 
  \end{matrix} 
\right), 
$$

\noindent
thereby also establishing non-convergence of the SPACE algorithm in
the case when the weights $w_i=1$. Note that some of the elements in
the two matrices above are vastly different. The sparsity pattern is also 
different, thereby yielding two different partial correlation graphs.

\newpage

\bibliographystylesec{apalike}
\bibliographysec{library}


\end{document}